\documentclass[aps,amsmath,amssymb,nofootinbib]{revtex4} 
\usepackage{graphicx} 
\usepackage{amsmath}
\usepackage{amsfonts,amsbsy}
\usepackage{amssymb}
\usepackage{color}
\usepackage{mathtools}

\usepackage{epsfig,epsf}
\usepackage{graphicx,color}

\def\be{\begin{equation}}
\def\ee{\end{equation}}
\def\bea{\begin{eqnarray}}
\def\eea{\end{eqnarray}}

\newcommand{\CGCket}{|{\rm CGC}\rangle}
\newcommand{\CGCbra}{\langle {\rm CGC}|}

\begin{document}

\title{Exploring correlations in the  CGC wave function: 
	odd azimuthal anisotropy.
}

\author{Alex Kovner}
\affiliation{Physics Department, University of Connecticut, 2152 Hillside Road, Storrs, CT 06269, USA}

\author{Michael Lublinsky}
\affiliation{Physics Department, Ben-Gurion University of the Negev, Beer Sheva 84105, Israel}

\author{Vladimir Skokov}
\affiliation{RIKEN/BNL, Brookhaven National Laboratory, Upton, NY 11973}

\begin{abstract}
We extend the CGC approach to calculation of the double inclusive gluon
production by including high density effect in the CGC wave function of the
projectile (proton). Our main result is that these effects lead to the
appearance of odd harmonics in the two particle correlation $C(k,p)$. 
We find that in the high momentum limit, $|k|,|p|\gg Q_s$, this results in a
positive $c_1\{2\}$. Additionally when the magnitudes of the two momenta are
approximately equal, $|k|/|p|\approx 1$,  the density effects also generate a
positive third harmonic $c_3\{2\}$, which translates into a non-vanishing $v_3$
when the momenta of the trigger and associated particle are in the same momentum
bin. The sign of $c_3\{2\}$ becomes negative when $|k|/|p|>1.1$ suggesting an
interesting experimental signature.
\end{abstract}

\maketitle
\section {Introduction.}

Currently one of the outstanding questions of strong interactions is the origin of
long range rapidity correlations observed in p-p and p-Pb collisions at LHC.
Starting with the observation of the so-called ridge correlations in high
multiplicity p-p collisions by CMS~\cite{Khachatryan:2010gv}, 
this phenomenon generated a lot of interest. Subsequent observation of the same
effect in p-Pb collisions by all three big experiments at LHC sharpened the
questions even further~\cite{Abelev:2012ola,Abelev:2014mda,Aad:2012gla,Aad:2013fja,ATLAS:2014dha,CMS:2012qk,Chatrchyan:2013nka}. In particular the observation that the triangular ``flow''
coefficient $v_3$ is practically identical in p-Pb and Pb-Pb collisions at the
same total multiplicity strongly suggests the  origin of the correlations is
due to some collective or quasi-collective behavior. The recent experimental
results are even more surprising~\cite{Aad:2015gqa,Aaboud:2016yar}; with an improved subtraction of
hard component, the ridge type correlations are now seen  in the minimal bias
p-p data,  and even in events with lower than average multiplicity.

One possibility is that the collectivity is the result of strong final state
interactions, as is suggested by good hydrodynamics  fits to the data~\cite{Werner:2013ipa,Bozek:2013ska,Bozek:2014wpa,Kozlov:2014fqa}. This
is surprising in view of the fact that the correlations are observed up to
relatively high transverse momenta $k\sim 10$ GeV, and also in events where
the number of produced particles is small.
The transport model results presented in Ref.~\cite{Bzdak:2014dia} show that
final state interactions  even  with modest parton-parton 
cross section describe well some aspects of the data. 
Another possibility is that the
correlated features of production are due to preexisting correlations in the
wave function of the colliding hadrons.  This last possibility has been
intensively studied during the last several years in the framework of the Color
Glass
Condensate~\cite{Dumitru:2010iy,Levin:2010dw,Kovner:2010xk,Kovner:2012jm,
Kovchegov:2012nd,Dusling:2012iga,Dusling:2013qoz,Dumitru:2014dra}.

The CGC-based calculations have successfully described the ``ridge''
data~\cite{Dusling:2012iga,Dusling:2013qoz}. Nevertheless the CGC approach faces
serious challenges in describing other aspects of data. Perhaps the most
challenging aspect is description of multi-particle correlations, in particular
$v_2\{4\}$~\footnote{Here we follow the notation of Ref.~\cite{Borghini:2001vi}: 
\begin{align}
	\notag c_n\{2\} \equiv v_n^2\{2\} &=  \langle \exp\left[i n  (\phi_1 - \phi_2) \right] \rangle ,\\ 
	\notag c_n\{4\} \equiv - v_n^4\{4\} &= 
	\langle \exp\left[i n (\phi_1 +\phi_2 - \phi_3 -\phi_4) \right] \rangle 
	-2 
	\langle \exp\left[i n (\phi_1 - \phi_3) \right] \rangle 
	\langle \exp\left[i n (\phi_2 - \phi_4) \right] \rangle . 
\end{align}
}. The mechanism utilized in the numerical work of 
Refs.~\cite{Dusling:2012iga,Dusling:2013qoz} is
not capable of producing this quantity as has been shown in
Refs.~\cite{Dumitru:2014yza,Skokov:2014tka}. This is
currently an open question, and we do not have anything to add to this part of
discussion.

Another 
enduring problem 
of the CGC-based approaches to correlations has
been a non-vanishing value of the triangular ``flow'' coefficient $v_3$ clearly
observed in the data. All the approaches in Refs.~\cite{Dumitru:2010iy,Levin:2010dw,Kovner:2010xk,Kovner:2012jm,Kovchegov:2012nd,Dusling:2012iga,Dusling:2013qoz,Dumitru:2014dra}
predict that the
double inclusive production is symmetric under reversal of the direction of one of
the transverse momenta~\footnote{
Here and thereafter, in order to simplify the notation, 
we denote the transverse two-dimensional vectors by $k$, that is 
$k \equiv \vec{k}_\perp$  and  $|k| \equiv |\vec{k}_\perp|$. For the scalar
and the cross 
products we use $k\cdot q \equiv k_i q_i$ and $k\times q = \epsilon_{ij} k_i q_j$
correspondingly, 
where $1\le i,j\le 2 $.   
}
\begin{equation}
	\label{Eq:dilute_xsec}
\sigma(k,p)=\sigma(k,-p)\;
\end{equation}
even on the configuration-by-configuration basis. 
Such symmetry precludes existence of odd harmonics. Although this is not a
fundamental symmetry of QCD in any way, it proved to be very stubborn and
difficult to avoid in CGC without including final state interactions. It was
recently shown analytically  in Ref.~\cite{McLerran:2016snu} that odd harmonics are indeed
generated in the double inclusive production when classical evolution of the
Yang-Mills fields in the final state is accounted for. This conclusion is
consistent with earlier numerical work~\cite{Schenke:2015aqa}, although in the
latter case there is some uncertainty as to whether the results reflect
evolution to asymptotically large 
times.~\footnote{
We mention  recent papers~\cite{Gotsman:2016fee,Gotsman:2016wtq,Gotsman:2017zoq}, 
which purportedly obtain nonvanishing $v_3$ without invoking final state interactions. We failed to 
understand the theoretical framework of Ref.~\cite{Gotsman:2016fee} and thus cannot comment on
it.  On the other hand 
 Ref.~\cite{Gotsman:2016wtq} 
 has  focused on production of identical gluons (same color and same polarization). It thus did not include some of the  graphs which generally contribute to total double inclusive production.  In particular it does not  include a subset of  graphs 
with the amplitude in a color singlet state near the produced gluon rapidity
(which  appear as  diffractive graphs within the rapidity bin that includes the
two observed gluons, see Ref.~\cite{Dumitru:2008wn}). 
It is well known that within the glasma graph calculations these ``diffractive'' contributions generate
a correlation at $\Delta\phi=\pi$ identical in strength to that generated by
``non-diffractive'' graphs at $\Delta\phi=0$, see for example
Ref.~\cite{Dumitru:2008wn}. Thus in the framework of the glasma graphs
inclusion of these additional contributions renders all odd harmonics
to vanish.  { After our present manuscript was published in its preprint form, the authors of Ref.~\cite{Gotsman:2016fee,Gotsman:2016wtq}
released a new paper \cite{Gotsman:2017zoq}, which properly accounts for the  ``diffractive'' contributions. 
Ref.~\cite{Gotsman:2017zoq} argues that 
Sudakov
suppression leads to appearance of odd harmonics.}
}

The purpose of the present paper is to  point out that in fact final state
interactions are not essential to generate odd harmonics within the CGC
approach. Our central point is that all the CGC-based calculations so far have
used (some implicitly) the form of the CGC wave function which is appropriate
for description of dilute projectile only. This ``dilute CGC'' wave function leads to
the accidental symmetry alluded to earlier. Corrections to this dilute limit on
the wave function level  have been calculated a while ago~\cite{Kovner:2007zu,Altinoluk:2009je}. In
particular it was shown in Refs.~\cite{Kovner:2007zu,Altinoluk:2009je} that
these corrections are essential to
reproduce the JIMWLK evolution equation (see
Refs.~\cite{JalilianMarian:1997jx,JalilianMarian:1997dw,Kovner:2000pt,Kovner:1999bj,Weigert:2000gi}
and \cite{Iancu:2000hn,Iancu:2001ad,Ferreiro:2001qy}) when the
evolution is generated by boosting the dense hadron. In the present paper we use
this improved CGC wave function to calculate the double inclusive gluon
distribution. 

We were inspired to consider this departure from the commonly employed dilute
CGC state by an old work by Kharzeev, Levin and McLerran (KLM) \cite{Kharzeev:2004bw}. The
idea suggested in Ref.~\cite{Kharzeev:2004bw} is that a wave function of a dense object can
naturally incorporate nontrivial correlations. Consider a fluctuation in such a
wave function which contains a high $p$ parton. The transverse momentum of
this parton has to be balanced in this component of a wave function. In a dilute
system it is most likely to be balanced by another hard  parton with momentum
$-p$. However in a dense environment it is more likely that  the balancing
transverse momentum is shared by several semi hard partons. Although the
original suggestion in Ref.~\cite{Kharzeev:2004bw}  was that the transverse momentum is
distributed between semi hard partons at different rapidities, the extent of
this rapidity spread may be not too large. The momentum distribution in such a
component of the  wave function looks somewhat like a directed flow, with the
direction defined by the momentum of the hard fluctuation. If the number of the
balancing semi hard partons is large enough the two particle correlation
function should exhibit a maximum for same sign transverse momenta. This picture
suggests a positive $c_1\{2\}$, and possibly a non-vanishing $c_3\{2\}$ in this
type of components of a dense wave function. In fact with ``directional flow'' of
the kind described above one may expect a non-vanishing contribution to
$c_2\{4\}$ as well, although this effect will be presumably smaller. Of course
it is a quantitative question whether these signatures survive the averaging
over all components of the wave function, and later the production process, and
how large the net effect is. 

In any case, it appears that the dilute CGC wave function does not encode the
KLM-type correlations. It is not all that surprising, given that the effect
requires the partonic system to be dense enough. Hence the motivation to
consider high density corrections to the dilute CGC limit.
 
We find that when the improved  wave function is used to calculate the double
inclusive gluon production the accidental symmetry present in the dilute limit
disappears. This leads to appearance of the odd harmonics in the two particle
correlation function. We find (as in Ref.~\cite{McLerran:2016snu}) that
the odd harmonics ($c_1\{2\}$ and $c_3\{2\}$) are
suppressed relative to the even (e.g. $c_2\{2\}$) parametrically by a
factor of $\alpha_s$. This is roughly consistent with the experimentally
observed hierarchy between $v_2$ and $v_3$. We calculate $c_1\{2\}$ and
$c_3\{2\}$ in the limit of high transverse momenta.

The paper is structured as follows. In Sec.~2 we describe the improved CGC wave
function, and calculate the antisymmetric (in $(k,p)\rightarrow (k,-p)$) piece
in the gluon pair density in the wave function. We show that this piece does not
vanish. The sign of the first harmonic is positive, while the sign of the third
harmonic is negative.
In Sec.~3 we calculate the double inclusive gluon production for eikonal
scattering on a target. We then expand the general expressions in the limit of
large transverse momentum of produced particles. Interestingly, we find that to
leading power in $Q_s^2/p^2$ the antisymmetric part of the correlation function
is saturated by a charge conjugation odd ``condensate'', i.e. the Odderon. Since
the contribution of the Odderon is expected to be subleading at high energy, we
calculate the next order in the expansion, and estimate it using a simple model
for target averaging. We find that    $c_1\{2\}$ is positive, while $c_3\{2\}$
is negative when $|k|/|p| >1.1$ and  $|p|/|k| >1.1$, but is positive for
$1.1>|k|/|p|>1/1.1$. 

We close by discussing our results and their implications in Sec.~4.

\section{Gluon pair density in the CGC wave function.}
\subsection{The CGC wave function.}
We start with discussing the wave function of the vacuum of the soft gluons in
the background of strong valence color charge density.

 The CGC calculations so far have relied on a simplified version of this  wave
 function valid in the situation when the color charge density is weak and
 perturbative. 
In this regime one can diagonalize the QCD Hamiltonian pertubatively with the
result that the vacuum of the soft modes is a coherent state of the form
\cite{Kovner:2005nq}
\begin{equation}\label{dilute}
	|{\rm CGC}\rangle_{\rm dilute}=
	{\cal C}
	|0\rangle\;,
\end{equation}
where ${\cal C}$ is the displacement operator defined as 
\begin{equation}
	{\cal C} = e^{i \sqrt{2} \int_k  b_{\alpha i }(-k) \left[ a^\dagger_{\alpha i } (k) \ +\  a_{\alpha i } (-k)  \right]  }.
	\label{Eq:D}
\end{equation}
Here $|0\rangle$ is the light cone vacuum of the soft modes, and the soft gluon creation and
annihilation operators represent the rapidity independent (rapidity averaged)
mode of the soft gluon field over the soft rapidity interval, that is 
\begin{equation}
 a_{\alpha i } (k)\equiv\frac{1}{\sqrt{Y}}\int \frac{d\eta}{2\pi}a_{\alpha i}(\eta, k)\;.
\end{equation}
The Weizs\"acker-Williams field $b_{\alpha i}$~\footnote{The Fourier transform of the classical field is defined as
	$b^\alpha_i(k)=\int d^2x e^{-ikx}b^\alpha_i(x)$.}
is generated by valence color charges of a hadronic projectile  
\begin{equation}
	\partial_i  b_{\alpha i} (x) = \rho_\alpha (x).
	\label{b}
\end{equation}
The field is two-dimensional pure-gauge:
$$
b_{\alpha i} (x) = -\frac{1}{g} f_{\alpha \beta \delta} U^+_{\beta \gamma} (x) \partial_i U_{ \gamma\delta}(x), 
$$
where $U$ is a SU(3) element in the adjoint representation.

It has been known for a while that Eq.~(\ref{dilute})  is not an
appropriate wave function for a dense system. In particular it was shown by
direct calculation in Refs.~\cite{Kovner:2007zu,Altinoluk:2009je} that in the soft
gluon wave function the coherent state as in Eq.~(\ref{dilute}) is accompanied by
a  Gaussian factor, so that the state is a coherent Bogoliubov (squeezed) state. The
squeezing  is crucial to reproduce the JIMWLK evolution equation
of a dense object at high energy, see
Refs.~\cite{JalilianMarian:1997jx,JalilianMarian:1997dw,Kovner:2000pt,Kovner:1999bj,Weigert:2000gi} and \cite{Iancu:2000hn,Iancu:2001ad,Ferreiro:2001qy}.

 The original presentation in Refs.~\cite{Kovner:2007zu,Altinoluk:2009je} is
 quite complicated. It involves calculation of the wave function of all soft
 gluon rapidity modes in the soft rapidity interval. On the other hand it is
 clear that only the rapidity independent mode is large, and it is this mode
 that is responsible for high energy evolution. This is  obvious in
 Eq.~(\ref{dilute}). One therefore expects that it should be possible to
 ``integrate out'' all the modes of the gluon field perturbatively  except for the
 rapidity independent one. 

Such a first principle calculation has not been done. However there is a simplified
way to infer the reduced wave function of the rapidity integrated gluon mode
using the results of Refs.~\cite{Kovner:2007zu,Altinoluk:2009je}. 
The important point is that the general wave function derived in
Refs.~\cite{Kovner:2007zu,Altinoluk:2009je} is Gaussian and integrating
over part of the field modes is bound to lead to a Gaussian shape of the
reduced wave function as well. Therefore in order to find the reduced wave
function we  ask what is the Gaussian wave function that depends only on
the rapidity integrated field and which reproduces the JIMWLK evolution
equation. The answer is that the following reduced wave function fits the bill~\footnote{
	We reiterate that to tie loose ends it would be desirable to derive this
	wave function directly from the results of Refs.~\cite{Kovner:2007zu,Altinoluk:2009je} by integrating out the rapidity
dependent modes of the soft gluon field. In the present paper we will not
undertake this endeavour  but will rather content ourselves with the simplified
argument given above.}
\begin{equation}
	\Psi_{\rm CGC}[\phi] = 
\langle \phi |  \Omega	|0\rangle
=
	\,e^{i 2 \int_k  b_{\alpha i }(-k) 
	\phi_{\alpha i} (k) } 
	\langle \phi |  {\cal B}\rangle\;,
	\label{Eq:CGC}
\end{equation}
where $\Omega = {\cal C B}$ is a unitary operator 
and
for convenience we separately defined  the Gaussian state $|{\cal B}\rangle$ 
as~\footnote{Using the relation~\eqref{Eq:PhiPi_1}
one can recognise $\langle \phi | {\cal C}|0\rangle$ in the prefactor in  Eq.~\eqref{Eq:CGC}.}  
\begin{equation}
	 |  {\cal B}\rangle\equiv  {\cal B}|0\rangle;\ \ \ \ \ \ \ \   \langle \phi |  {\cal B}\rangle
	={\cal N}
	e^{-
	\frac12 \int_{k, p}  B^{-1}_{\alpha \beta i j} (k, p ) 
	\phi_{\alpha i}(k)
	\phi_{\beta j}(p)
	}\;.
	\label{Eq:CalB}
\end{equation}
Here the field $\phi$ is defined as
\begin{equation}\label{field}
\phi_{i\alpha} (k)=\frac{1}{\sqrt{2}} \left(  a_{i \alpha}^\dagger(k)+ a_{i \alpha}(- k) \right), \\
\end{equation}

The operator ${\cal B}$ is a unitary operator,  exponential of a quadratic function of the creation and annihilation operators $a(k)$ and $a^\dagger(k)$. We do not write it out explicitly as the knowledge of the wave function in field space is sufficient for our purposes.

The constant ${\cal N}$ is the normalization factor and the operator $B$  is given by
\begin{equation}
	B=(1-\mathfrak{l}-L)^2 = 1 - \mathfrak{l} - L + [\mathfrak{l},L]_+\;,
\label{B}
\end{equation}
where the longitudinal projector  in coordinate space is  
\begin{equation}
\mathfrak{l}_{ij}^{\alpha\beta}(x,y)\equiv \delta^{\alpha\beta}\frac{\partial_i\partial_j}{\partial^2}(x,y)\label{B0}
\end{equation}
and 
\begin{equation}
L_{ij}^{\alpha\beta}(x,y)=U^{\alpha\gamma}(x)\frac{\partial_i\partial_j}{\partial^2}(x,y)U^{\dagger\,\gamma\beta}(y)=D_i^{\alpha\gamma}\left[\frac{1}{D^2}\right]^{\gamma\lambda}D_j^{\lambda\beta}\;.
\label{B1}
\end{equation}
Here the covariant derivative is given by
\begin{equation}
	D_i^{\alpha \beta} (x )  = \delta^{\alpha \beta} \partial_i  -g f^{\alpha \beta \gamma} b^\gamma_i (x ) \;.
\end{equation}

Note that the CGC state in Eq.~\eqref{Eq:CGC} is not specified in terms of the  light cone vacuum ket , but rather the wave function is written in the ``field representation'', meaning that it
is a function of the field $\phi(k)$.  
The calculation of
various expectation values is performed by a functional integration over the
field $\phi$.
In the weak field limit, where the eikonal factor is trivial $U(x)=1$, we have
{  $B(x-y)=\delta^2(x-y)$, or in momentum space $B(k,p)= (2\pi)^2 \delta^2(k+p)$}, and the
Gaussian state $| {\cal B} \rangle $ in Eq.~(\ref{Eq:CGC}) becomes the light cone vacuum state, 
see also Eq.~\eqref{dilute}.
In the dense regime $b\sim 1/g$ and the correction due to a
nontrivial squeezing parameter is an order one effect.

\subsection{The pair density}
Our first task is to see what is the effect of the nontrivial squeezing on the gluon pair density in the CGC wave function.

The single gluon number density in the CGC wave function is given by the following formal expression
\begin{equation}
	f(k) 
	=  
	{\cal N}  \int D \rho W_{\rm P} [\rho] \;  
	f_\rho(k) = 
	{\cal N}  \int D \rho W_{\rm P} [\rho] \;  
	\CGCbra a^\dagger_{i \alpha} (k)  a_{i \alpha} (k)   \CGCket\;,
	\label{Eq:f}
\end{equation}
where 
${\cal N}$ is the normalisation factor and we 
explicitly show the ensemble average with 
a weight functional $W_{\rm P}[\rho]$
characterising the distribution of the projectile color sources, $\rho$.  

Similarly, we define the gluon pair density as  
\begin{equation}
	f(k, p) 
	=    
	{\cal N}  \int D \rho W_{\rm P} [\rho] \;  
	f_\rho (k, p) 
	=
	{\cal N}  \int D \rho W_{\rm P} [\rho] \; 
	\CGCbra 
	a^\dagger_{i \alpha} (k)  
	a^\dagger_{j \beta} (p)  
	a_{i \alpha} (k)  
	a_{j \beta} (p)  
	\CGCket\;.
	\label{Eq:f2}
\end{equation}
 Our normalisation of creation and annihilation operators is such that 
\begin{equation}
[a_{i\alpha}(k),a^\dagger_{j\beta}(p)]=(2\pi)^2 \delta_{ij}\delta_{\alpha\beta}\delta^2(k-p)\;.
\end{equation}

We define the field variable and its conjugate momentum
\begin{eqnarray}
	\label{Eq:PhiPi_1}
	\phi_{i\alpha} (k)&=&\frac{1}{\sqrt{2}} \left(  a_{i \alpha}^\dagger(k)+ a_{i \alpha}(- k) \right), \nonumber\\
	\pi_{i\alpha} (k)&=&\frac{i}{\sqrt{2}} \left(   a_{i \alpha}^\dagger(k) -  a_{i \alpha}(- k) \right),
	\label{Eq:PhiPi}
\end{eqnarray}
so that
\begin{equation}  
	[\phi_{i\alpha}(k),\pi_{j\beta}(p)]=i(2\pi)^2\delta_{ij}\delta_{\alpha\beta}\delta^2(k+p)
\end{equation}
and the inverse 
\begin{eqnarray}
	a_{i \alpha} (k) &=&  \frac{1}{\sqrt{2}} \left( 	\phi_{i\alpha} (-k)  + i 	\pi_{i\alpha} (-k)   \right), \nonumber\\
	a^\dagger_{i \alpha} (k) &=&  \frac{1}{\sqrt{2}} \left( 	\phi_{i\alpha} (k)  - i 	\pi_{i\alpha} (k)   \right).
	\label{Eq:aa}
\end{eqnarray}

To calculate the expectation values in Eq.~\eqref{Eq:f2} we use the following properties of the 
displacement operator~\eqref{Eq:D}:
\begin{eqnarray}
	{\cal C}^\dagger  a_{\alpha i } (q) {\cal C} &=&  a_{\alpha i } (q) + i \sqrt{2}  b_{\alpha i } (-q), \nonumber\\
	{\cal C}^\dagger  a^\dagger_{\alpha i } (q) {\cal C} &=&  a^\dagger_{\alpha i } (q) - i \sqrt{2}  b_{\alpha i } (q).
	\label{Eq:Dact}
\end{eqnarray}


The expectation values in the Gaussian state $| {\cal B} \rangle$ 
 are
\begin{eqnarray}
	\langle \phi_{i \alpha} (k)  \phi_{j\beta} (k^\prime)  \rangle_{{\cal B}}  &=& \frac{1}{2} B_{\alpha \beta i j}(k, k^\prime),\nonumber \\
	\langle \phi_{i \alpha} (k)  \pi_{j\beta} (k^\prime)  \rangle_{{\cal B}}  &=&
	-i \langle \phi_{i \alpha} (k)  \frac{\delta}{\delta \phi_{j\beta} (-k^\prime) }   \rangle 
	= \frac{i}{2}  (2\pi)^2\delta_{ij} \delta_{\alpha \beta} \delta^2(k + k^\prime)  ,\nonumber \\
	\langle \pi_{i \alpha} (k)  \phi_{j\beta} (k^\prime)  \rangle_{{\cal B}}  &=& 
	\langle \phi_{j \beta} (k^\prime)  \pi_{i \alpha} (k)  \rangle_{{\cal B}}   -  \langle [\phi_{j \beta} (k^\prime)  ,\pi_{i \alpha} (k) ] \rangle  =  -  \frac{i}{2} (2\pi)^2\delta_{ij} \delta_{\alpha \beta} \delta^2(k + k^\prime)  ,\nonumber  \\
	\langle \pi_{i \alpha} (k)  \pi_{j\beta} (k^\prime)  \rangle_{{\cal B}}  &=& \frac{1}{2} B^{-1}_{\alpha \beta i j}(k, k^\prime).	\label{Eq:phiphi}
\end{eqnarray}

In Eq.~\eqref{Eq:f2} we will keep only terms of order $1/g^4$ and $1/g^2$, neglecting all other subleading terms
\begin{eqnarray}
	f_\rho(k, p) &\approx& 
	4 
	b_{i\alpha}(-k) 
	b_{i\alpha}(k) 
	b_{j\beta}(-p) 
	b_{j\beta}(p) 
	+ 
	2 \left(
	b_{i\alpha} (-k) b_{i\alpha} (k)
	\langle a^\dagger_{j\beta} (p) a_{j\beta} (p)\rangle_{{\cal B}}
	+
	b_{j\beta} (-p) b_{j\beta} (p)
	\langle a^\dagger_{i\alpha} (k) a_{i\alpha} (k) \rangle_{{\cal B}}
	\right) \notag
	\\ \notag 
	&+&  
	2 
 	\left(
	b_{i\alpha} (k) b_{j\beta} (-p)
	\langle a^\dagger_{j\beta} (p) a_{i\alpha} (k)\rangle_{{\cal B}}
	+
	b_{j\beta} (p) b_{i\alpha} (-k)
	\langle a^\dagger_{i\alpha} (k) a_{j\beta} (p)\rangle_{{\cal B}}
	\right) \\ 
	&-&
	2 
 	\left(
	b_{i\alpha} (k) b_{j\beta} (p)
	\langle a_{i \alpha} (k) a_{j\beta} (p)\rangle_{{\cal B}}
	+
	b_{i\alpha} (-k) b_{j\beta} (-p)
	\langle a^\dagger_{i\alpha} (k) a^\dagger_{j\beta} (p)\rangle_{{\cal B}}
	\right). 
	\label{Eq:f2approx}
\end{eqnarray}

Recall that we are interested in the odd moments of the correlation function. The only piece that can give a non-vanishing contribution to this quantity is:
\begin{eqnarray}
	\tilde{f}_\rho(k, p) &=& 
	2 
 	\left(
	b_{i\alpha} (k) b_{j\beta} (-p)
	\langle a^\dagger_{j\beta} (p) a_{i\alpha} (k)\rangle_{{\cal B}}
	+
	b_{j\beta} (p) b_{i\alpha} (-k)
	\langle a^\dagger_{i\alpha} (k) a_{j\beta} (p)\rangle_{{\cal B}}
	\right)\nonumber \\ 
	&-&
	2 
 	\left(
	b_{i\alpha} (k) b_{j\beta} (p)
	\langle a_{i \alpha} (k) a_{j\beta} (p)\rangle_{{\cal B}}
	+
	b_{i\alpha} (-k) b_{j\beta} (-p)
	\langle a^\dagger_{i\alpha} (k) a^\dagger_{j\beta} (p)\rangle_{{\cal B}}
	\right)\nonumber \\ &=&
	\frac12  
	b_{i\alpha}  (k) 
	\left[ 
		B_{\alpha \beta i j} (-k, p)  
		+ B^{-1}_{\alpha \beta i j} (-k, p)
		- 2 (2\pi)^2\delta_{ij} \delta_{\alpha \beta} \delta^2(p-k)
	\right]
	b_{j\beta}  (-p) 
	\nonumber\\
	&+& 
	\frac12  
	b_{i\alpha}  (-k) 
	\left[ 
		B_{\alpha \beta i j} (k, -p)  
		+ B^{-1}_{\alpha \beta i j} (k, -p)
		- 2 (2\pi)^2 \delta_{ij} \delta_{\alpha \beta} \delta^2(p-k)
	\right]
	b_{j\beta}  (p) 
	\nonumber\\
	&+& 
	\frac12  
	b_{i\alpha}  (k) 
	\left[ 
		- B_{\alpha \beta i j} (- k, -p)  
		+ B^{-1}_{\alpha \beta i j} (-k, -p)
	\right]
	b_{j\beta}  (p) 
	\nonumber\\
	&+& 
	\frac12  
	b_{i\alpha}  (-k) 
	\left[ 
		- B_{\alpha \beta i j} ( k, p)  
		+ B^{-1}_{\alpha \beta i j} (k, p)
	\right]
	b_{j\beta}  (-p),  \label{anti}
\end{eqnarray}
where $f_\rho(k, p)  =  \tilde{f}_\rho(k, p) + \{\text {explicitly even part}\}$. 
Note that the purely ``classical'' term - the first term in
Eq.~(\ref{Eq:f2approx}) is of order $1/g^4$, the antisymmetric piece
Eq.~(\ref{anti}) is of order $1/g^2$. Since the classical term contributes to
$v_2$ (see e.g. Ref.~\cite{Kovner:2010xk}), we may expect the ratio $v_3/v_2$ to be of order $\alpha_s$, which is
phenomenologically reasonable.

It is easy to see that terms with $B^{-1}$ are even under $p \to - p$.
The potential contribution to odd cumulants will be given by the asymmetric part of $\tilde{f}$, i.e. 
\begin{eqnarray}
	\frac{1}{2} \left(	\tilde{f}_\rho(k, p) - 	\tilde{f}_\rho(k,- p)  \right) &=& 
	\frac{1}{2} 
	\left( 
	b(k) \tilde{B}(-k, p)   b(-p) + 
	b(-k) \tilde{B}(k, - p)   b(p)\right)\nonumber \\  &-&
	\frac12 \left(
	b(k) \tilde{B}(-k, - p)   b(p) +
	b(-k) \tilde{B}(k, p)   b(-p) \right),\label{anti1}
\end{eqnarray}
where the matrix convolution is implied and  $\tilde{B}_{\alpha \beta i j} ( k, p)  = {B}_{\alpha \beta i j} ( k, p) - 
(2\pi)^2\delta_{ij} \delta_{\alpha \beta} \delta^2(p+k)
$. 

Equation~(\ref{anti1}) is fairly complicated, as the squeezing parameter $B$ is a non-local function of the classical field $b$. To get some insight into this expression we will evaluate it in the  limit of large $k$ and $ p$.
\subsection{The high momentum limit.}

Henceforth we will assume that the color charge density is averaged over using
the McLerran-Venugopalan (MV) model~\cite{McLerran:1993ni,McLerran:1993ka} with the width $\mu$, that is 
\begin{equation}
	W_{\rm P}[\rho]={\cal \bar N}\exp\left \{-\frac{1}{2}\int_k\frac{\rho^a(k)\rho^a(-k) }{\mu^2(k)}\right \}\;	.
	\label{Eq:W_P}
\end{equation}
where ${\cal \bar N}$ is a normalization factor ensuring the normalization of $W_P$ as the probability density distribution.
For the most part we take $\mu^2(k)=\mu^2$, but whenever necessary we will assume that $\mu^2(k)_{k^2/Q_s^2\rightarrow 0}\rightarrow 0$, which implements the global color neutrality constraint on the MV ensemble.

The high momentum limit then corresponds to the limit $\mu^2/k^2\ll 1$
and $\mu^2/p^2\ll 1$. 
Therefore we need to expand our expression to leading order in $\mu^2$.
 
First we find the leading contribution to $ \tilde{B} ( k, p) $ in the expansion in powers of $\rho$. 
We start with the key ingredients for the operator  $L$. The square of the covariant derivative 
is given by 
\begin{equation}
	\left[D^2\right]^{\alpha \beta}  = \delta^{\alpha \beta} \partial^2  - gf^{\alpha \beta \gamma} [\partial_i, b^\gamma_i ]_+ + 
	g^2f^{\alpha \lambda \gamma} 	f^{\lambda \beta \gamma^\prime}   b^\gamma_i  b^{\gamma^\prime}_i = 
	\delta^{\alpha \beta} \partial^2 -   g[\partial_i, b^{\alpha \beta}_i ]_+  +g^2  b^{\alpha \gamma}_i  b^{\gamma \beta}_i \;, 
\end{equation}
where we introduced the adjoint representation for the field  $b_i$: 
$b^{\alpha\beta}_i\equiv f^{\alpha\beta\gamma}b_i^\gamma$.
To second order in the field we get 
\begin{equation}\label{1/D}
	\frac{1}{D^2}  \approx  \left[ 1 +g \frac{1}{\partial^2} [\partial_i, b_i]_+ - g^2\frac{1}{\partial^2} b^2 + g^2\frac{1}{\partial^2} 
 [\partial_i, b_i]_+  \frac{1}{\partial^2}  [\partial_i, b_i]_+\right] \frac{1}{\partial^2} \;. 
\end{equation}
Thus 
\begin{equation}
	\tilde{B}_{ij} =g^2\left\{ - b_i \frac{1}{\partial^2} b_j - \partial_i \frac{1}{\partial^2} b^2 \frac{1}{\partial^2} \partial_j 
	+ \frac{\partial_i}{\partial^2} (\vec{\partial} \cdot \vec{b})  \frac{1}{\partial^2} b_j  
	+ b_i \frac{1}{\partial^2} (\vec{b} \cdot \vec{\partial})  \frac{\partial_j}{\partial^2} 
	+ \frac{\partial_i}{\partial^2} \left[ 
		(\vec{b} \cdot \vec{\partial}) \frac{1}{\partial^2} 	(\vec{\partial} \cdot \vec{b})  
		-
		(\vec{\partial} \cdot \vec{b}) \frac{1}{\partial^2} 	(\vec{b} \cdot \vec{\partial})  
	\right] \frac{\partial_j}{\partial^2}\right\}\;.
\end{equation}
Note that there are no linear  terms in $b_i$.  
{In  momentum space this gives 
\begin{equation}\label{btilde}
	\tilde{B}_{ij}(k,p)\equiv\int_{x,y}e^{-ikx}\tilde B_{ij}(x,y)e^{-ipy} =
	g^2\int_q
	b_n(k-q) \left[ 
		\mathfrak{t}_{i n}(k) 
		\frac{1}{q^2} 
		\mathfrak{t}_{mj}(p) -
		\frac{k_{ i}}{k^2} 
		\mathfrak{t}_{nm}(q) 
		\frac{p_{ j}}{p^2} 
	\right] 
	b_m(p+q), 
\end{equation}
where the transverse projector in momentum space reads  
\begin{equation}
	\mathfrak{t}_{ij}(q) = \delta_{ij} - \frac{q_{ i} q_{ j}}{q^2} .
\end{equation}
}
We also have to expand the classical field $b$ in powers of $\rho$. To leading order we have 
\begin{equation}
	b_i(k)  = k_{ i} c(k)
	\label{Eq:c}
\end{equation}
with $c(k) \sim \rho(k)/k^2$. 
Then
\begin{equation}\label{btilde1}
	\tilde{B}_{ij} (k, p)=-g^2
	\int_q
	c(k-q) \left[
		\mathfrak{t}_{i n}(k) 
		\mathfrak{l}_{n m}(q) 
		\mathfrak{t}_{mj}(p) + 
		\mathfrak{l}_{i n}(k) 
		\mathfrak{t}_{nm}(q) 
		\mathfrak{l}_{j m}(p) 
	\right] 
	c(p+q), 
\end{equation}
where the longitudinal projector of Eq.~\eqref{B0} in momentum space  
\begin{equation}
	\mathfrak{l}_{ij}(q) = \frac{q_{ i} q_{ j}}{q^2} .
\end{equation}
Note that under the approximation \eqref{Eq:c}, we will be interested in the combination
$k_{ i}  \tilde{B}_{ij} (k, p)	p_{ j}$, to which the first term of Eq.~\eqref{btilde1}
does not contribute:   
\begin{equation}
	k_{ i}  \tilde{B}_{ij} (k, p)	p_{ j}   = -g^2
	\int_q
	c(k-q) \left[ 
		k_{ n}  	\mathfrak{t}_{nm}(q)  	p_{ m}  
	\right] 
	c(p+q). 
\end{equation}
Therefore we obtain the following for the antisymmetric part 
\begin{eqnarray}
	&&\frac{1}{2} \left(	\tilde{f}_\rho(k, p) - 	\tilde{f}_\rho(k,- p)  \right) =\\ && 
	g^2\Bigg[\frac{1}{2} 
	c^\alpha(k) \int_q
	c^{\alpha \gamma}(-k-q) \left[ 
		k_{ n}  	\mathfrak{t}_{nm}(q)  	p_{ m}  
	\right] 
	c^{\gamma \beta}(p+q) 
	c^\beta(-							 p)  	\nonumber
	\\&&+\frac{1}{2} 
	c^\alpha(-k) 
	\int_q
	c^{\alpha \gamma}(k-q) \left[ 
		k_{ n}  	\mathfrak{t}_{nm}(q)  	p_{ m}  
	\right] 
	c^{\gamma \beta}(-p+q)
	c^\beta(p)\nonumber \\  
	&&+
	\frac12
	c^\alpha(k)
	\int_q
	c^{\alpha \gamma}(-k-q) \left[ 
		k_{ n}  	\mathfrak{t}_{nm}(q)  	p_{ m}  
	\right] 
	c^{\gamma \beta}(-p+q)
	c^\beta(p) \nonumber\\ 
	&&+
	\frac12
	c^\alpha(-k) 
		\int_q
		c^{\alpha \gamma}(k-q) \left[ 
		k_{ n}  	\mathfrak{t}_{nm}(q)  	p_{ m}  
	\right] 
	c^{\gamma \beta}(p+q)
	c^\beta(-p)\Bigg].\nonumber
\end{eqnarray}
Here to avoid confusion the color indices were explicitly shown. 
As before $c^{\alpha \beta} = f^{\alpha\beta\gamma} c^\gamma$.

The above expression does not vanish configuration-by-configuration 
in contrast to what one would get by using Eq.~\eqref{Eq:dilute_xsec} of the dilute limit.
This is encouraging enough to continue by computing 
an average with respect to the projectile field using the 
Gaussian ensemble 
\begin{equation}
	\langle c^\alpha(k) c^\beta(k^\prime) \rangle_\rho = - \delta^{\alpha \beta} 
	 (2\pi)^2 \delta^2(k + k^\prime) 
	\gamma(k ), 
	\label{Eq:cc} 
\end{equation}
where $\gamma$ is defined by usual 
\begin{equation}
	\langle \frac{1}{-\partial^2} \rho^\alpha(x)  \frac{1}{-\partial^2} \rho^\beta(y) \rangle_\rho 
	= \delta^{\alpha \beta} \gamma(x-y). 
\end{equation}
In the MV model, we have  $\gamma(k)=\frac{\mu^2}{k^4}$.


We finally get 
\begin{eqnarray}
	\frac{1}{2} 	\langle \tilde{f}_\rho(k, p) - 	\tilde{f}_\rho(k,- p) \rangle_\rho  & =&
	-g^2 N_c (N_c^2-1)  (2\pi)^2 S_\perp \gamma(k) \int_q \gamma(q) p_{ n} \mathfrak{t}_{nm}(q-p)  p_{ m}
	\left( \delta^2(p-k) -  \delta^2(p+k)   \right)
	\nonumber\\ 
	&+& 
	g^2 N_c (N_c^2-1)   S_\perp \gamma(k) \gamma(p) k_{ n} \left[  \mathfrak{t}_{nm}(k-p)  +  \mathfrak{t}_{nm}(k+p)    \right]  p_{ m} , 
\end{eqnarray}
where  $S_\perp$ is the transverse area of the projectile.

The first term describes  the back-to-back component of the gluon correlation
at exactly the same momentum. It has to be treated as the ``hard'' component to be subtracted from the correlation  and does not represent any particular interest for the current study. 
The combination appearing in the last term can be simplified 
\begin{align}
	 k_{ n} \left[  \mathfrak{t}_{nm}(k-p)  +  \mathfrak{t}_{nm}(k+p)    \right]  p_{ m} = 
\frac{4 k \cdot p} {|k-p|^2|k+p|^2} \left( k \times p \right)^2
=
	 \frac{4 k \cdot p} {|k-p|^2|k+p|^2} \left( k^2 p^2 - (k \cdot p)^2 \right). 
\end{align}

Note that for $k=p$ 
\begin{equation}
	k_{ n} \left[  \mathfrak{t}_{nm}(k-p)  +  \mathfrak{t}_{nm}(k+p)    \right]  p_{ m} |_{p=k} = 
	k^2 \cos(\phi),   
\end{equation}
where $\phi$ is the angle between $k$ and  $p$.

In general we get  
\begin{align}
	k_{ n} \left[  \mathfrak{t}_{nm}(k-p)  +  \mathfrak{t}_{nm}(k+p)    \right]  p_{ m}  = 
	\frac{4 k^3 p^3 \cos(\phi) \sin^2(\phi)}{k^4 + p^4 - 2 k^2 p^2 \cos(2\phi)}  = k^2 C(z,\phi), 
\end{align}
where 
\begin{equation}
	C(z, \phi) = \frac{2 z^3 \sin(\phi) \sin(2 \phi)}{z^4 - 2z^2 \cos(2\phi) +1}; \ \ \ \ \ \ \ \ \ \ \ \ z\equiv \frac{|p|}{|k|}.
\end{equation}
We plot this function in Fig.~\ref{fig:cfn}.

As a proxy to $c_3\{2\}$, our goal is to evaluate 
\begin{equation}
\int d\phi e^{ i 3\phi} C(z, \phi)  = 
\frac{z^3}{2i} \oint_{|\zeta|=1}   d\zeta  
\frac{\zeta(1+\zeta^2)(1-\zeta^2)^2} {z^2 \zeta^4 - (1+z^4)\zeta^2 + z^2}, 
\end{equation}
where $\zeta = e^{i\phi}$ was introduced. 
The poles are located at $\pm z$ and $\pm 1/z$. Therefore for $z<1$, 
\begin{equation}
\int d\phi e^{ i 3\phi} f(z, \phi)  = 
\pi z^3 (z^2-1)
\end{equation}
and for $z>1$ 
\begin{equation}
\int d\phi e^{ i 3\phi} f(z, \phi)  = 
\pi  \frac{1-z^2}{z^3} .
\end{equation}
We plot the third harmonic  in Fig.~\ref{fig:cum}.
We also plot the first harmonic $c_1\{2\}$, which can be readily 
evaluated. We will apply (but not explicitly detail) a similar procedure for the case of particle production in the next section.

We conclude that in the high momentum limit, 
the first correction to the dilute limit introduces a non-trivial odd azimuthal anisotropy. 
The first harmonic is positive,
but the third harmonic is negative. Since the flow coefficient $v_3$ is defined
as $v_3=\sqrt{c_3\{2\}}$, our calculation does not yield a real $v_3$.

We note however, that the calculation in this section is not a proper
calculation of particle production, but rather of correlations in the
projectile wave function. In the next section we perform a more  appropriate
calculation, namely that of particle production.

\section{Correlations in particle production.}
In this section we calculate the double inclusive production in scattering of
the CGC state Eq.~(\ref{Eq:CGC}).

Like in the previous section we will be
forced into  a high momentum approximation to be able to extract some usable
information from the general formulae. For the moment we start with the formal
derivation.
\subsection{The double inclusive production.}  
According to Refs.~\cite{Kovner:2006wr,Kovner:2006ge} the single inclusive gluon production is given by
\begin{equation}
\label{Eq:SingleInc}
\frac{d^3N}{d^2k d\eta} =
 \int D \rho W_{\rm P} [\rho] \;  
\int D \alpha_{\rm T} W_{\rm T} [\alpha_{\rm T}] \;  
\langle 0|\Omega^\dagger \hat S^\dagger \Omega
\left[a^\dagger (k)  a(k)\right]
\Omega^\dagger \hat S\Omega|0\rangle .
\end{equation}
and similarly for the double inclusive 
\begin{equation}
\label{Eq:DoubleInc}
\frac{d^6N}{d^2k d^2pd\eta_k d\eta_p} = 
  \int D \rho W_{\rm P} [\rho] \;  
\int D \alpha_{\rm T} W_{\rm T} [\alpha_{\rm}] \;  
\langle 0|\Omega^\dagger \hat S^\dagger \Omega \left[a^\dagger (k)a^\dagger(p) a(p)a(k)\right]\Omega^\dagger \hat S\Omega|0\rangle .
\end{equation}
Here $W_{\rm T} [\alpha_{\rm T}]$ is a probability density distribution for the target color field. 
We remind that $\Omega$ is the unitary operator
which diagonalizes the QCD Hamiltonian~\eqref{Eq:CGC},
and $\hat S$ is the second-quantized eikonal S-matrix
\begin{equation}
	\hat S = {\cal P} \exp \left\{ 
		i \int d^2 x \hat \rho^a (x) \hat \alpha_{\rm T} (x)
	\right\} . 
	\label{Eq:Smat}
\end{equation}
Since $\Omega(\phi,\pi)$ is a Gaussian operator, we 
can  find the transformation of the fields and momenta under the action of
$\Omega$.  We read off these expressions from the expectation values of
quadratic operators Eq.~(\ref{Eq:phiphi}) calculated in the previous
section~\footnote{In the following we discard the terms involving commutators
	of the field $b$ which can be calculated at relevant
	order\cite{Kovner:2007zu,Altinoluk:2009je}. In principle these terms are
not suppressed by $\alpha_s$ relative to the terms we keep. However they
correspond to processes where a gluon rescatters several times on the same
valence parton, which do not seem to be physically important. Including these
terms would be very cumbersome and not particularly illuminating. }
\begin{eqnarray}\label{pitrans}
	&&\Omega \phi(k)\Omega^\dagger=(1-\mathfrak{l}-L)^{-1}(\mathfrak{t}-\mathfrak{l})(k,p)\phi(-p),\nonumber
\\ 
&&\Omega \pi(k)\Omega^\dagger=(1-\mathfrak{l}-L)(\mathfrak{t}-\mathfrak{l})(k,p)\{\pi(-p)-2b(-p)\}.
\end{eqnarray}
Let us simplify our notations by using the index $i$ cumulatively for the rotational, color indices and spatial coordinate, and by defining 
\begin{equation}
	\Gamma\equiv (\mathfrak{t}-\mathfrak{l})(1-\mathfrak{l}-L); 
	\ \ \bar b\equiv b[S\rho]; \ \ \ \ \ \ \bar L=L[\bar b]\;.
\end{equation}
All the matrix products in the following expressions are understood in coordinate space. Transition to momentum space is done eventually by the Fourier transform.
 We then have 
 \begin{eqnarray}\label{Eq:rules}
	 && {\cal C} b_i {\cal C}^\dagger=b_i;
	 \ \ \ \ {\cal C}\Gamma[b]{\cal C}^\dagger=\Gamma[b];  
\nonumber\\
&&  {\cal C} ^\dagger b_i {\cal C} =b_i; \ \ \ \ {\cal C}^\dagger\Gamma[b] {\cal C}=\Gamma[b]; \nonumber\\
&&{\cal C}\pi_i {\cal C}^\dagger=\pi_i-2b_i; \ \ \ {\cal C}^\dagger\pi_i {\cal C}=\pi_i+2b_i;\\
&&{\cal B}^\dagger \phi_i{\cal B}=\Gamma_{ij}\phi_j; \ \ {\cal B}^\dagger \pi_i{\cal B}=\Gamma^{-1T}_{ij}\pi_j;   \ \ \ {\cal B} \phi_i{\cal B}^\dagger=\Gamma^{-1}_{ij}\phi_j; \ \ {\cal B} \pi_i{\cal B}^\dagger=\Gamma^T_{ij}\pi_j;  \nonumber
\end{eqnarray}
and finally
\begin{eqnarray}
&&\Omega^\dagger\hat S^\dagger\Omega \phi_i\Omega^\dagger \hat S\Omega=\bar\Gamma^{-1}_{ij}S_{jk}\Gamma_{kl}\phi_l \,,\\
&&\Omega^\dagger\hat S^\dagger\Omega \pi_i\Omega^\dagger \hat S\Omega=2\bar \Gamma^T_{ij}(S_{jk}b_k-\bar b_j)  + \bar\Gamma^T_{ij}S_{jk}\Gamma^{-1T}_{kl}\pi_l\,,  \nonumber
\end{eqnarray}
where 
$\bar \Gamma[b]\equiv \Gamma[\bar b]$\,.
Only  
the following operators will be of relevance   
\begin{eqnarray}\label{abc}
	\mathbb{A}  &  \overset{\rm def}{=}  & \bar\Gamma^{-1} S \Gamma \,, \nonumber\\  
	\mathbb{B}  &  \overset{\rm def}{=}  & \bar\Gamma^T S \Gamma^{-1T}  \,, \nonumber\\  
	\mathbb{C}  &  \overset{\rm def}{=}  & 2\bar \Gamma^T (S b -\bar b )\, .   
\end{eqnarray} 
More explicitly in momentum space
\begin{eqnarray}
	\mathbb{A}^{ad}_{im}(k,l)  &  \overset{\rm def}{=}  & \bar\Gamma^{-1\,ab}_{ij}(k,p) S^{bc}(q-p) \Gamma^{cd}_{jm}(-q,l) \,, \nonumber \\  
	\mathbb{B}^{ad}_{im}(k,l) &  \overset{\rm def}{=}  &  \bar\Gamma^{Tab}_{ij}(k,p) S^{bc}(q-p) \Gamma^{-1T\,cd}_{jm}(-q,l)  \,, \nonumber\\  
	\mathbb{C}_i^a(k)  &  \overset{\rm def}{=}  & 2\bar \Gamma^{Tab}_{ij}(k,p) [S^{bc}(q-p) b_j^c(-q) -\bar b_j^b(-p) ]\,   
\end{eqnarray} 
with
\begin{equation}
S^{ab}(p-q)=\int_x S^{ab}(x) e^{-i(p-q) x}\,.
\end{equation}
For the defined combinations 
the following property obviously holds
\begin{equation}
	\mathbb{A} \mathbb{B}^T = 1. 
\end{equation}

Using this definitions we can find the transformation of the creation/annihilation operators under the unitary operator $ \Omega^\dagger\hat S \Omega$:
\begin{eqnarray}
	\label{Eq:PaP}
	( \Omega^\dagger\hat S \Omega )^\dagger a(k)  ( \Omega^\dagger\hat S \Omega )  &=& \frac{1}{\sqrt{2}} 
	\left( 
	(\mathbb{A} \phi) (-k) + i (\mathbb{B} \pi) (-k) + i \mathbb{C}(-k)
	\right)\, ,\\
	( \Omega^\dagger\hat S \Omega )^\dagger a^\dagger (k)  ( \Omega^\dagger\hat S \Omega ) 
	&=& \frac{1}{\sqrt{2}} 
	\left( 
	(\mathbb{A} \phi) (k) - i (\mathbb{B} \pi) (k) - i \mathbb{C}(k)
	\right). 
	\label{Eq:PaDP}
\end{eqnarray}

Now it is easy to find  the expression for the single inclusive gluon production, see Eq.~\eqref{Eq:SingleInc}, 
\begin{align}
&	\langle 0| ( \Omega^\dagger\hat S \Omega )^\dagger   a^\dagger (k) a(k) ( \Omega^\dagger\hat S \Omega )     |0\rangle  = \notag \\ & {1\over 4}\left\{  {\rm tr}_{\rm L,C}[\mathbb{A}(k,p)\mathbb{A}^T(-k,-p)] + {\rm tr}_{\rm L,C}[\mathbb{B}(k,p)\mathbb{B}^T(-k,-p)]-
	{\rm tr}_{\rm L,C}[\mathbb{A}(k,p)\mathbb{B}^T(-k,-p)]\right.\nonumber \\
	&-\left. {\rm tr}_{\rm L,C}[\mathbb{B}(k,p)\mathbb{A}^T(-k,-p)]\right\} + {1\over 2} \mathbb{C}(k)\mathbb{C}(-k) \nonumber  \\
	&= {1\over 4}\left\{ {\rm tr}_{\rm L,C}[\mathbb{A}(k,p)\mathbb{A}^T(-k,-p)] + {\rm tr}_{\rm L,C}[\mathbb{B}(k,p)\mathbb{B}^T(-k,-p)]\right \} - {1\over 2} C_A\delta(0)+ {1\over 2} \mathbb{C}(k)\mathbb{C}(-k)\;,
\end{align}
where ${\rm tr}_{\rm L,C}$ represents trace over both color and Lorentz
indices. In the above the transposition refers only to color and Lorentz
indices. The result is obviously symmetric under $k\rightarrow -k$ without any
reference to averaging over the projectile or target fields.
The $\mathbb{C}(k)\mathbb{C}(-k)$ agrees with the corresponding expression in
Ref.~\cite{Altinoluk:2009jf}. { Note that although the structure of this term
	is that of the production off a classical field, the ``field'' $\mathbb{C}$
	is not of entirely classical origin. It is affected by the  Bogoliubov part
	of the CGC vacuum wave function through the factor $\bar\Gamma$ in
	Eq.~(\ref{abc}). In the high density limit, $\rho\sim 1/g$, this is an order
one correction to the classical result.}

Now consider the double inclusive production~\eqref{Eq:DoubleInc} before averaging with respect to projectile/target configurations 
\begin{equation}
F(k,p)\equiv	
\langle 0| ( \Omega^\dagger\hat S \Omega )^\dagger 
a^\dagger (k)a^\dagger(p) a(p)a(k) 
( \Omega^\dagger\hat S \Omega ) 
|0\rangle . 
\end{equation}
By inserting $ 1 = ( \Omega^\dagger\hat S \Omega )   ( \Omega^\dagger\hat S \Omega )^\dagger$
after each creation and anihilation operator 
and using Eqs.~\eqref{Eq:PaP} and \eqref{Eq:PaDP} 
the expectation value can be readily found. 

{ The leading term is $\mathbb{C}^4$.  As the analogous term in the single
gluon inclusive production, this term is affected at $O(1)$ by the Bogoliubov
component of the CGC wave function. Without the Bogoliubov correction this is
just the ``glasma graph'' term frequently discussed in the literature. Thus we
see that the high density corrections affect the glasma graph term at leading
order.}

However
this term is  symmetric under $k \to - k$. 
As before we are interested only in extracting the 
piece that is antisymmetric under this transformation. 
We  will systematically neglect all symmetric terms in our analysis.  We  thus keep only the NLO (in $b$) term only. 
For this operation instead of the equality sign  we will use ``$\models$''.
\begin{eqnarray}
\label{Eq:CorrelF}
	{1\over 2}\left (F(k, p)-F(k,-p)\right) &\models&
	\frac{\mathbb{C} (k) } {2}  
	\frac{\mathbb{A} 
	\mathbb{A}^T (-k, p) - (2\pi)^2 \delta^2(p -k)  }{2}
	\frac{\mathbb{C} ( - p) } {2}  \nonumber
	\\ 
	&+& 
	\frac{\mathbb{C} (-k) } {2}  
	\frac{\mathbb{A}
	 \mathbb{A}^T (k, -p) -(2\pi)^2 \delta^2(p - k)  }{2}
	\frac{\mathbb{C} (  p) } {2} \nonumber
	\\ 
	&-& 
	\frac{\mathbb{C} (k) } {2}  
	\frac{\mathbb{A} 
	\mathbb{A}^T (-k, -p) - (2\pi)^2\delta^2(p +k)  }{2}
	\frac{\mathbb{C} (  p) } {2}\nonumber
	\\ 
	&-& 
	\frac{\mathbb{C} (-k) } {2}  
	\frac{\mathbb{A} 
	\mathbb{A}^T (k, p) - (2\pi)^2\delta^2(p +k)  }{2}
	\frac{\mathbb{C} ( - p) } {2}\,. 
\end{eqnarray}
{ Note that this term is entirely absent from the dilute limit of the
glasma graph contribution. In the dense limit discussed here it is
nonvanishing, albeit suppressed with respect to the leading symmetric term by a
single power of $\alpha_s$.}
\subsection{High momentum expansion.}
We now expand this expression in $\mu^2/p^2$. First,  expanding to order $b^2$ (see Appendix \ref{App:AA_exp} for details) we get: 
\begin{eqnarray}
	\label{Eq:AAA}
&&\left[\mathbb{A}\mathbb{A}^T-1\right]^{ab}_{ij}(k,p)\approx\\
	&&	g^2\int_{q,l,m}\Bigg[\left[f^{acd}S^{de}( k- l) \rho^e(l-q) 	\right]\left[f^{cbf}S^{fg}( p- m)\rho^g(m+q)\right]\times\nonumber\\
&&\frac{1}{(k-q)^2(p+q)^2}\left[
		\mathfrak{t}_{i n}(k) 
		\mathfrak{l}_{n m}(q) 
		\mathfrak{t}_{mj}(p) + 
		\mathfrak{l}_{i n}(k) 
		\mathfrak{t}_{nm}(q) 
		\mathfrak{l}_{j m}(p) 
	\right] \nonumber\\
&&-	\left[S^{ac}( k-l)f^{cde} \rho^e(l-q) 	\right]\left[S^{bf}( p- m)f^{dfg}\rho^g(m+q)\right]\times\nonumber\\
	&&\frac{1}{(l-q)^2(m+q)^2}	 \left[
		\mathfrak{t}_{i n}(l) 
		\mathfrak{l}_{n m}(q) 
		\mathfrak{t}_{mj}(m) + 
		\mathfrak{l}_{i n}(l) 
		\mathfrak{t}_{nm}(q) 
		\mathfrak{l}_{j m}(m) 
	\right] \Bigg]\nonumber.
		\end{eqnarray}
Although we do not favor this classification, we nevertheless comment that the first term 
of Eq.~\eqref{Eq:AAA} in the square brackets corresponds to  the so-called final state interaction -- 
emission and possible rescattering of gluons after the scattering of the valence charges off the target.

Also to leading order we have
\begin{equation}
\mathbb{C}_i^a (k)\approx i\int_u\left[\frac{u_i}{ u^2}-\frac{k_i}{k^2}\right]S^{as}(k- u)\rho^s( u)\,.
\end{equation}
Here the second term also represents the final state interactions, as defined above.

Now we have to put it all together. We will consider one term, the rest will be restored by inspection.
\begin{eqnarray}
	\label{Eq:AA_part}
	&&\frac{\mathbb{C} (-k) } {2}  
	\left[\mathbb{A}  \mathbb{A}^T (k, p) - (2\pi)^2 \delta^2(k+ p)\right]
	\frac{\mathbb{C} ( -p) } {2}
	=g^2\int_{q,l,m,u,v} \\
&&  \Bigg[\left[f^{acd}S^{de}( k- l) 
	\underbracket{\rho^e(l-q)}_{1}
\right]\left[f^{cbf}S^{fg}( p- m)
	\underbracket{\rho^g(m+q)}_2
\right]S^{as}(- k- u)
\underbracket{\rho^s( u)}_3
S^{bt}(- p- v)
\underbracket{\rho^t( v)}_4
\times\nonumber\\
&&\frac{1}{(k-q)^2(p+q)^2}\left[\frac{u_i}{ u^2}+\frac{k_i}{k^2}\right]\left[
		\mathfrak{t}_{i n}(k) 
		\mathfrak{l}_{n m}(q) 
		\mathfrak{t}_{mj}(p) + 
		\mathfrak{l}_{i n}(k) 
		\mathfrak{t}_{nm}(q) 
		\mathfrak{l}_{j m}(p) 
	\right] \left[\frac{v_j}{ v^2}+\frac{p_j}{p^2}\right]\nonumber\\
&&-	\left[S^{ac}( k- l)f^{cde} 
\underbracket{\rho^e(l-q)}_{1}
\right]\left[S^{bf}( p- m)f^{dfg}
\underbracket{\rho^g(m+q)}_2
\right]S^{as}(- k- u)
\underbracket{\rho^s( u)}_3
S^{bt}(- p- v)
\underbracket{\rho^t( v)}_4
\times\nonumber\\
	&&\frac{1}{(l-q)^2(m+q)^2}\left[\frac{u_i}{ u^2}+\frac{k_i}{k^2}\right]	 \left[
		\mathfrak{t}_{i n}(l) 
		\mathfrak{l}_{n m}(q) 
		\mathfrak{t}_{mj}(m) + 
		\mathfrak{l}_{i n}(l) 
		\mathfrak{t}_{nm}(q) 
		\mathfrak{l}_{j m}(m) 
	\right] \left[\frac{v_j}{ v^2}+\frac{p_j}{p^2}\right] \Bigg]\nonumber\,.
\end{eqnarray}

\begin{figure}
	\centerline{
	\includegraphics[width=0.45\linewidth]{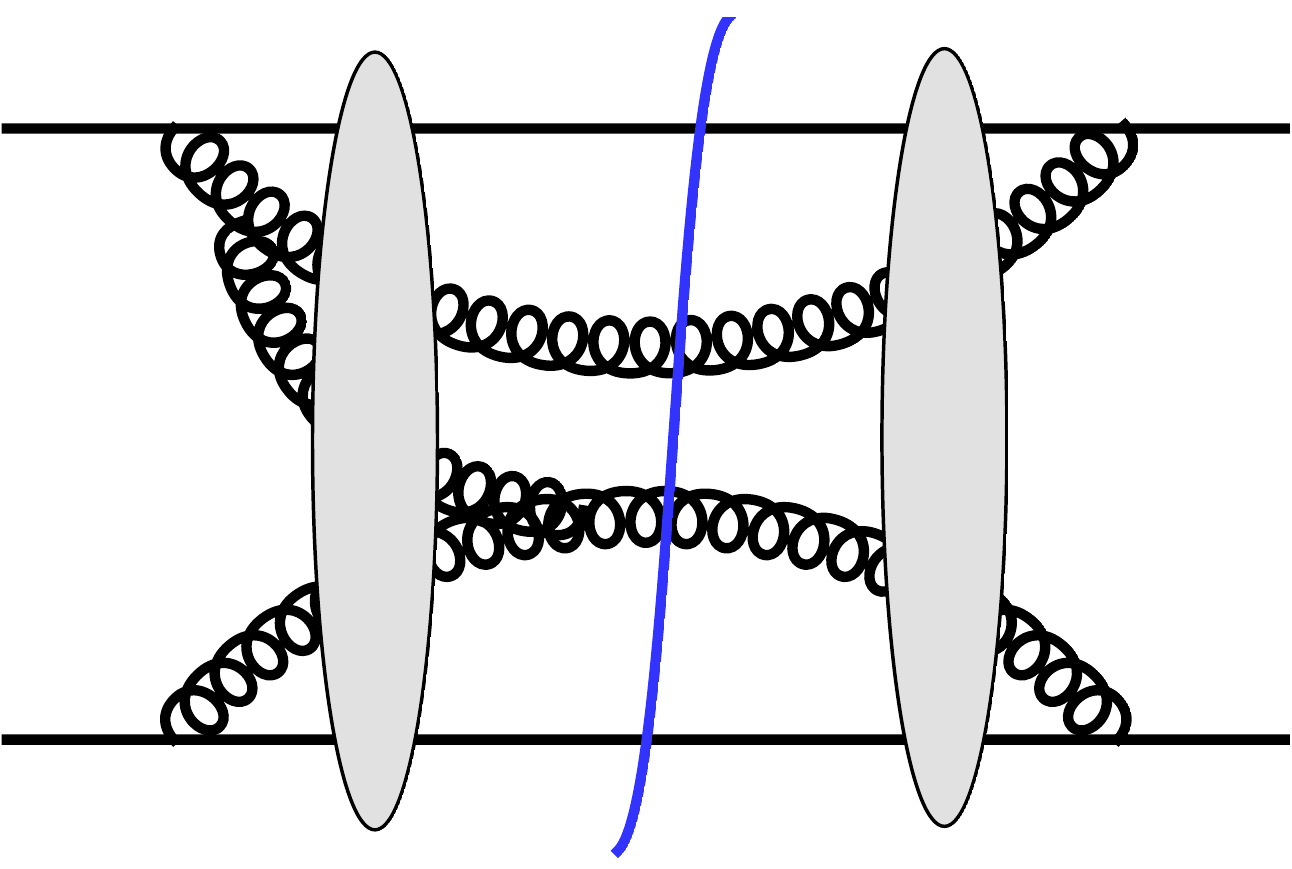}
	\includegraphics[width=0.45\linewidth]{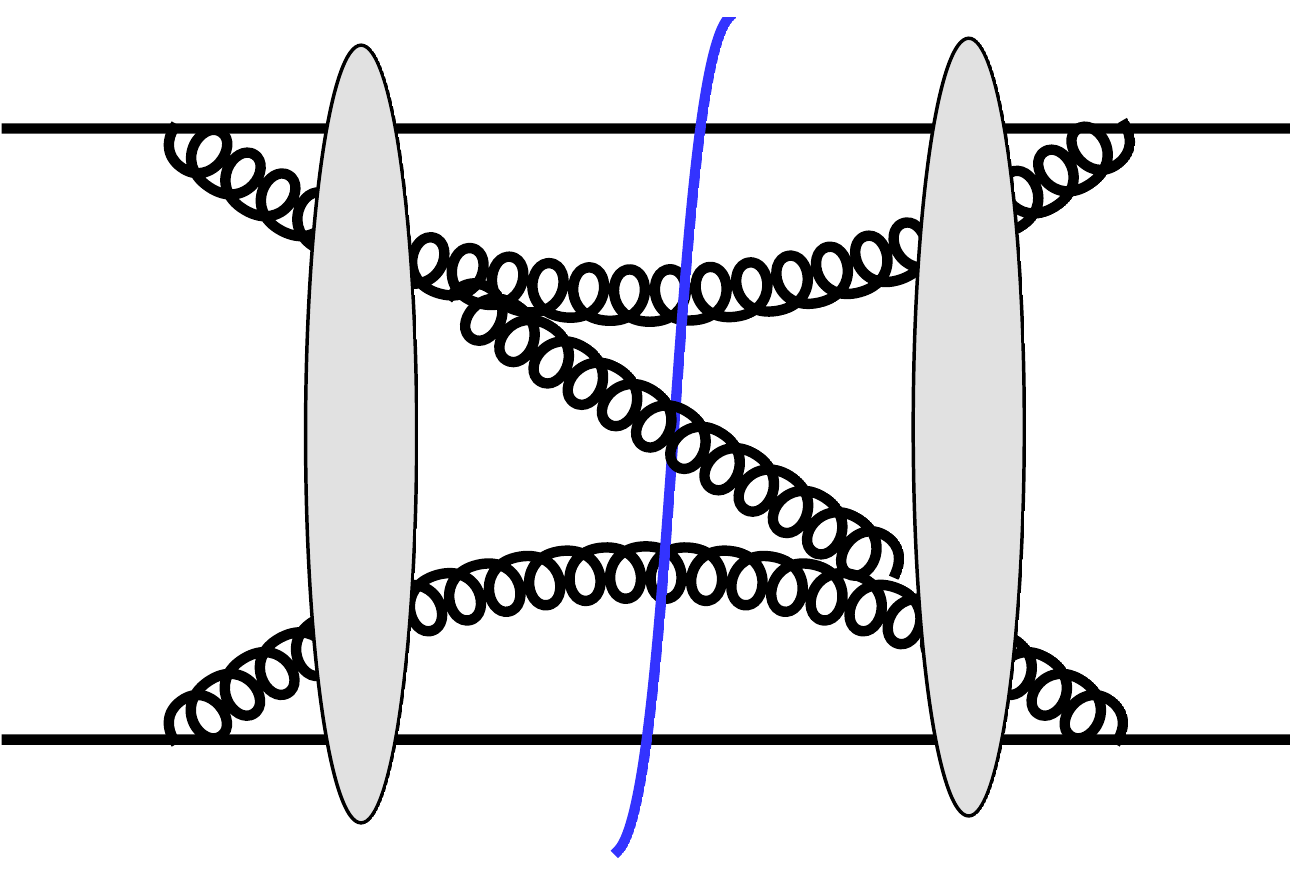}
}
	\caption{Schematic representation of the types of  diagrams that contribute to the double inclusive production at order $g^6\rho^4$.  The horizontal lines symbolize the valence gluons constituting the valence color charge density $\rho$. The vertical line denotes the final state. Momenta of two final sate gluons are fixed.  In the case of the three gluon final state, one of the gluons is summed over inclusively. The blobs symbolize the scatterings on the target. The final state gluons can be emitted either before or after these scatterings, and all the combinations have to be accounted for.}
	\label{fig:Diagrams}
\end{figure}

{ Note that although our original expressions for production resum infinite
	number of diagrams in the external field, any fixed order in expansion in
	powers of $\rho$ corresponds to a finite number of diagrams. Although it is
	not entirely straightforward to 
identify directly the  diagrams that correspond to eq.(\ref{Eq:AA_part}), the
systematics of expansion in powers of $\rho$ and the discussion in Appendix  A
suggest that those are the diagrams illustrated  schematically in
Fig.~\ref{fig:Diagrams}. Thus it should be possible to check the results of
this section directly by 
computing the corresponding diagrams in  the framework of the light cone perturbation theory. 
}

The next step is to contract the charge densities using the MV model. 
This generates three distinct contractions  in each one of the two terms
$$
\left \langle 
\frac{\mathbb{C} (-k) } {2}  
	\left[\mathbb{A}  \mathbb{A}^T (k, p) -  (2\pi)^2  \delta^2( k+ p)\right]
	\frac{\mathbb{C} ( -p) } {2} 
\right \rangle_\rho 
= g^2\left[
\langle 12 \rangle \langle 34 \rangle+
\langle 13 \rangle \langle 24 \rangle+
\langle 14 \rangle \langle 23 \rangle\right],  
$$
where the integers in the angular brackets denote the corresponding contractions of the sources 
identified in Eq.~\eqref{Eq:AA_part} by the  numbers located below the rectangular brackets,  so that   
\begin{eqnarray}\label{mvprojectile}
&&\langle 12 \rangle \langle 34 \rangle =S^{\dagger sa}(-k-u)f^{acd}S^{de}(k-l)S^{\dagger ef}(p+l)f^{fcb}S^{bs}(-p+u)\mu^2(u)\mu^2(q-l)\times\\
&&\frac{1}{(k-q)^2(p+q)^2}\left[\frac{u_i}{ u^2}+\frac{k_i}{k^2}\right]\left[
		\mathfrak{t}_{i n}(k) 
		\mathfrak{l}_{n m}(q) 
		\mathfrak{t}_{mj}(p) + 
		\mathfrak{l}_{i n}(k) 
		\mathfrak{t}_{nm}(q) 
		\mathfrak{l}_{j m}(p) 
	\right] \left[-\frac{u_j}{u^2}+\frac{p_j}{p^2}\right] \,,\nonumber\\
	&&+N_c {\rm Tr}\left[S^\dagger(-k-u)S(k-l)S^\dagger(p+l)S(-p+u)\right]\mu^2(u)\mu^2(q-l)\times\nonumber\\
&&\frac{1}{(l-q)^4}\left[\frac{u_i}{ u^2}+\frac{k_i}{k^2}\right]	 \left[
		\mathfrak{t}_{i n}(l) 
		\mathfrak{l}_{n m}(q) 
		\mathfrak{t}_{mj}(l) + 
		\mathfrak{l}_{i n}(l) 
		\mathfrak{t}_{nm}(q) 
		\mathfrak{l}_{j m}(l) 
	\right] \left[-\frac{u_j}{ u^2}+\frac{p_j}{p^2}\right]\,,\nonumber\\
	&&\langle 13 \rangle \langle 24 \rangle =f^{acd}S^{de}(k-l)S^{\dagger ea}(-k+l-q)f^{cbf}S^{fg}(p-m)S^{\dagger gb}(-p+m+q)\mu^2(q-l)\mu^2(m+q)\times\nonumber\\
	&&\frac{1}{(k-q)^2(p+q)^2}\left[\frac{(q-l)_i}{ (q-l)^2}+\frac{k_i}{k^2}\right]\left[
		\mathfrak{t}_{i n}(k) 
		\mathfrak{l}_{n m}(q) 
		\mathfrak{t}_{mj}(p) + 
		\mathfrak{l}_{i n}(k) 
		\mathfrak{t}_{nm}(q) 
		\mathfrak{l}_{j m}(p) 
	\right] \left[-\frac{(m+q)_j}{(m+q)^2}+\frac{p_j}{p^2}\right] \nonumber\\
	&&-S^{ac}(k-l)f^{cde}S^{\dagger ea}(-k+l-q)S^{bf}(p-m)f^{dfg}S^{\dagger gb}(q-p+m)\mu^2(q-l)\mu^2(m+q)\times\nonumber\\
	&&\frac{1}{(l-q)^2(m+q)^2}\left[\frac{(q-l)_i}{ (q-l)^2}+\frac{k_i}{k^2}\right]	 \left[
		\mathfrak{t}_{i n}(l) 
		\mathfrak{l}_{n m}(q) 
		\mathfrak{t}_{mj}(m) + 
		\mathfrak{l}_{i n}(l) 
		\mathfrak{t}_{nm}(q) 
		\mathfrak{l}_{j m}(m) 
	\right] \left[-\frac{(m+q)_j}{ (m+q)^2}+\frac{p_j}{p^2}\right]\,, \nonumber\\
	&&\langle 14 \rangle \langle 23 \rangle =f^{acd}S^{de}(k-l)S^{\dagger eb}(-p+l-q)f^{cbf}S^{fg}(p-m)S^{\dagger ga}(-k+q+m)\mu^2(l-q)\mu^2(m+q)\times\nonumber\\
	&&\frac{1}{(k-q)^2(p+q)^2}\left[-\frac{(q+m)_i}{ (q+m)^2}+\frac{k_i}{k^2}\right]\left[
		\mathfrak{t}_{i n}(k) 
		\mathfrak{l}_{n m}(q) 
		\mathfrak{t}_{mj}(p) + 
		\mathfrak{l}_{i n}(k) 
		\mathfrak{t}_{nm}(q) 
		\mathfrak{l}_{j m}(p) 
	\right] \left[\frac{(q-l)_j}{(q-l)^2}+\frac{p_j}{p^2}\right] \nonumber\\
	&&- S^{ac}(k-l)f^{cde}S^{\dagger eb}(-p+l-q)S^{bf}(p-m)f^{dfg}S^{\dagger ga}(-k+q+m)\mu^2(l-q)\mu^2(m+q)\times	\nonumber\\
	&&\frac{1}{(l-q)^2(m+q)^2}\left[-\frac{(q+m)_i}{ (q+m)^2}+\frac{k_i}{k^2}\right]	 \left[
		\mathfrak{t}_{i n}(l) 
		\mathfrak{l}_{n m}(q) 
		\mathfrak{t}_{mj}(m) + 
		\mathfrak{l}_{i n}(l) 
		\mathfrak{t}_{nm}(q) 
		\mathfrak{l}_{j m}(m) 
	\right] \left[\frac{(q-l)_j}{ (q-l)^2}+\frac{p_j}{p^2}\right] \nonumber	\,.
\end{eqnarray}

	\subsection{Operator product expansion.}
The last step in the calculation of double inclusive production is the averaging over the target fields. Before restricting ourselves to an explicit model for this averaging, we consider the case where the target fields are much softer than $p, \ k$, that is  
$Q_s/p \ll 1$
and 
$Q_s/k \ll 1$.
In this case  we can perform what can be called an operator product expansion in the general 
model-independent framework.  The technical details are presented in Appendix B.

	\subsubsection{Odderon} 
To the leading order of the operator product expansion of Eqs.~\eqref{mvprojectile}  we get 
\begin{eqnarray}
	\label{odderon_1}
&&\langle 13\rangle\langle 24\rangle
=
	-\int_q \frac{\mu^4}{p^4k^4}\left[
		\mathfrak{t}_{s n}(k) 
		\mathfrak{l}_{n m}(q) 
		\mathfrak{t}_{mj}(p) + 
		\mathfrak{l}_{i n}(k) 
		\mathfrak{t}_{nm}(q) 
		\mathfrak{l}_{t m}(p) 
	\right]\times\nonumber\\
	&&\Bigg[\Big[f^{acd}S^{de}\partial_sS^{\dagger ea}\Big](-q)\Big[f^{cbf}S^{fg}\partial_tS^{\dagger gb}\Big](q)-\Big[S^{ac}f^{cde}\partial_sS^{\dagger ea}\Big](-q)	\Big[	S^{bf}f^{dfg}\partial_tS^{\dagger gb}\Big](q)\Bigg]\,,
\\ 
&&\langle 14\rangle\langle 23\rangle
=
	-\int_q \frac{\mu^4}{p^4k^4}\left[
		\mathfrak{t}_{s n}(k) 
		\mathfrak{l}_{n m}(q) 
		\mathfrak{t}_{mj}(p) + 
		\mathfrak{l}_{i n}(k) 
		\mathfrak{t}_{nm}(q) 
		\mathfrak{l}_{t m}(p) 
	\right]\times\nonumber\\
	&&\Bigg[\Big[f^{acd}S^{de}\partial_tS^{\dagger eb}\Big](-q)\Big[f^{cbf}S^{fg}\partial_sS^{\dagger ga}\Big](q)-\Big[S^{ac}f^{cde}\partial_tS^{\dagger eb}\Big](-q)	\Big[	S^{bf}f^{dfg}\partial_sS^{\dagger ga}\Big](q)\Bigg]\,.
	\label{odderon}
\end{eqnarray}

 The sum of these expressions 
 is odd under the charge conjugation transformation $S\rightarrow S^\dagger$
 and thus is due to the Odderon exchange. In principle this contribution does
 not have to vanish. However we do not have a well motivated model  how to
 incorporate the Odderon in the target probability 
 distribution.\footnote{
 The Odderon contribution averages to zero   in the MV model.}
 We therefore have no way of determining the sign of this contribution without
 a specific model. Additionally, the Odderon contribution becomes subleading at
 high energy, see e.g.~\cite{Bartels:1999yt, Kovchegov:2003dm}.
 We therefore consider the next order term in the operator
 product expansion which must dominate at high enough energy.

\subsubsection{Non-vanishing contribution in the charge conjugation even ensemble.} 
The next order expressions analogous to Eqs.~(\ref{odderon_1}) and
\eqref{odderon} are long and cumbersome. We therefore will not present them
here in full generality. Instead  we directly present the result of averaging
using a simple averaging model.

The averaging over the target fields is performed in the following way. We
first note that 
 since we are probing the correlators of the eikonal factors $S$ on short
 distance scales,  we can write without any loss of generality\footnote{{
	 The reasoning behind this is the following. We will need to calculate
	 averages of products for the  eikonal factors of the type $\langle S(x_1)...S(x_n) \rangle $
	 where all distances are smaller than the inverse saturation momentum of
	 the target, $|x_i-x_j|<1/Q_s$. 
For translationally invariant target, which we assume here, we can shift all
the coordinates to vicinity of zero, so that $|x_i|<1/Q_s$. We can now fix the
convenient gauge $S(x=0)=1$. This can always be done by using a residual $x^-$
independent gauge transformation without leaving the light cone gauge.
Eq.~(\ref{model}) is just expansion of the phase of the eikonal factor to
leading order in $x$, which is a good approximation since in the regime of
interest $x$ is smaller than the correlation length of the color electric
fields. 
We note that this form of the eikonal factor yields the Golec-Biernat--Wusthoff
model~\cite{GolecBiernat:1998js,GolecBiernat:1999qd} of the  dipole cross
section. The McLerran-Venugopalan model at short distances corresponds to an
additional slow dependence of the field $E$ on $x$ in Eq.~(\ref{model}). This
can be easily incorporated in our formulae by considering $\lambda^2$ to be a
slowly (logarithmically) varying function of the external momenta $p$ and $k$.
Although this may be important for quantitative comparisons to the data,  it
does not affect qualitative features of our results and at the level of
accuracy of the present paper is clearly irrelevant. }}
\begin{equation}\label{model}
S(x)=\exp \{iT^aE_i^ax_i\}\,.
\end{equation}

To calculate averages of local products we need to take a maximum of three
derivatives  on each factor of $S$, and then set $x=0$. Under this procedure
\begin{equation}\label{derivatives}
\partial_sS(x)\rightarrow iT^aE^a_s; \ \ \ \ \partial_r\partial_sS(x)\rightarrow -\frac{1}{2}\{T^a,T^b\}E^a_sE^b_t;\ \ \ \ \partial_t\partial_r\partial_sS(x)\rightarrow -i\frac{1}{6}\{T^a,T^b,T^c\}E^a_sE^b_tE^c_r;
\end{equation}
where $\{T^a,T^b,T^c\}$ denotes the sum of all permutations of $a,b,c$. To average over the color electric field $E$ we then use a simple Gaussian ensemble~\cite{Skokov:2014tka} with the basic two point function of the form
\begin{equation}
	\label{Eq:Target_Ave}
\langle E_i^aE_j^b\rangle_\alpha=\lambda^2\delta^{ab}\delta_{ij}\,.
\end{equation}

The details of the calculation are presented in Appendix~\ref{Ap:Op}. 
The final expressions are
\begin{equation}
A_1(k,p)\equiv
\langle 
\langle 12\rangle\langle 34\rangle \rangle_\alpha =
7 S_\perp
N_c^3 (N_c^2-1) \frac{\lambda^4 \mu^4 }{k^6} 
\delta^2(p+k)
\int_q 
\frac{1}{(k-q)^4} 
\Bigg(
\frac32 - 2 \frac{(k\cdot q)^{2}}{ k^2 q^2} - 5 \frac{(k\times q)^2 k\cdot(k-q) } {(k-q)^2 q^2 k^2}
\Bigg), 
\end{equation}

\begin{eqnarray}
	&&A_2(k,p)\equiv  \langle \langle 13\rangle\langle 24\rangle  \rangle_\alpha =-\frac{S_\perp}{4}N_c^3(N_c^2-1)\frac{\mu^4\lambda^4}{p^6k^6}(k\cdot p)\label{169_last},
\end{eqnarray}
and finally 
\begin{eqnarray}
&&A_3\equiv  \langle  \langle 14\rangle\langle 23\rangle\rangle_\alpha =
N_c^3(N_c^2-1)S_\perp \frac{\mu^4\lambda^4}{k^4p^4}\Bigg\{\\
&&-7\frac{k\cdot p}{k^2p^2}-3\frac{(k\cdot p)^3}{k^4p^4}+\frac{15}{2}\frac{k\cdot (k-p)p\cdot(k-p)}{k^2p^2(k-p)^2}+\frac{k\cdot p(p\cdot(k-p))^2}{k^2p^4(k-p)^2}+\frac{k\cdot p(k\cdot(k-p))^2}{k^4p^2(k-p)^2}\nonumber\\
&&+\frac{1}{4}\frac{(k\cdot(k-p))^2}{k^2(k-p)^2}\left(\frac{5}{k^2}-\frac{7}{p^2}\right)+\frac{1}{4}\frac{(p\cdot(k-p))^2}{p^2(k-p)^2}\left(\frac{5}{p^2}-\frac{7}{k^2}\right)+\frac{7}{2}\left(\frac{1}{k^2}+\frac{1}{p^2}\right)\frac{k\cdot pk\cdot(k-p)p\cdot(k-p)}{k^2p^2(k-p)^2}\nonumber\\
&&+\frac{3}{8}\left[\frac{k\cdot(k-p)}{k^2(k-p)^2}-\frac{p\cdot(k-p)}{p^2(k-p)^2}\right]\nonumber\Bigg\}.
\end{eqnarray}

The expression $A_1$ yields ``hard'' back to back production. This is the analog of a similar term in our calculation of pair density in the projectile wave function, and we neglect it for the same reason.
We now combine all the terms to obtain
\begin{equation}
\frac12 \left[
	\frac{d^6N}{d^2k d^2pd\eta_k d\eta_p} (k,p) 
	-  
	\frac{d^6N}{d^2k d^2pd\eta_k d\eta_p} (k,-p) 
\right]
=
2g^2
\left[A_2(k,-p)+A_3(k,-p)-A_2(k,p)-A_3(k,p)\right]\,.
\end{equation}

\begin{figure}
	\centering
	\includegraphics[width=0.49\textwidth]{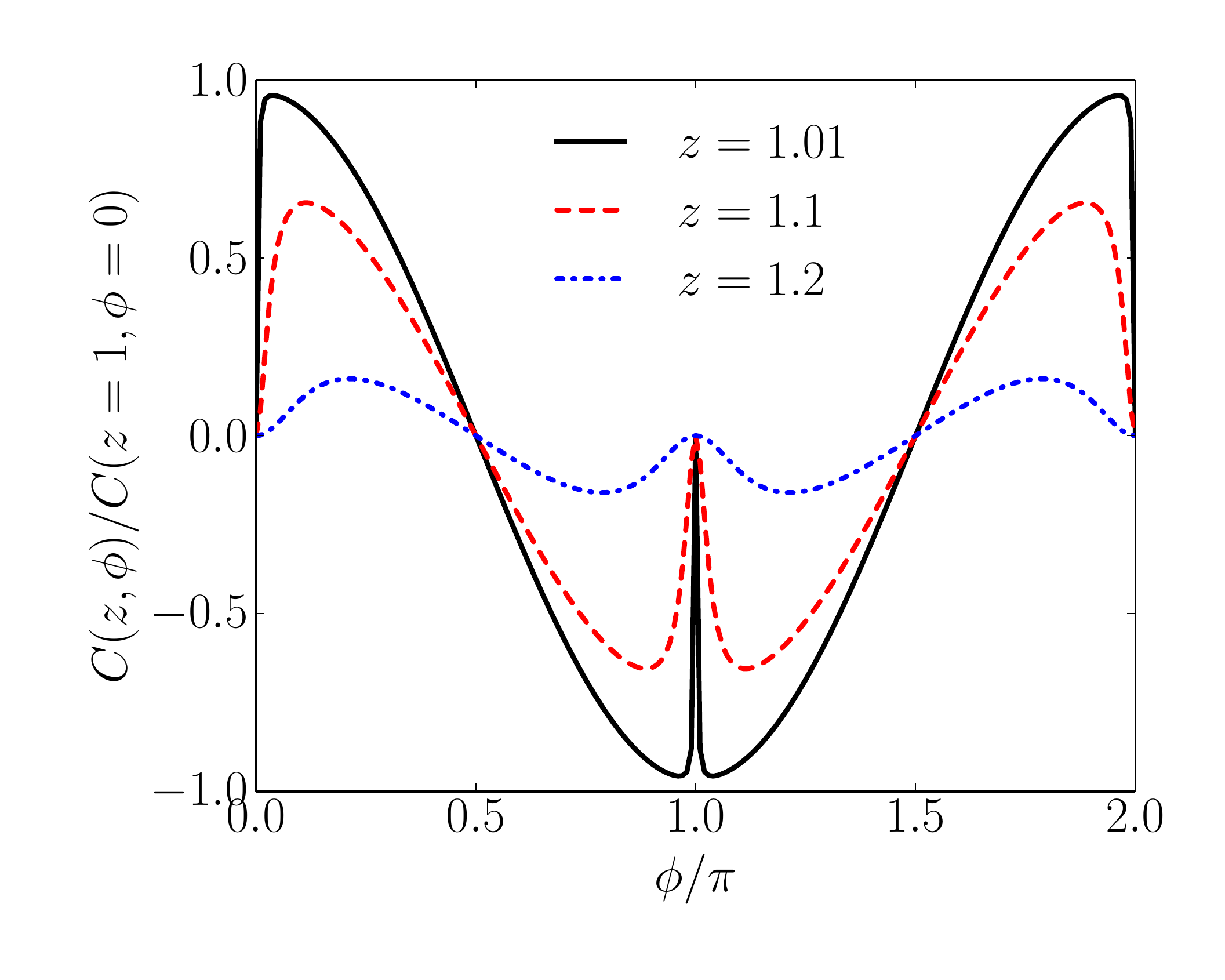}
	\includegraphics[width=0.49\textwidth]{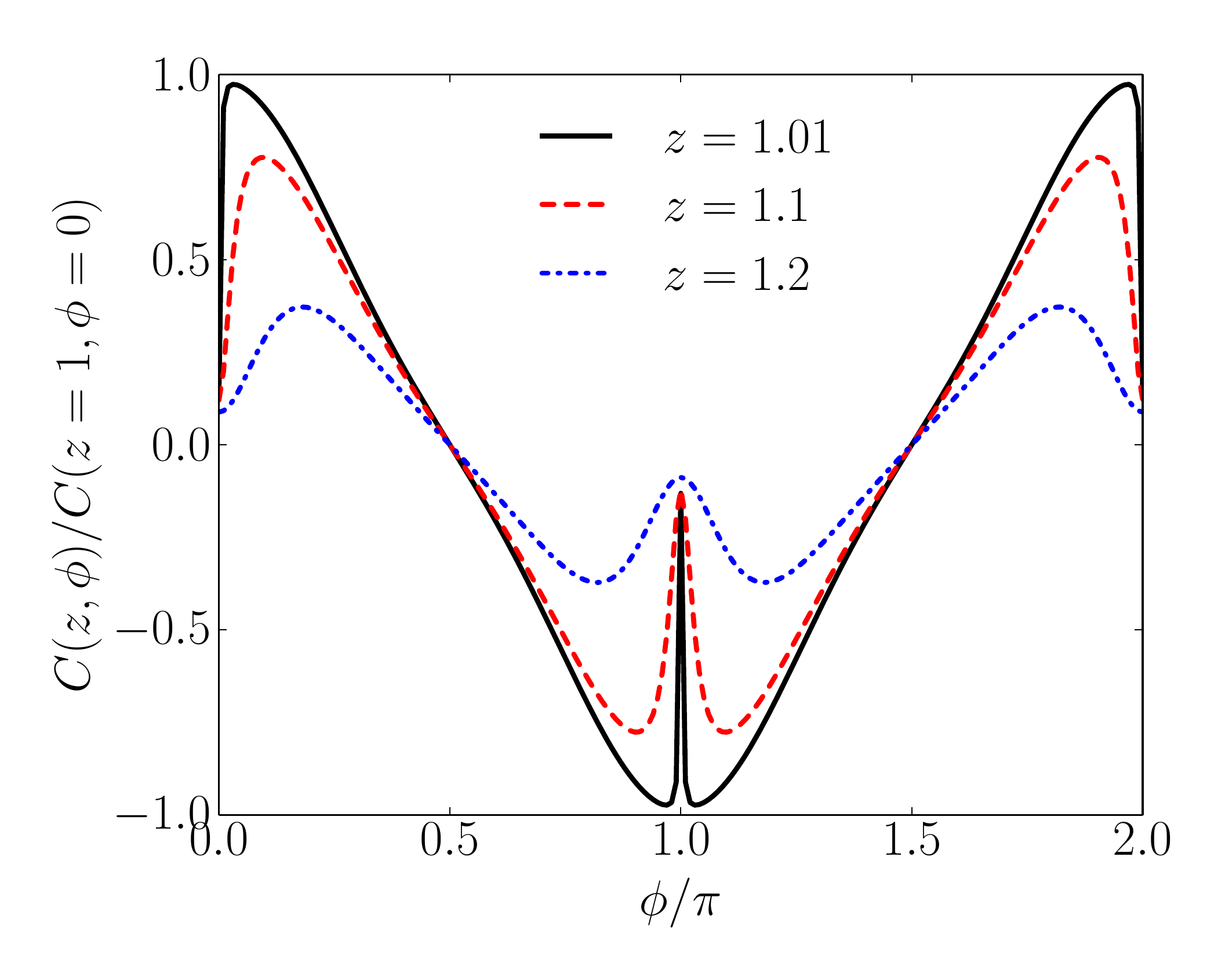}
	\caption{The correlation function as a function of the azimuthal angle, $\phi$ 
	for different values of $z = p/k$. The correlation functions are defined 
	in the text and normalized by $C(z=1,\phi=0)$. Left panel - correlation in the projectile wave function. Right panel - correlation in particle production.
	}
	\label{fig:cfn}
\end{figure}

In Fig.~\ref{fig:cfn} we show the correlation functions
for produced gluons  (right) 
and for gluons in the projectile wave function (left).  
For the former, we defined the correlation function using Eq.~\eqref{Eq:CorrelF} and 
normalizing by the uncorrelated piece $S_\perp^2 \mu^4 \lambda^4 /(k^4 p^4)$ 
\begin{equation}
\frac12 \frac{ 	\frac{d^6N}{d^2k d^2pd\eta_kd\eta_p} (k,p) 
	-  
	\frac{d^6N}{d^2k d^2pd\eta_kd\eta_p} (k,-p)   } {\frac{N_c^4S_\perp^2 \mu^4 \lambda^4}{k^4 p^4}}   
	= \frac{\alpha_sN_c}{S_\perp p^2} C(z = p/k, \phi).  
\label{Eq:CtoPlot}
\end{equation}
For the latter, the correlations function is defined similarly. 

\begin{figure}
	\centering
	\includegraphics[width=0.49\textwidth]{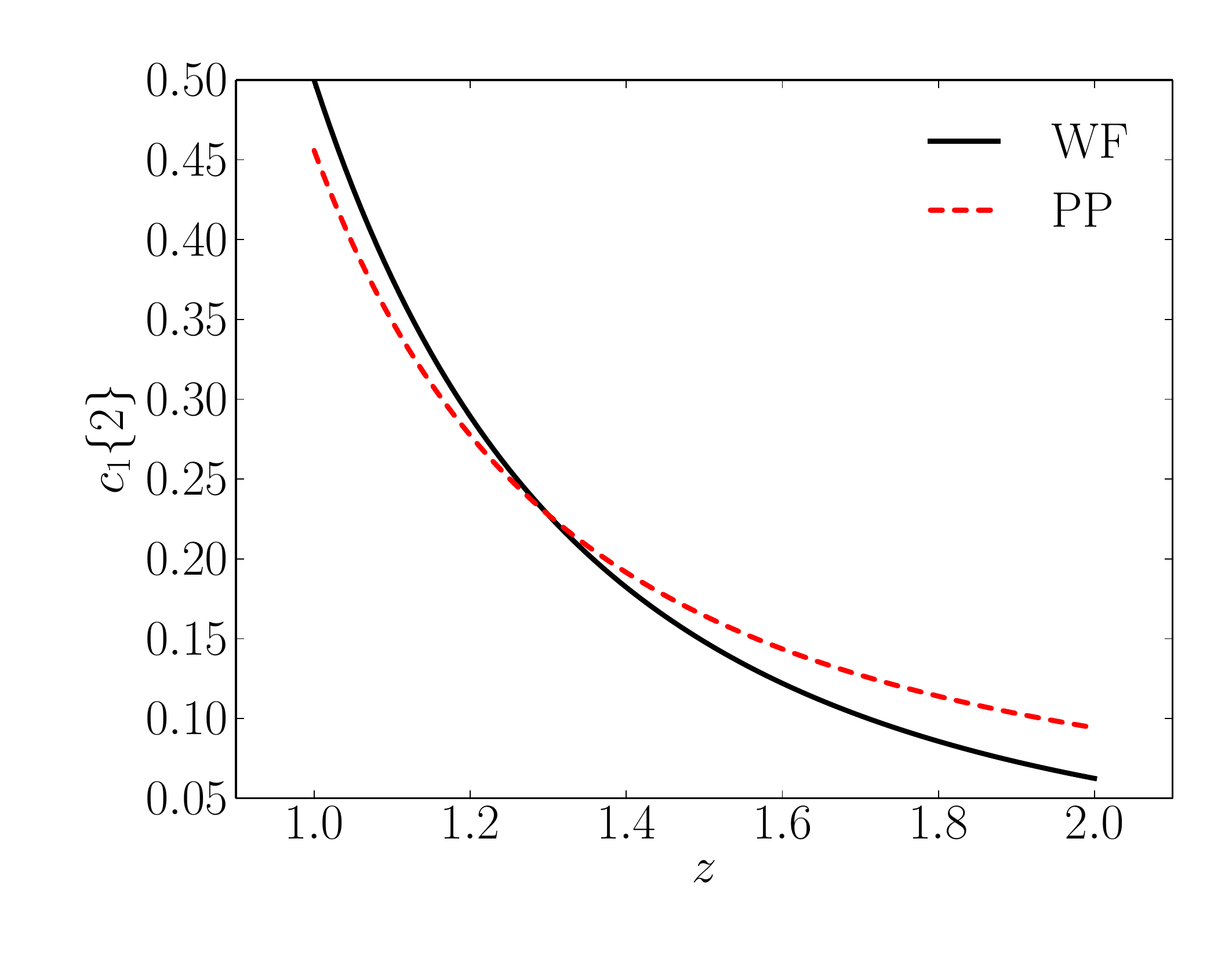}
	\includegraphics[width=0.49\textwidth]{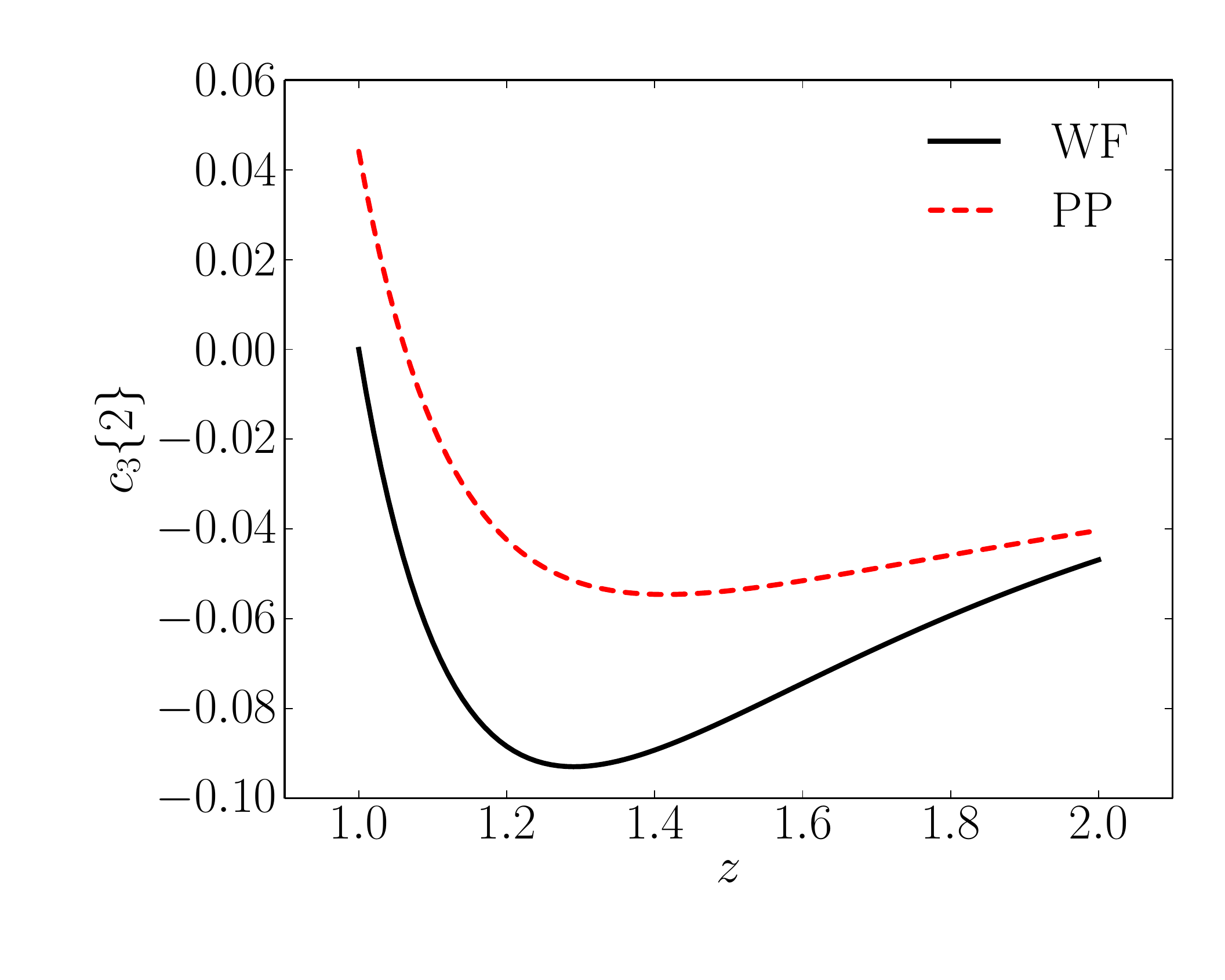}
	\caption{The first and the third cumulants  as a function of $z = p/k$
		obtained from the correlation function the projectile wave function (WF) and 
		from double inclusive particle production (PP).
	}
	\label{fig:cum}
\end{figure}

The first and the third harmonics of the correlations functions are shown in 
Fig.~\ref{fig:cum}.

\section{Discussion and Conclusions.}

 Note that the odd part of the production cross section is proportional to
 $\mu^4\lambda^4$. This means that if either the projectile or the target is
 dilute, this contribution to the correlated production vanishes. In this
 respect it is similar to the even contribution from ``glasma graphs''. As is
 clear from Eq.~(\ref{Eq:CtoPlot}) the leading contribution to the odd part of
 the correlation function is the same order in the color charge density  as
 that to the even part, and is just suppressed by one power of $\alpha_s$.

Another important point is that just like the glasma graph contribution, the
odd contribution is long range in rapidity. In fact our result
Eq.~(\ref{Eq:CtoPlot}) is rapidity independent. Some dependence on rapidity
separation between the gluons will undoubtedly appear once the rapidity
difference is large enough $|\eta_k-\eta_p|\sim 1/\alpha_s$. This effect is not
accounted for in our calculation.

The present calculation does not include contributions to two particle
production arising from a single Pomeron exchange. This contribution ( at the
leading order in weak field expansion) is proportional to
$\alpha_s\mu^2\lambda^2$ and obviously does not appear in our formulae. For
strong fields these contributions are subleading and for that reason they are
not contained in the CGC wave function.
This single Pomeron mechanism leads predominantly to back-to-back minijet
production which can  in principle contribute to nonvanishing odd azimuthal
anisotropy. Such back-to-back jets are however subtracted in the experimental
analysis and are of no interest to us here. 

{ Our calculation is based on the CGC wave function derived in the limit of
	the dense projectile. Recall that in Ref.~\cite{Altinoluk:2009je} this wave function was
	obtained in the leading order of light cone perturbation theory in strong
	background with $\rho\sim 1/g$. This wave function differs from the "dilute
	CGC" coherent state by the Bogoliubov squeezing prefactor which is an $O(1)$
	correction in this parametric regime.  In this sense the squeezing 
	provides the most important correction to the wave function. All other
	corrections not included in this Bogoliubov factor lead to terms suppressed
	by powers of the QCD coupling on the level of the wave function.

We have indeed  shown that the full account of the squeezing leads to
an $O(1)$ correction to the ``glasma graph''  results both in the single
inclusive and the double inclusive production cross sections.
 This $O(1)$ correction unfortunately preserves the accidental symmetry $(k,p)\rightarrow
 (k,-p)$ observed in the dilute regime. However the expression for the double
 inclusive production also contains a term odd under this transformation. This
 term is suppressed by a single power of $\alpha_s$ relative to the glasma
 graphs contribution. Our strategy in this paper was to take this odd contribution at its
 face value and explore its consequences. 

 In order to get our numerical estimates we had to expand the production cross
 section to leading order in powers of $\rho$.  Excluding the odderon, the
 leading contribution is of order $g^6\rho^4$.  In this order our calculation
 should be interpretable in terms of a finite number of Feynman diagrams. We have
 tentatively identified the relevant graphs in Fig.~\ref{fig:Diagrams}. We note  that the
 procedure employed here does not sum all the Feynman diagram contribution to
 the double inclusive cross section at order $g^6\rho^4$. For example the
 running coupling correction to the glasma graphs is absent. However the
 physical  feature that allows  for appearance of the odd contribution in our
 calculation is the nonfactorizable production of the two gluons (configuration
 by configuration at fixed color charge density). We believe that including of
 the additional factorizable terms at the same order in $\alpha_s$ (like the running coupling correction) will have no
 effect on the calculation of $v_3$.

  }

We now discuss qualitative features of our results.

Consider first the shape of the  (scaled, see Eq.~(\ref{Eq:CtoPlot}))
correlation function in Fig.~\ref{fig:cfn}. Qualitatively, the gluon pair density in the projectile CGC
wave function is consistent with the expectations based on KLM argument
outlined in the introduction. Indeed the pair density has a strong peak in
forward direction.  At $z\approx 1$ the peak is close to $\phi=0$. As $z$ grows
the peak decreases in hight and moves to larger angles,  but always stays at
$\cos\phi>1/\sqrt{3}$. These properties remain practically unaltered in the
double inclusive gluon production. The overall shape of the production cross
section resembles closely the form of the gluon pair density in the projectile
wave function.

 Note the overall normalization of the double inclusive production  amplitude.
As expected, it is suppressed by the factor of $S_\perp p^2$, reflecting the fact that
the sources of correlation are local in the coordinate space. If we were able
to calculate production for $p^2,k^2\sim Q_s^2$, the suppression factor would
presumably be $S_\perp Q_s^2$. This is exactly the same as that of the local
anisotropy \cite{Kovner:2010xk,Kovner:2012jm,Dumitru:2014dra} and the ``glasma
graph'' \cite{Dumitru:2010iy}, or Bose enhancement \cite{Altinoluk:2015uaa}
contributions.  Our result has an additional suppression by a factor of
$\alpha_s$ relative to those contributions, however it is leading at large
$N_c$ whereas both the glasma graphs and the local anisotropy are order
$1/N_c^2$~\cite{Dumitru:2010iy,Dumitru:2014vka}. At $N_c=3$ and
$\alpha_s\sim .2$ the relative importance of these contributions is determined by a
numerical factors of order one, which may well be model-dependent. It does
however raise an interesting possibility that additional contribution to $v_2$
that should arise from using the amended CGC wave function can be competitive
with the contributions hitherto considered, or even larger than those  at least
close to the  saturation momentum. We have not calculated here the even part of
the correlation function which gives rise to $v_2$, but such calculation is
certainly feasible and will be reported elsewhere.
 
Next consider Fig.~\ref{fig:cum} which shows the coefficients of the first and third
harmonics in the correlation function. The coefficient $c_1\{2\}$ is positive
both for the pair density in the wave function and for pair production in
scattering, consistent with the KLM expectation. There is no significant
difference between the two, although it drops slower with $z$ for production.
The situation is different for $c_3\{2\}$. It is always negative for the pair
density in the wave function. For particle production $c_3\{2\}$ is negative
for most of $z$ interval, but it changes sign and is positive for $z$ close to
one, approximately for $.9<z<1.1$.

We do not have a good understanding why $c_3\{2\}$ changes sign due to
scattering. We can nevertheless argue why the form of the pair density should
be  distorted the most when the two momenta are close to each other in
magnitude. Remember that we are working in the regime where the momenta of
produced particles are much larger than the momentum transfer from the target.
In this regime most of the momentum of the produced gluons is inherited from
the projectile wave functions. The role of the scattering is then mostly to
decohere these gluons from the incoming wave function and put them on shell. If
the relative momentum of the gluons is large, they sit close to each other (in
coordinate space) in the projectile wave function, and it is difficult to
decohere them. This is always the case if $z\gg 1$ (or $z\ll 1$). However the
``bin'' with $z\approx 1$, contains two gluon configurations which have small
relative momentum and thus are well separated in the coordinate space. Such
gluons  probe widely separated regions of the target and acquire significantly
different eikonal phase during the propagation. The scattering is therefore
much more efficient in decohering such configurations from the incoming wave
function. Thus we expect that in the $z\approx 1$ ``bin'' the scattering skews
the distribution of pairs towards the ones which have same sign transverse
momenta. This may be the mechanism that produces a positive $c_3\{2\}$ in this
transverse momentum bin.

	The fact that we obtain positive $c_3\{2\}$ in a narrow range of $z$ is rather
interesting. When $v_3$ is determined by taking the
square root of $c_3\{2\}$ keeping both the ``trigger'' and the ``associated''
particle in the same momentum bin,  our results yield a
positive $c_3\{2\}$ and therefore real $v_3$.  Increasing the momentum  of the
bin $p$ as per our results would lead to a rather slow decrease of $v_3$ as
$1/p$. However if one samples the trigger and associated particles from
different momentum bins, the $v_3$ should decrease much faster as seen in experiment, see e.g.~Ref.~\cite{Aad:2014lta}. In fact it
should cease being real when the two bins are separated by about $10-20\%$ of
the central value. It would be very interesting to see if such trend can be
traced in the data at lower multiplicities and higher trigger momentum.

Finally we want to comment on the range of applicability of the improved CGC
wave function that we have utilized in the current calculation. In Refs.~\cite{Kovner:2007zu,Altinoluk:2009je}
it was derived as the wave function containing the bulk of the probability
density in the field space. The tails of the wave function at large values of
the field $\phi$ however are not described well by this Gaussian shape. For
that reason it should be used with caution. For example one cannot use this
wave function to describe nucleus-nucleus scattering. In the latter case the
scattering modifies the field $b$ by a ``factor'' of order one. To calculate the
scattering amplitude one would then need to take an overlap of two Gaussians
with central values displaced by an amount  of order $\phi_1-\phi_2\sim
1/\alpha_s$. Such an overlap is dominated by the tails of the two wave
functions which are not under control in our approximation. Similarly, by a
straightforward counting of powers of $\alpha_s$ on can see that the accuracy
of the present approximation is not sufficient to calculate particle production
when the number of particles produced  in the collision is $O(1/\alpha_s)$. Even
though the wave function Eq~.(\ref{Eq:CGC}) is more accurate than the dilute
limit Eq.~(\ref{dilute}), it is not accurate enough for processes where the
number of produced particles is parametrically large. The natural habitat of
Eq.~(\ref{Eq:CGC}) is in situations where the number of produced particles in
the collision  is $O(1)$. Note that this is still parametric improvement over
Eq.~(\ref{dilute}) which is strictly speaking valid only in the dense-dilute
regime  where the number of produced particles is $O(\alpha_s)$.
Moreover, this regime is of relevance for minimal bias p-p collisions where the
correlations have been recently observed~\cite{Aad:2015gqa,Aaboud:2016yar}. With some measure of
optimism we can hope that at least qualitatively we can trust our results also
in p-p the events with higher than average multiplicity, although probably not
for very high multiplicity events.

\begin{acknowledgements}
V.S. thanks A. Bzdak for comments and discussions. 
We are indebted to the organizers of Workshop on QCD 
and Diffraction - ``Saturation 1000+'', during which the 
manuscript was finalized.  M.L. thanks  the University of Connecticut  while A.K thanks the Ben-Gurion University of 
the Negev for the hospitality  when parts of this work were done. 
The research was supported by 
the NSF Nuclear Theory grant 1614640 (A.K.); the  Israeli Science Foundation grants \# 1635/16 and \# 147/12 (M.L.);  the
BSF grants \#2012124 and \#2014707 (A.K., M.L.);  the People Program (Marie Curie Actions) of the European Union's Seventh 
Framework under REA grant agreement \#318921 (M. L) and the COST Action CA15213 THOR (M.L.).
\end{acknowledgements}

\appendix 
\section{The CGC wave-function}
\label{CGCwtf}

In this appendix we sketch the derivation of the Gaussian CGC ground state. For details of the derivation the reader should consult the original work \cite{Kovner:2007zu, Altinoluk:2009je}, while here we only provide the main logical steps of the derivation.

First, we note that the CGC ground state is the ground state of the soft gluons in the presence of eikonal coupling to valence modes represented by the color charge density $\rho(x)$. We thus have to diagonalize the light cone Hamiltonian
\begin{equation}
H_{\rm soft}=H_{\rm LFQCD}+\int d^2x\rho^a(x)\int d\eta \partial^i A^a_i(x,\eta)\, ,
\end{equation}
where $H_{\rm LFQCD}$ is the QCD Hamiltonian on the light front, and $\eta$ is rapidity.
It is obvious that shifting the field  one can get rid of the last term in the
Hamiltonian since it is linear in the field $A_i$. This shift is affected by the
coherent operator ${\cal C}=\,e^{i 2 \int_k  b_{\alpha i }(-k) 
	\phi_{\alpha i} (k) }$ so that
\begin{equation}
{\cal C}H_{\rm soft}{\cal C^\dagger}=H_0+\delta H_2+\delta \tilde H\,.
\end{equation}
Here $\delta H_2$ is quadratic in the soft gluon operators, while $\delta\tilde H$ contains third and fourth power of $A_i$.	

When the color source $\rho$ is small, i.e. $\rho^a=O(g)$, this shift
diagonalizes the Hamiltonian up to terms of order $g$, i.e. $\delta H_2=O(g^2)$
and $\delta \tilde H=O(g^3)$. Thus in the weak filed limit the CGC ground state
(up to higher order perturbative corrections)  is the coherent state 
\begin{equation}
	|{\rm CGC}\rangle_{\rm dilute}={\cal C}|0\rangle\,,
\end{equation}
where $|0\rangle$ is the light cone vacuum of the soft gluons.

However in the strong field limit, where $\rho=O(1/g)$, the situation is  different,
since $\delta H_2=O(1)$ while $\delta\tilde H=O(g)$. Thus in order to
diagonalize the total Hamiltonian at order one, it is still necessary to
diagonalize its  nontrivial quadratic part  $H_0+\delta H_2$. It can be achieved by a Bogoliubov
transformation owing to quadratic dependence on the field.  The action of this Bogoliubov transformation on the soft gluon
creation and annihilation operators was found in Ref.~\cite{Kovner:2007zu}. 
The transformation is generated by  unitary operator of the Gaussian form, schematically
\begin{equation}
{\cal B}=\exp\{-\frac{1}{2}(a+a^\dagger)\Lambda (a+a^\dagger)\}\,.
\end{equation} 
Thus at the end of the day in the strong field limit the diagonalizing operator can be written in the form
 $\Omega = {\cal C B}$ and the wave function of the CGC ground state
\begin{equation}
	\Psi_{\rm CGC}[\phi] = 
\langle \phi |  \Omega	|0\rangle
=
	\,e^{i 2 \int_k  b_{\alpha i }(-k) 
	\phi_{\alpha i} (k) } 
	\langle \phi |  {\cal B}\rangle\;,
	\label{Eq:CGC_1}
\end{equation}
where for convenience we separately defined  the Gaussian state $|{\cal B}\rangle$ 
as  
\begin{equation}
	 |  {\cal B}\rangle\equiv  {\cal B}|0\rangle;\ \ \ \ \ \ \ \   \langle \phi |  {\cal B}\rangle
	={\cal N}
	e^{-
	\frac12 \int_{k, p}  B^{-1}_{\alpha \beta i j} (k, p ) 
	\phi_{\alpha i} (k)
	\phi_{\beta j} (p)
	}\;.
	\label{Eq:CalB1}
\end{equation}
As discussed in Ref.~\cite{Altinoluk:2009je} this calculation can be given a
diagrammatic interpretation. In particular the action of the coherent operator
on the soft gluon vacuum is represented by (exponentiating) the sum of the tree
level diagrams  and corresponding virtual corrections, see Fig.3.

\begin{figure}[hbt]
\begin{center}
\vspace*{-0.2cm}
\includegraphics[width=8cm]{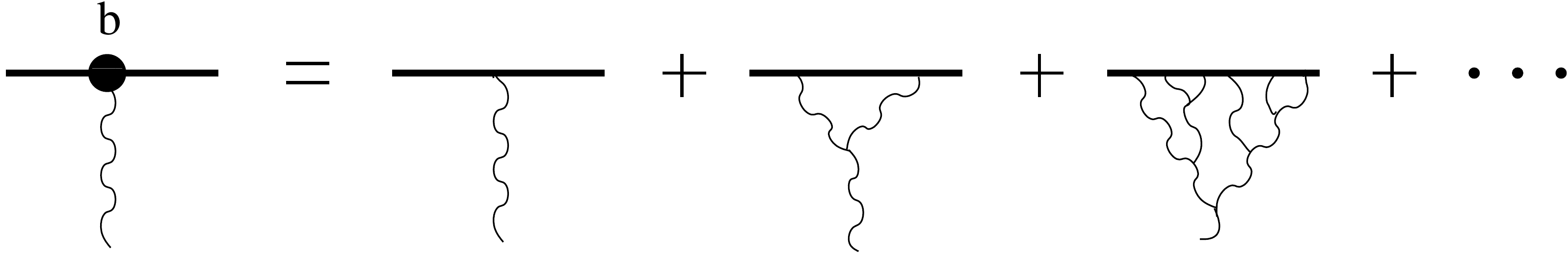}
\end{center}
\label{fig3}
\caption{The tree level diagrams representing the action of the coherent operator ${\cal C}$ on the soft gluon vacuum.}
\end{figure}

The action of the Bogoliubov operator is represented by (exponentiating) the
sum in Fig. 4, including appropriate virtual corrections. 

\begin{figure}[hbt]
\begin{center}
\vspace*{-0.2cm}
\includegraphics[width=8cm]{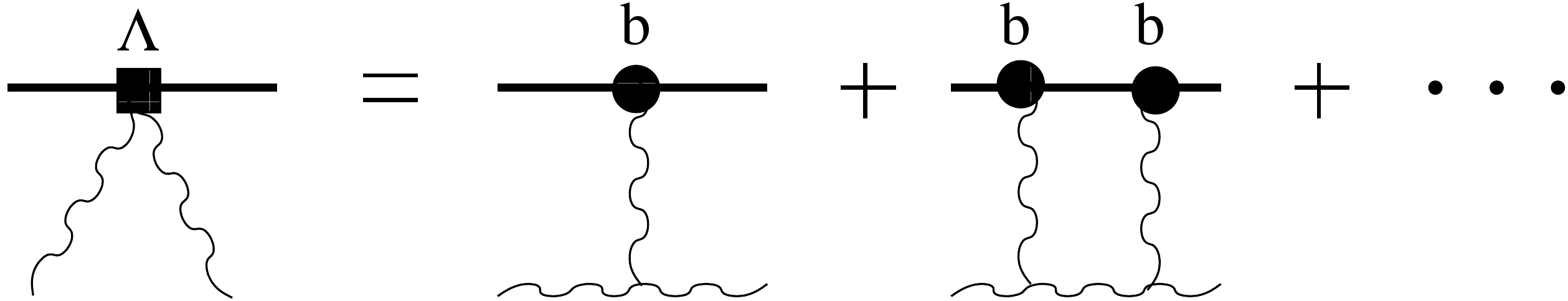}
\end{center}
\label{fig4}
\caption{ The diagrams representing the action of the Bogoliubov operator on the soft gluon vacuum.}
\end{figure}

Thus the diagrammatics of the CGC vacuum wave function Eq.~\eqref{Eq:CalB1} is
that of independent emission of single gluons and pairs of gluons  with
subsequent rescattering of these gluons on the valence charge  density.

It was explicitly shown  in Ref.~\cite{Kovner:2007zu, Altinoluk:2009je} that
the presence of the Bogoliubov operator ${\cal B}$ is crucial for derivation of
the JIMWLK evolution equation. 
Although the derivation in Ref.~\cite{Kovner:2007zu, Altinoluk:2009je}  is fairly
complicated, as it deals explicitly with the contributions of all the rapidity
modes of the soft gluon field, it was recognized in those papers that only
limited amount of information is necessary
in order to reproduce the JIMWLK equation. In particular it was shown that all that is required of the Gaussian state $|{\cal B}\rangle$ is that it correctly encodes the correlation function
\begin{equation}
	\langle {\cal B}|\phi(x)\phi(y)|{\cal B}\rangle=(1-  \mathfrak{l}-L)^2(x,y)\,,
\end{equation}
where $\phi(x)$ is the integrated over rapidity field Eq.~(\ref{field}). In the present paper, 
this property of the Gaussian state was used  to infer the form of
the state after integration over all the rapidity modes except for  the constant mode $\phi(x)$.





\section{Expansion in the classical field.}
\label{App:AA_exp}
In this Appendix we derive Eq.~(\ref{Eq:AAA}) -  the expansion of the main calculational block for the  double inclusive production cross section to leading order in the projectile charge density (or equivalently $b$). {
	For notational simplicity we rescale the classical field $b$ via $b\rightarrow \frac{1}{g} b$.}
 Using Eq.~(\ref{1/D}) we have 
\begin{equation}
	1-\mathfrak{l}-L\approx \mathfrak{t}-\mathfrak{l}+b\frac{1}{\partial^2}\partial+\partial\frac{1}{\partial^2}b-\frac{\partial}{\partial^2}[\partial b]_+\frac{\partial}{\partial^2}-b\frac{1}{\partial^2}b+b\frac{1}{\partial^2}[\partial b]_+\frac{1}{\partial^2}\partial+\frac{\partial}{\partial^2}[\partial b]_+\frac{1}{\partial^2}b+\frac{\partial}{\partial^2}b^2\frac{\partial}{\partial^2}-\frac{\partial}{\partial^2}[\partial b]_+\frac{1}{\partial^2}[\partial b]_+\frac{\partial}{\partial^2}
\end{equation}
and
\begin{equation}
\Gamma=
1+b\frac{1}{\partial^2}\partial-\partial\frac{1}{\partial^2}b-\frac{\partial}{\partial^2}[\partial b]_-\frac{\partial}{\partial^2}-b\frac{1}{\partial^2}b+b\frac{1}{\partial^2}[\partial b]_+\frac{1}{\partial^2}\partial+\frac{\partial}{\partial^2}[\partial b]_-\frac{1}{\partial^2}b-\frac{\partial}{\partial^2}b^2\frac{\partial}{\partial^2}-\frac{\partial}{\partial^2}[\partial b]_-\frac{1}{\partial^2}[\partial b]_+\frac{\partial}{\partial^2}
\end{equation}
with the inverse 
\begin{equation}
\Gamma^{-1}=
1-b\frac{1}{\partial^2}\partial+\partial\frac{1}{\partial^2}b+\frac{\partial}{\partial^2}[\partial b]_-\frac{\partial}{\partial^2}-b\frac{1}{\partial^2}\partial b\frac{1}{\partial^2}\partial+\frac{\partial}{\partial^2}b\partial\frac{1}{\partial^2}b-\frac{\partial}{\partial^2}b\partial\frac{1}{\partial^2}b\partial\frac{\partial}{\partial^2}+\frac{\partial}{\partial^2}\partial b\frac{1}{\partial^2}\partial b\frac{\partial}{\partial^2} .
\end{equation}
The expression we need is 
\begin{equation}\label{aat}
\mathbb{A}\mathbb{A}^T=\bar\Gamma^{-1}S\Gamma\Gamma^TS^T\bar \Gamma^{-1}=\bar\Gamma^{-1}S(\mathfrak{t}-\mathfrak{l})B(\mathfrak{t}-\mathfrak{l})S^T\bar \Gamma^{-1T}
\end{equation}
so 
\begin{equation}
(\mathfrak{t}-\mathfrak{l})B(\mathfrak{t}-\mathfrak{l})=1- b_i \frac{1}{\partial^2} b_j - \partial_i \frac{1}{\partial^2} b^2 \frac{1}{\partial^2} \partial_j 
	+ \frac{\partial_i}{\partial^2} (\vec{\partial} \cdot \vec{b})  \frac{1}{\partial^2} b_j  
	+ b_i \frac{1}{\partial^2} (\vec{b} \cdot \vec{\partial})  \frac{\partial_j}{\partial^2} 
	+ \frac{\partial_i}{\partial^2} \left[ 
		(\vec{b} \cdot \vec{\partial}) \frac{1}{\partial^2} 	(\vec{\partial} \cdot \vec{b})  
		-
		(\vec{\partial} \cdot \vec{b}) \frac{1}{\partial^2} 	(\vec{b} \cdot \vec{\partial})  
	\right] \frac{\partial_j}{\partial^2} .
\end{equation}
We only need expression in Eq.~(\ref{aat}) to order $b^2$
\begin{eqnarray}
&&\mathbb{A}\mathbb{A}^T-1\approx \bar\Gamma^{-1}\bar \Gamma^{-1T}+S(\mathfrak{t}-\mathfrak{l})\tilde B(\mathfrak{t}-\mathfrak{l})S^\dagger-1\\
&&= \bar b_i \frac{1}{\partial^2} \bar b_j + \partial_i \frac{1}{\partial^2} \bar b^2 \frac{1}{\partial^2} \partial_j 
	-\frac{\partial_i}{\partial^2} (\vec{\partial} \cdot \vec{\bar b})  \frac{1}{\partial^2}\bar b_j  
	-\bar  b_i \frac{1}{\partial^2} (\vec{\bar b} \cdot \vec{\partial})  \frac{\partial_j}{\partial^2} 
	- \frac{\partial_i}{\partial^2} \left[ 
		(\vec{\bar b} \cdot \vec{\partial}) \frac{1}{\partial^2} 	(\vec{\partial} \cdot \vec{\bar b})  
		-
		(\vec{\partial} \cdot \vec{\bar b}) \frac{1}{\partial^2} 	(\vec{\bar b} \cdot \vec{\partial})  
	\right] \frac{\partial_j}{\partial^2}\nonumber\\
&&- Sb_i \frac{1}{\partial^2} b_jS^\dagger - S\partial_i \frac{1}{\partial^2} b^2 \frac{1}{\partial^2} \partial_jS^\dagger 
	+S \frac{\partial_i}{\partial^2} (\vec{\partial} \cdot \vec{b})  \frac{1}{\partial^2} b_jS^\dagger  
	+ Sb_i \frac{1}{\partial^2} (\vec{b} \cdot \vec{\partial})  \frac{\partial_j}{\partial^2} S^\dagger
	\nonumber \\&&+S \frac{\partial_i}{\partial^2} \left[ 
		(\vec{b} \cdot \vec{\partial}) \frac{1}{\partial^2} 	(\vec{\partial} \cdot \vec{b})  
		-
		(\vec{\partial} \cdot \vec{b}) \frac{1}{\partial^2} 	(\vec{b} \cdot \vec{\partial})  
	\right] \frac{\partial_j}{\partial^2}S^\dagger\nonumber. 
\end{eqnarray}
In order to perform Fourier transformation we use Eq.~(\ref{btilde})
\begin{eqnarray}
&&\left[\mathbb{A}\mathbb{A}^T-1\right](k,p)\approx\int_q
	\bar b_n(k-q) \left[ -
		\mathfrak{t}_{i n}(k) 
		\frac{1}{q^2} 
		\mathfrak{t}_{mj}(p) +
		\frac{k_{ i}}{k^2} 
		\mathfrak{t}_{nm}(q) 
		\frac{p_{ j}}{p^2} 
	\right] 
	\bar b_m(p+q)\\
&&+\int_{q,l,m} S(k-l)
	 b_n(l-q) \left[ 
		\mathfrak{t}_{i n}(l) 
		\frac{1}{q^2} 
		\mathfrak{t}_{mj}(m) -
		\frac{l_{ i}}{l^2} 
		\mathfrak{t}_{nm}(q) 
		\frac{m_{ j}}{m^2} 
	\right] 
	 b_m(m+q)		S^\dagger(p-m)\nonumber .
		\end{eqnarray}

Using Eq.~(\ref{btilde1}) we get
\begin{eqnarray}
&&\left[\mathbb{A}\mathbb{A}^T-1\right](k,p)\approx\int_q
	\bar c(k-q) \left[
		\mathfrak{t}_{i n}(k) 
		\mathfrak{l}_{n m}(q) 
		\mathfrak{t}_{mj}(p) + 
		\mathfrak{l}_{i n}(k) 
		\mathfrak{t}_{nm}(q) 
		\mathfrak{l}_{j m}(p) 
	\right] 
	\bar c(p+q)\\
	&&-
	\int_{q,l,m}
	S( k- l) c(l-q) \left[
		\mathfrak{t}_{i n}(l) 
		\mathfrak{l}_{n m}(q) 
		\mathfrak{t}_{mj}(m) + 
		\mathfrak{l}_{i n}(l) 
		\mathfrak{t}_{nm}(q) 
		\mathfrak{l}_{j m}(m) 
	\right] 
	c(m+q)S^\dagger( p- m)\nonumber .
	\end{eqnarray}
	Finally we have to substitute 
	$c(k)=\frac{1}{k^2}\rho(k)$ and  $\bar c(k)=\frac{1}{k^2}\int_lS(k-l)\rho(l)$ to obtain 
\begin{eqnarray}\label{AAT}
&&\left[\mathbb{A}\mathbb{A}^T-1\right]^{ab}_{ij}(k,p)\approx\int_{q,l,m}\\
&&\Bigg[\frac{1}{(k-q)^2(p+q)^2}
	[fS( k- q- l)\rho(l)] \left[
		\mathfrak{t}_{i n}(k) 
		\mathfrak{l}_{n m}(q) 
		\mathfrak{t}_{mj}(p) + 
		\mathfrak{l}_{i n}(k) 
		\mathfrak{t}_{nm}(q) 
		\mathfrak{l}_{j m}(p) 
	\right] 
	[fS(p+q- m)\rho( m)]-\nonumber\\
	&&
	\frac{1}{(l-q)^2(m+q)^2}
	[S( k- l)f \rho(l-q)] \left[
		\mathfrak{t}_{i n}(l) 
		\mathfrak{l}_{n m}(q) 
		\mathfrak{t}_{mj}(m) + 
		\mathfrak{l}_{i n}(l) 
		\mathfrak{t}_{nm}(q) 
		\mathfrak{l}_{j m}(m) 
	\right] 
	[f\rho(m+q)]S^\dagger( p- m)\Bigg]\nonumber , 
\end{eqnarray}
where we have indicated explicitly the position of the structure constant tensor $f$ for the purposes of color algebra. After shifting the momentum integration variable in the first term, { and restoring the original normalization of $b$ and $\rho$ used in the main text}  we can finally write
\begin{eqnarray}
&&\left[\mathbb{A}\mathbb{A}^T-1\right]^{ab}_{ij}(k,p)\approx g^2\int_{q,l,m}\\
	&&	\Bigg[\left[f^{acd}S^{de}( k- l) \rho^e(l-q) 	\right]\left[f^{cbf}S^{fg}( p- m)\rho^g(m+q)\right]\times\nonumber\\
&&\frac{1}{(k-q)^2(p+q)^2}\left[
		\mathfrak{t}_{i n}(k) 
		\mathfrak{l}_{n m}(q) 
		\mathfrak{t}_{mj}(p) + 
		\mathfrak{l}_{i n}(k) 
		\mathfrak{t}_{nm}(q) 
		\mathfrak{l}_{j m}(p) 
	\right] \nonumber\\
&&-	\left[S^{ac}( k- l)f^{cde} \rho^e(l-q) 	\right]\left[S^{bf}( p- m)f^{dfg}\rho^g(m+q)\right]\times\nonumber\\
	&&\frac{1}{(l-q)^2(m+q)^2}	 \left[
		\mathfrak{t}_{i n}(l) 
		\mathfrak{l}_{n m}(q) 
		\mathfrak{t}_{mj}(m) + 
		\mathfrak{l}_{i n}(l) 
		\mathfrak{t}_{nm}(q) 
		\mathfrak{l}_{j m}(m) 
	\right] \Bigg]\nonumber .
		\end{eqnarray}

		\section{Operator product expansion.} 
		\label{Ap:Op}
In this Appendix we derive the operator product expansion, i.e. expansion of
the double inclusive cross section in powers of the derivatives $\partial S$.
The physical parameter of this expansion is $Q_s^2/k^2$ or $Q_s^2/p^2$  where $Q_s$ is
the saturation momentum of the target.

	In order to organize the expansion 
	we will use the following notation $X_{[n,m,l]}$ 
	where $n$, $m$ and $l$ are integers corresponding to the orders 
 	in the expansion of the terms originating from $\mathbb{C}(k)$, 
	$\mathbb{A}\mathbb{A}^T$, and $\mathbb{C}(p)$ correspondingly.
	For example, the notation $[1,0,2]$ means the expansion of 
	$\mathbb{C}(k)$ to the first order, $\mathbb{A}\mathbb{A}^T$ to the leading order,
	$\mathbb{C}(p)$ to the second order.  

	Some preliminary formulae are in order. Define $w=-k-u$, then
	the expansion of $\mathbb{C}(k)$ involves the following or, similar 
	modulo redefinition of the variables, 
	expansion 
	\begin{eqnarray}
	\frac{u_i}{u^2}+\frac{k_i}{k^2}=\frac{(-w-k)_i}{(k+w)^2}+\frac{k_i}{k^2}
	&=&-\left[\frac{w_i}{k^2}-2\frac{k_i(k\cdot w)}{k^4}\right]\\
	&&+\left[ \frac{k_iw^2}{k^4}-4\frac{k_i(k\cdot w)^2}{k^6}+2\frac{w_i(k\cdot w)}{k^4}\right]\\
	&&-\left[-\frac{w_iw^2}{k^4}+4\frac{w_i(k\cdot w)^2}{k^6}+4\frac{k_i w^2(k\cdot w)}{k^6}-8\frac{k_i(k\cdot w)^3}{k^8}\right]\\
	&&\equiv\frac{1}{k^2}\left[-w_sC_{si}(k)+w_sw_tD_{sti}(k)-w_sw_tw_rE_{stri}(k)\right] .
	\end{eqnarray}
	Similarly for $\mathbb{C}(p)$  define $v=-p+u$. Then
	\begin{eqnarray}
	-\frac{u_i}{u^2}+\frac{p_i}{p^2}=-\frac{(p+v)_i}{(p+v)^2}+\frac{p_i}{p^2}&=&
	-\left[\frac{v_i}{p^2}-2\frac{p_i(p\cdot v)}{p^4}\right]\\
	&&+\left[ \frac{p_iv^2}{p^4}-4\frac{p_i(p\cdot v)^2}{p^6}+2\frac{v_i(p\cdot v)}{p^4}\right]\\
	&&-\left[-\frac{v_iv^2}{p^4}+4\frac{v_i(p\cdot v)^2}{p^6}+4\frac{p_i v^2(p\cdot v)}{p^6}-8\frac{p_i(p\cdot v)^3}{p^8}\right]\\
	&&\equiv\frac{1}{p^2}\left[-v_sC_{si}(p)+v_sv_tD_{sti}(p)-v_sv_tv_rE_{stri}(p)\right] .
	\end{eqnarray}
	Here
	\begin{eqnarray}
	&&C_{si}(k)=(\mathfrak{t}-\mathfrak{l})_{si}(k)\\
	&&D_{sti}(k)=\left[\delta_{st}-4\mathfrak{l}_{st}(k)\right]\frac{k_i}{k^2}+2\delta_{si}\frac{k_t}{k^2}=\delta_{st}\frac{k_i}{k^2}+2(\mathfrak{t}-\mathfrak{l})_{si}(k)\frac{k_t}{k^2}, \\
	&&E_{stri}(k)=\frac{1}{k^2}\Big[-\delta_{si}[\delta_{tr}-4\mathfrak{l}_{tr}(k)]+4\mathfrak{l}_{si}(k)(\mathfrak{t}-\mathfrak{l})_{tr}(k)\Big] . 
	\end{eqnarray}
	
	Now the expansion of 	$\mathbb{A}\mathbb{A}^T$  
	will involve 
	\begin{equation}
	\frac{1}{(l-q)^4}=\left[\frac{1}{(k-q-a)^2}\right]^2=\frac{1}{(k-q)^4}+4\frac{a\cdot(k-q)}{(k-q)^6}-2\frac{a^2}{(k-q)^6}+12\frac{[a\cdot(k-q)]^2}{(k-q)^8}
	\end{equation}
	and 
	\begin{eqnarray}
		&&\Big[\mathfrak{t}(l)\mathfrak{l}(q)\mathfrak{t}(l)+\mathfrak{l}(l)\mathfrak{t}(q)\mathfrak{l}(l)\Big]_{ij}
		=\Big[\mathfrak{l}(q)+\mathfrak{l}(l)-\mathfrak{l}(l)\mathfrak{l}(q)-\mathfrak{l}(q)\mathfrak{l}(l)\Big]_{ij}\\
		&&=\Big[\mathfrak{l}(q)+\mathfrak{l}(k)-\mathfrak{l}(k)\mathfrak{l}(q)-\mathfrak{l}(q)\mathfrak{l}(k)\Big]_{ij}\nonumber\\
	&&-\frac{1}{k^2} \Big[a_ik_j+a_jk_i-2(a\cdot k)\frac{k_ik_j}{k^2}\Big]
	\nonumber \\ &&
	+\frac{1}{k^2q^2}\Big[(q\cdot a)\left[q_ik_j+q_jk_i\right]+(q\cdot k)\left[q_ia_j+q_ja_i\right]-2(a\cdot k)(q\cdot k)\frac{q_ik_j+k_iq_j}{k^2}\Big]\nonumber\\
	&&+\frac{1}{k^2}\Big[a_ia_j-2(a\cdot k)\frac{k_ia_j+k_ja_i}{k^2}+k_ik_j\frac{4(a\cdot k)^2-k^2a^2}{k^4}\Big]\nonumber\\
	&&-\frac{1}{k^2q^2}\Bigg[(a\cdot q)[a_iq_j+q_ia_j]-
	\frac{2(a\cdot k)(k\cdot q)}{k^2}[q_ia_j+a_iq_j]
	\nonumber \\ && - 
	\frac{2(a\cdot k)(a\cdot q)}{k^2}[q_ik_j+k_iq_j]+\frac{[4(a\cdot k)^2-k^2a^2](q\cdot k)}{k^4}[q_ik_j+k_iq_j]\Bigg]\nonumber , 
	\end{eqnarray}
	where
	we used 
	\begin{eqnarray}
		\mathfrak{l}_{in}(l)=\mathfrak{l}_{in}(k-a)&=&\mathfrak{l}_{in}(k)-\frac{1}{k^2} \Big[a_ik_n+a_nk_i-2(a\cdot k)\frac{k_ik_n}{k^2}\Big] \nonumber\\
	&& +\frac{1}{k^2}\Big[a_ia_n-2(a\cdot k)\frac{k_ia_n+k_na_i}{k^2}+k_ik_n\frac{4(a\cdot k)^2-k^2a^2}{k^4}\Big] .
	\end{eqnarray}
Combining these together we get 	
	\begin{eqnarray}
		\nonumber &&	\frac{1}{(l-q)^4}\Big[\mathfrak{t}(l)\mathfrak{l}(q)\mathfrak{t}(l)+\mathfrak{l}(l)\mathfrak{t}(q)\mathfrak{l}(l)\Big]_{ij}=\\
	&&=\frac{1}{(k-q)^4}A^0_{ij}(k,q)+\frac{1}{k^2(k-q)^4}a_sA^1_{sij}(k,q)+\frac{1}{k^2(k-q)^4}a_sa_tA^2_{stij}(k,q)
	, 
	\end{eqnarray}
where we defined 
	\begin{equation}
		A^0_{ij}(k,q)=\Big[\mathfrak{l}(q)+\mathfrak{l}(k)-\mathfrak{l}(k)\mathfrak{l}(q)-\mathfrak{l}(q)\mathfrak{l}(k)\Big]_{ij},
	\end{equation}
	\begin{eqnarray}
	A^1_{sij}(k,q)&=& \mathfrak{t}_{si}(k)k_j \mathfrak{t}_{sj}(k)k_i+(\mathfrak{t}-\mathfrak{l})_{sk}(k)\Big[\mathfrak{l}_{ki}(q)k_j+\mathfrak{l}_{kj}(q)k_i\Big]+\delta_{si} \mathfrak{l}_{jk}(q)k_k
+\delta_{sj}  \mathfrak{l}_{ik}(q)k_k\nonumber\\	&+&4\frac{k^2}{(k-q)^2}(k-q)_s \Big[\mathfrak{l}(q)+\mathfrak{l}(k)-\mathfrak{l}(k)\mathfrak{l}(q)-\mathfrak{l}(q)\mathfrak{l}(k)\Big]_{ij},
	\end{eqnarray}
	\begin{eqnarray}
&&	A^2_{stij}(k,q)=-2\frac{k^2}{(k-q)^2}(1-6l)_{st}(k-q)\Big[\mathfrak{l}(q)+\mathfrak{l}(k)-\mathfrak{l}(k)\mathfrak{l}(q)-\mathfrak{l}(q)\mathfrak{l}(k)\Big]_{ij}\nonumber\\
	&+&
	[(\mathfrak{t}-\mathfrak{l})_{si}(k)(\mathfrak{t}-\mathfrak{l})_{tj}(k) - \delta_{st}\mathfrak{l}_{ij}(k)]
	\nonumber	\\ 
&-& [ (\mathfrak{t}-\mathfrak{l})_{si}(k)(\mathfrak{t}-\mathfrak{l})_{tn}(k) - \delta_{st}\mathfrak{l}_{in}(k)]\mathfrak{l}_{nj}(q)
-[ (\mathfrak{t}-\mathfrak{l})_{sn}(k)(\mathfrak{t}-\mathfrak{l})_{tj}(k) - \delta_{st}\mathfrak{l}_{jn}(k)]\mathfrak{l}_{ni}(q) \nonumber\\
&+&4\frac{(k-q)_t}{(k-q)^2}  \left( \mathfrak{t}_{si}(k)k_j \mathfrak{t}_{sj}(k)k_i+(\mathfrak{t}-\mathfrak{l})_{sk}(k)\Big[\mathfrak{l}_{ki}(q)k_j+\mathfrak{l}_{kj}(q)k_i\Big]+\delta_{si}\mathfrak{l}_{jk}(q)k_k
+\delta_{sj}\mathfrak{l}_{ik}(q)k_k \right)
.
	\end{eqnarray}
With these expressions we can now proceed to reduce term by term Eq.~(\ref{mvprojectile}). 
We start with the first term.

\subsection{$\langle 12\rangle\langle 34\rangle$.}
 Here we know that the terms that do not involve expansion of $l$ 
cancel between the two lines. Thus we only write the terms that involve the coefficients $A$ and $B$.
\begin{eqnarray}\label{1234,111}
	\langle 12\rangle\langle 34\rangle_{[1,1,1]}&=&N_c {\rm Tr}\left[S^\dagger(w)S(a)S^\dagger(p+k-a)S(v)\right]\mu^4w_sv_ta_r\frac{1}{p^2k^4(k-q)^4}C_{si}(k)A^1_{rij}(k,q)C_{tj}(p)\nonumber\\
	&=&iN_c\int_{a,w}{\rm Tr}\left[\partial_sS^\dagger(w)\partial_rS(a)S^\dagger(p+k-a)\partial_tS(-p-k-w)\right]\mu^4\frac{1}{p^2k^4(k-q)^4}C_{si}(k)A^1_{rij}(k,q)C_{tj}(p)\nonumber\\
	&=&iN_c {\rm Tr}\left[\{\partial_rSS^\dagger\}(p+k)\{\partial_tS\partial_sS^\dagger\}(-p-k)\right]\mu^4\frac{1}{p^2k^4(k-q)^4}C_{si}(k)A^1_{rij}(k,q)C_{tj}(p) .
\end{eqnarray}
The quantity under trace is a function of one momentum $(p+k)$, which is sharply peaked around $p+k=0$. We therefore can expand it in derivatives of a delta function. In general we have:
\begin{eqnarray}
f(p)g(-p)&=&\delta^2(p)\int_kf(k)g(-k)-\partial_i\delta^2(p)\int_k k _if(k)g(-k)+\frac{1}{2}\partial_i\partial_j\delta^2(p)\int_kk_ik_jf(k)g(-k)\nonumber\\
&=&\delta^2(p)\int_xf(x)g(x)+i\partial_i\delta^2(p)\int_x[\partial_if(x)]g(x)-\frac{1}{2}\partial_i\partial_j\delta^2(p)\int_x[\partial_i\partial_jf(x)]g(x) .
\end{eqnarray}
In Eq.~(\ref{1234,111}) only the derivative of the delta function survives if we assume rotational invariance of the target averages, and we therefore obtain
\begin{equation}
	\langle 12\rangle\langle 34\rangle_{[1,1,1]}=-\partial_n\delta^2(p+k)N_c\int_x {\rm Tr}\left[\{\partial_n[\partial_rSS^\dagger]\partial_tS\partial_sS^\dagger\}(x)\right]\int_q\frac{\mu^4}{p^2k^4(k-q)^4}C_{si}(k)A^1_{rij}(k,q)C_{tj}(p), 
\end{equation}

\begin{equation}
	\langle 12\rangle\langle 34\rangle_{[1,2,1]}=\delta^2(p+k)N_c\int_x {\rm Tr}\left[\{[\partial_n\partial_rS]S^\dagger\partial_tS\partial_sS^\dagger\}(x)\right]\int_q\frac{\mu^4}{p^2k^4(k-q)^4}C_{si}(k)A^2_{nrij}(k,q)C_{tj}(p), 
\end{equation}
\begin{equation}
	\langle 12\rangle\langle 34\rangle_{[2,1,1]}=-\delta^2(p+k)N_c\int_x {\rm Tr}\left[\{[\partial_rS]S^\dagger\partial_tS[\partial_n\partial_sS^\dagger]\}(x)\right]\int_q\frac{\mu^4}{p^2k^4(k-q)^4}D_{nsi}(k)A^1_{rij}(k,q)C_{tj}(p), 
\end{equation}
\begin{equation}
	\langle 12\rangle\langle 34\rangle_{[1,1,2]}=-\delta^2(p+k)N_c\int_x {\rm Tr}\left[\{[\partial_rS]S^\dagger[\partial_n\partial_tS]\partial_sS^\dagger\}(x)\right]\int_q\frac{\mu^4}{p^2k^4(k-q)^4}C_{si}(k)A^1_{rij}(k,q)D_{ntj}(p).  
\end{equation}

Substituting the condensates (see  Appendix~\ref{App:Target})  we get 
\begin{equation}
	\langle 
\langle 12\rangle\langle 34\rangle_{\rm sum}
\rangle_\alpha
=
7
N_c^3 (N_c^2-1) \frac{\lambda^4 \mu^4 }{k^6} 
\delta^2(p+k)
\int_q 
\frac{1}{(k-q)^4} 
\Bigg(
\frac32 - 2 \frac{(k\cdot q)^{2}}{ k^2 q^2} - 5 \frac{(k\times q)^2 k\cdot(k-q) } {(k-q)^2 q^2 k^2}
\Bigg) . 
\end{equation}
	
\subsection{$\langle 13\rangle\langle 24\rangle$.}
To calculate this term we define
\begin{equation}
-k+l-q=w; \ \ k-l=a; \ \ \ p-m=b;\ \ \ -p+m+q=v
\end{equation}
which gives
\begin{equation}
q-l=-(k+w); \ \ \ m+q=p+v; \ \ \ q=-(w+a)=v+b .
\end{equation}
Now
\begin{eqnarray}
&&\langle 13\rangle\langle 24\rangle_{[1,0,1]}=\frac{\mu^4}{p^4k^4}C_{si}(k)C_{tj}(p)(-k+l-q)_s(-p+m+q)_t\left[
		\mathfrak{t}_{i n}(k) 
		\mathfrak{l}_{n m}(q) 
		\mathfrak{t}_{mj}(p) + 
		\mathfrak{l}_{i n}(k) 
		\mathfrak{t}_{nm}(q) 
		\mathfrak{l}_{j m}(p) 
	\right]\times\nonumber\\
&&\Bigg[f^{acd}S^{de}(k-l)S^{\dagger ea}(-k+l-q)f^{cbf}S^{fg}(p-m)S^{\dagger gb}(-p+m+q) \nonumber \\ &&	- S^{ac}(k-l)f^{cde}S^{\dagger ea}(-k+l-q)S^{bf}(p-m)f^{dfg}S^{\dagger gb}(q-p+m)\Bigg]\nonumber\\
&&=-\frac{\mu^4}{p^4k^4}C_{si}(k)C_{tj}(p)\left[
		\mathfrak{t}_{i n}(k) 
		\mathfrak{l}_{n m}(q) 
		\mathfrak{t}_{mj}(p) + 
		\mathfrak{l}_{i n}(k) 
		\mathfrak{t}_{nm}(q) 
		\mathfrak{l}_{j m}(p) 
	\right]\times\nonumber\\
	&&\Bigg[\Big[f^{acd}S^{de}\partial_sS^{\dagger ea}\Big](-q)\Big[f^{cbf}S^{fg}\partial_tS^{\dagger gb}\Big](q)-\Big[S^{ac}f^{cde}\partial_sS^{\dagger ea}\Big](-q)	\Big[	S^{bf}f^{dfg}\partial_tS^{\dagger gb}\Big](q)\Bigg]
	\nonumber \\
	&&=-\frac{\mu^4}{p^4k^4}\left[
		\mathfrak{t}_{s n}(k) 
		\mathfrak{l}_{n m}(q) 
		\mathfrak{t}_{mj}(p) + 
		\mathfrak{l}_{i n}(k) 
		\mathfrak{t}_{nm}(q) 
		\mathfrak{l}_{t m}(p) 
	\right]\times\nonumber\\
	&&\Bigg[\Big[f^{acd}S^{de}\partial_sS^{\dagger ea}\Big](-q)\Big[f^{cbf}S^{fg}\partial_tS^{\dagger gb}\Big](q)-\Big[S^{ac}f^{cde}\partial_sS^{\dagger ea}\Big](-q)	\Big[	S^{bf}f^{dfg}\partial_tS^{\dagger gb}\Big](q)\Bigg].
		\end{eqnarray}
	
It is clear that this expression is odd under the charge conjugation transformation 
$S\rightarrow S^\dagger$, and therefore vanishes in any MV-like model. In principle
this contribution is nonzero, however it is impossible to determine its sign without 
a specific model. We therefore will not consider it any further.	
	\begin{eqnarray}
	&&\langle 13\rangle\langle 24\rangle_{[2,0,1]}=-\frac{\mu^4}{p^4k^4}D_{rsi}(k)C_{tj}(p)(-k+l-q)_r(-k+l-q)_s(-p+m+q)_t\times\nonumber\\
	&&\left[
		\mathfrak{t}_{i n}(k) 
		\mathfrak{l}_{n m}(q) 
		\mathfrak{t}_{mj}(p) + 
		\mathfrak{l}_{i n}(k) 
		\mathfrak{t}_{nm}(q) 
		\mathfrak{l}_{j m}(p) 
	\right]\times\nonumber\\
&&\Bigg[f^{acd}S^{de}(k-l)S^{\dagger ea}(-k+l-q)f^{cbf}S^{fg}(p-m)S^{\dagger gb}(-p+m+q)	\nonumber\\
&&-S^{ac}(k-l)f^{cde}S^{\dagger ea}(-k+l-q)S^{bf}(p-m)f^{dfg}S^{\dagger gb}(q-p+m)\Bigg]\nonumber\\
&&=i\frac{\mu^4}{p^4k^4}D_{rsi}(k)C_{tj}(p)\left[
		\mathfrak{t}_{i n}(k) 
		\mathfrak{l}_{n m}(q) 
		\mathfrak{t}_{mj}(p) + 
		\mathfrak{l}_{i n}(k) 
		\mathfrak{t}_{nm}(q) 
		\mathfrak{l}_{j m}(p) 
	\right]\times\nonumber\\
	&&\Bigg[\Big[f^{acd}S^{de}[\partial_r\partial_sS^{\dagger ea}]\Big](-q)\Big[f^{cbf}S^{fg}\partial_tS^{\dagger gb}\Big](q)-\Big[S^{ac}f^{cde}[\partial_r\partial_sS^{\dagger ea}]\Big](-q)	\Big[	S^{bf}f^{dfg}\partial_tS^{\dagger gb}\Big](q)\Bigg],
		\end{eqnarray}
		
\begin{eqnarray}
	&&\langle 13\rangle\langle 24\rangle_{[1,0,2]}=i\frac{\mu^4}{p^4k^4}C_{si}(k)D_{rtj}(p)\left[
		\mathfrak{t}_{i n}(k) 
		\mathfrak{l}_{n m}(q) 
		\mathfrak{t}_{mj}(p) + 
		\mathfrak{l}_{i n}(k) 
		\mathfrak{t}_{nm}(q) 
		\mathfrak{l}_{j m}(p) 
	\right]\times\nonumber\\
	&&\Bigg[\Big[f^{acd}S^{de}\partial_sS^{\dagger ea}\Big](-q)\Big[f^{cbf}S^{fg}[\partial_r\partial_tS^{\dagger gb}]\Big](q)-\Big[S^{ac}f^{cde}\partial_sS^{\dagger ea}\Big](-q)	\Big[	S^{bf}f^{dfg}[\partial_r\partial_tS^{\dagger gb}]\Big](q)\Bigg].
		\end{eqnarray}
		Adding the two together we get 
		\begin{eqnarray}
	&&\langle 13\rangle\langle 24\rangle_{[2,0,1]}+\langle 13\rangle\langle 24\rangle_{[1,0,2]}=i\frac{\mu^4}{p^4k^4}\Big[D_{rsi}(k)C_{tj}(p)+C_{ti}(k)D_{rsj}(p)\Big]\left[
		\mathfrak{t}_{i n}(k) 
		\mathfrak{l}_{n m}(q) 
		\mathfrak{t}_{mj}(p) + 
		\mathfrak{l}_{i n}(k) 
		\mathfrak{t}_{nm}(q) 
		\mathfrak{l}_{j m}(p) 
	\right]\times\nonumber\\
	&&\Bigg[\Big[f^{acd}S^{de}[\partial_r\partial_sS^{\dagger ea}]\Big](-q)\Big[f^{cbf}S^{fg}\partial_tS^{\dagger gb}\Big](q)-\Big[S^{ac}f^{cde}[\partial_r\partial_sS^{\dagger ea}]\Big](-q)	\Big[	S^{bf}f^{dfg}\partial_tS^{\dagger gb}\Big](q)\Bigg].
	\end{eqnarray}
	This expression is also odd under charge conjugation. Additionally the integral over $q$ results in an object with five rotational indices. There is no invariant tensor with five indices in 2 dimensions, therefore for a rotationally invariant target charge distribution this term vanishes.
	\begin{eqnarray}
	&&\langle 13\rangle\langle 24\rangle_{[2,0,2]}=-\frac{\mu^4}{p^4k^4}D_{lsi}(k)D_{rtj}(p)\left[
		\mathfrak{t}_{i n}(k) 
		\mathfrak{l}_{n m}(q) 
		\mathfrak{t}_{mj}(p) + 
		\mathfrak{l}_{i n}(k) 
		\mathfrak{t}_{nm}(q) 
		\mathfrak{l}_{j m}(p) 
	\right]\times\nonumber\\
	&&\Bigg[\Big[f^{acd}S^{de}[\partial_l\partial_sS^{\dagger ea}]\Big](-q)\Big[f^{cbf}S^{fg}[\partial_r\partial_tS^{\dagger gb}]\Big](q)-\Big[S^{ac}f^{cde}[\partial_l\partial_sS^{\dagger ea}]\Big](-q)	\Big[	S^{bf}f^{dfg}[\partial_r\partial_tS^{\dagger gb}]\Big](q)\Bigg]	.
	\end{eqnarray}
	Writing $S^{ab}=S^{\dagger ba}$ in the second term, we see that this expression again is antisymmetric under $S\rightarrow S^\dagger$. Therefore again, it vanishes in a MV-like model. 
	
Going back to Eq.~(\ref{mvprojectile}) it is obvious that if we set $l=k,\ p=m$ in the second chain 
of projectors and in the ``propagators'', both the expressions in  
$\langle 13\rangle\langle 24\rangle$ and  $\langle 14\rangle\langle 23\rangle$  
are odd under the charge conjugation. Thus in both these expressions we only need consider terms 
where at least one of the momenta $l$ or $m$ is expanded around $k$ and $p$ respectively.
	
Again we expand
\begin{equation}\label{exp1}
\frac{1}{(k+w)^2(p+v)^2}=\frac{1}{k^2p^2}\Bigg[1-\frac{2p\cdot v}{p^2}-\frac{2k\cdot w}{k^2}-\frac{w^2}{k^2}-\frac{v^2}{p^2}+\frac{4(k\cdot w)^2}{k^4}+\frac{4(p\cdot v)^2}{p^4}+\frac{4(k\cdot w)(p\cdot v)}{k^2p^2}\Bigg],
\end{equation}	
\begin{eqnarray}\label{exp2}
&&\frac{1}{(k-q)^2(p+q)^2}=\frac{1}{k^2p^2}\times\\
&&\Bigg[1-\frac{2p\cdot (v+b)}{p^2}-\frac{2k\cdot (w+a)}{k^2}-\frac{(w+a)^2}{k^2}-\frac{(v+b)^2}{p^2}
\nonumber \\ &&+
\frac{4(k\cdot (w+a))^2}{k^4}+\frac{4(p\cdot (v+b))^2}{p^4}+\frac{4(k\cdot (w+a))(p\cdot (v+b))}{k^2p^2}\Bigg]\nonumber
\end{eqnarray}	
and 
\begin{eqnarray}
&&\Big[\mathfrak{l}(q)-\mathfrak{l}(k-a)\mathfrak{l}(q)-\mathfrak{l}(q)\mathfrak{l}(p-b)+\mathfrak{l}(k-a)\mathfrak{l}(p-b)\Big]_{ij}=\Big[\mathfrak{l}(q)-\mathfrak{l}(k)\mathfrak{l}(q)-\mathfrak{l}(q)\mathfrak{l}(p)+\mathfrak{l}(k)\mathfrak{l}(p)\Big]_{ij}\\
&&-\frac{1}{k^2}\Big[a_ik_n+a_nk_i-2(a\cdot k)\frac{k_ik_n}{k^2}\Big]\big[\mathfrak{l}(p)-\mathfrak{l}(q)\big]_{nj}-\frac{1}{p^2}\big[\mathfrak{l}(k)-\mathfrak{l}(q)\big]_{in}\Big[b_np_j+p_nb_j-2(b\cdot p)\frac{p_np_j}{p^2}\Big]\nonumber\\
&&+\frac{1}{k^2}\Big[a_ia_n-2(a\cdot k)\frac{k_ia_n+k_na_i}{k^2}+k_ik_n\frac{4(a\cdot k)^2-k^2a^2}{k^4}\Big]\big[\mathfrak{l}(p)-\mathfrak{l}(q)\big]_{nj}\nonumber\\
&&+\frac{1}{p^2}\big[\mathfrak{l}(k)-\mathfrak{l}(q)\big]_{in}
\Big[b_nb_j-2(b\cdot p)\frac{p_jb_n+p_nb_j}{p^2}+p_jp_n\frac{4(b\cdot p)^2-p^2b^2}{p^4}\Big]\nonumber\\
&&+\frac{1}{k^2p^2}\Big[a_ik_n+a_nk_i-2(a\cdot k)\frac{k_ik_n}{k^2}\Big]\Big[b_np_j+p_nb_j-2(b\cdot p)\frac{p_np_j}{p^2}\Big].
\end{eqnarray}
	
All of this will generate different terms where the derivatives will act on different factors of $S$. It thus seems prudent to organize things according to different powers of the four momenta.
The overall factor which will multiply all the following expressions is
\begin{equation}
	-S^{ac}(a)f^{cde}S^{\dagger ea}(w)S^{bf}(b)f^{dfg}S^{\dagger gb}(v)\frac{\mu^4}{k^2p^2}={\rm Tr}[S(a)f^{d}S^{\dagger}(w)]{\rm Tr}[S(b)f^{d}S^{\dagger }(v)]\frac{\mu^4}{k^2p^2} . 
	\end{equation}
These are the contributions that arise from using only Eq.~(\ref{exp1}):	

\begin{eqnarray}\label{1half}
\frac{1}{k^4p^2}w_sw_rw_ov_t\Big[C_{si}(k)[-\delta_{ro}+4\mathfrak{l}_{ro}(k)]  +2D_{sri}(k)k_0\Big]\Big[\mathfrak{l}(q)-\mathfrak{l}(k)\mathfrak{l}(q)-\mathfrak{l}(q)\mathfrak{l}(p)+\mathfrak{l}(k)\mathfrak{l}(p)\Big]_{ij}C_{jt}(p) ,
\end{eqnarray}
\begin{eqnarray}\label{2half}
\frac{1}{p^4k^2}w_sv_rv_ov_tC_{si}(k)\Big[\mathfrak{l}(q)-\mathfrak{l}(k)\mathfrak{l}(q)-\mathfrak{l}(q)\mathfrak{l}(p)+\mathfrak{l}(k)\mathfrak{l}(p)\Big]_{ij}\Big[C_{tj}(p)(-\delta_{ro}+4\mathfrak{l}_{ro}(p) ) +	2D_{trj}(p)p_0\Big],
\end{eqnarray}
\begin{eqnarray}\label{3half}
&&\frac{2}{k^2p^2}w_sw_rv_ov_t\left\{\Big[D_{sri}(k) +C_{si}(k)\frac{k_r}{k^2}\Big] C_{tj}(p)\frac{p_o}{p^2}+C_{si}(k)\frac{k_r}{k^2}\Big[D_{toj}(p)+C_{tj}(p)\frac{p_o}{p^2}\Big]\right\}
\nonumber \\&& \times \Big[\mathfrak{l}(q)-\mathfrak{l}(k)\mathfrak{l}(q)-\mathfrak{l}(q)\mathfrak{l}(p)+\mathfrak{l}(k)\mathfrak{l}(p)\Big]_{ij},
\end{eqnarray}
\begin{eqnarray}
\frac{1}{k^4p^2}w_sw_ra_ov_t\Big[D_{sri}(k) +2C_{si}(k)\frac{k_r}{k^2}\Big]\Big[ \mathfrak{t}_{io}(k)k_n+ \mathfrak{t}_{no}(k)k_i\Big]\big[\mathfrak{l}(p)-\mathfrak{l}(q)\big]_{nj}C_{tj}(p),
\end{eqnarray}
	\begin{eqnarray}
\frac{1}{k^4p^2}w_sa_rv_ov_tC_{si}(k)\Big[ \mathfrak{t}_{ir}(k)k_n+ \mathfrak{t}_{nr}(k)k_i\Big]\big[\mathfrak{l}(p)-\mathfrak{l}(q)\big]_{nj}\Big[D_{toj}(p) +2C_{tj}(p)\frac{p_o}{p^2}\Big],
\end{eqnarray}
\begin{eqnarray}
\frac{1}{2k^4p^2}w_sa_ra_ov_tC_{si}(k)\Big[\delta_{on}\big[\delta_{ri}-4\mathfrak{l}_{ri}(k)\big]+\delta_{oi}\big[\delta_{rn}-4\mathfrak{l}_{rn}(k)\big]-2 \mathfrak{l}_{in}(k)\big[\delta_{ro}-4\mathfrak{l}_{ro}(k)\big]\big[\mathfrak{l}(p)-\mathfrak{l}(q)\big]_{nj}\Big]C_{tj}(p), 
\end{eqnarray}
	\begin{eqnarray}
\frac{1}{k^2p^4}w_sw_rb_ov_t\Big[D_{sri}(k) +2C_{si}(k)\frac{k_r}{k^2}\Big]\big[\mathfrak{l}(k)-\mathfrak{l}(q)\big]_{in}\Big[ \mathfrak{t}_{jo}(p)p_n+ \mathfrak{t}_{no}(p)p_j\Big]C_{tj}(p), 
\end{eqnarray}
		\begin{eqnarray}
\frac{1}{k^2p^4}w_sb_rv_ov_tC_{si}(k)\big[\mathfrak{l}(k)-\mathfrak{l}(q)\big]_{ni}\Big[ \mathfrak{t}_{jr}(p)p_n+ \mathfrak{t}_{nr}(p)p_j\Big]\Big[D_{toj}(p) +2C_{tj}(p)\frac{p_o}{p^2}\Big], 
\end{eqnarray}
	\begin{eqnarray}
\frac{1}{2k^2p^4}w_sb_rb_ov_tC_{si}(k)\big[\mathfrak{l}(k)-\mathfrak{l}(q)\big]_{in}\Big[\delta_{on}\big[\delta_{rj}-4\mathfrak{l}_{rj}(p)\big]+\delta_{oj}\big[\delta_{rn}-4\mathfrak{l}_{rn}(p)\big]-2 \mathfrak{l}_{jn}(p)\big[\delta_{ro}-4\mathfrak{l}_{ro}(p)\big]\Big]C_{tj}(p), 
\end{eqnarray}
\begin{eqnarray}
\frac{1}{k^4p^4}w_sa_rb_ov_tC_{si}(k)\Big[ \mathfrak{t}_{ir}(k)k_n+ \mathfrak{t}_{nr}(k)k_i\Big]\Big[ \mathfrak{t}_{jo}(p)p_n+ \mathfrak{t}_{no}(p)p_j\Big]C_{tj}(p). 
\end{eqnarray}
From the above we have to subtract terms that arise from Eq.~(\ref{exp2}), but only when they are
not accompanied by expansion of the projector part. In other words, all the terms above that 
contain at least one factor of $a$ or $b$ remain as they are, but the rest get modified by the subtraction.
The additional terms that have to be added are 
\begin{eqnarray}
 &&
	-\frac{1}{k^4p^2}\Big[w_s(w+a)_r(w+a)_ov_tC_{si}(k)[-\delta_{ro}+4\mathfrak{l}_{ro}(k)]  +w_sw_r(w+a)_ov_t2D_{sri}(k)k_0\Big]
\nonumber \\&&
\times 
\Big[\mathfrak{l}(q)-\mathfrak{l}(k)\mathfrak{l}(q)-\mathfrak{l}(q)\mathfrak{l}(p)+\mathfrak{l}(k)\mathfrak{l}(p)\Big]_{ij}C_{jt}(p) ,
\end{eqnarray}
\begin{eqnarray}
&& -\frac{1}{p^4k^2}C_{si}(k)\Big[\mathfrak{l}(q)-\mathfrak{l}(k)\mathfrak{l}(q)-\mathfrak{l}(q)\mathfrak{l}(p)+\mathfrak{l}(k)\mathfrak{l}(p)\Big]_{ij}
\nonumber \\&&
\times 
\Big[w_s(v+b)_r(v+b)_ov_tC_{tj}(p)(-\delta_{ro}+4\mathfrak{l}_{ro}(p) ) +	w_sv_r(v+b)_ov_t2D_{trj}(p)p_0\Big],
\end{eqnarray}
\begin{eqnarray}
&&-\frac{2}{k^2p^2}\Big[\mathfrak{l}(q)-\mathfrak{l}(k)\mathfrak{l}(q)-\mathfrak{l}(q)\mathfrak{l}(p)+\mathfrak{l}(k)\mathfrak{l}(p)\Big]_{ij}\times\\
&&\Bigg\{\Big[w_sw_rD_{sri}(k) +w_s(w+a)_rC_{si}(k)\frac{k_r}{k^2}\Big] (v+b)_ov_tC_{tj}(p)\frac{p_o}{p^2}
\nonumber \\&&
+
w_s(w+a)_rC_{si}(k)\frac{k_r}{k^2}\Big[v_ov_tD_{toj}(p)+(v+b)_ov_tC_{tj}(p)\frac{p_o}{p^2}\Big]\Bigg\}\nonumber. 
\end{eqnarray}

In other words  Eqs.~(\ref{1half}), (\ref{2half}) and (\ref{3half}) have to be exchanged for 
\begin{eqnarray}\label{1full}
&&-\frac{1}{k^4p^2}w_sw_ra_ov_t\Big[C_{si}(k)[-\delta_{ro}+4\mathfrak{l}_{ro}(k)]  +2D_{sri}(k)k_0\Big]\Big[\mathfrak{l}(q)-\mathfrak{l}(k)\mathfrak{l}(q)-\mathfrak{l}(q)\mathfrak{l}(p)+\mathfrak{l}(k)\mathfrak{l}(p)\Big]_{ij}C_{jt}(p)\nonumber\\
&&-\frac{1}{k^4p^2}\Big[w_sw_oa_rv_t+ w_sa_oa_rv_t\Big]C_{si}(k)[-\delta_{ro}+4\mathfrak{l}_{ro}(k)] \Big[\mathfrak{l}(q)-\mathfrak{l}(k)\mathfrak{l}(q)-\mathfrak{l}(q)\mathfrak{l}(p)+\mathfrak{l}(k)\mathfrak{l}(p)\Big]_{ij}C_{jt}(p),
\end{eqnarray}
\begin{eqnarray}\label{2full}
&&-\frac{1}{p^4k^2}w_sv_rb_ov_tC_{si}(k)\Big[\mathfrak{l}(q)-\mathfrak{l}(k)\mathfrak{l}(q)-\mathfrak{l}(q)\mathfrak{l}(p)+\mathfrak{l}(k)\mathfrak{l}(p)\Big]_{ij}\Big[C_{tj}(p)(-\delta_{ro}+4\mathfrak{l}_{ro}(p) ) +	2D_{trj}(p)p_0\Big]\nonumber\\
&&-\frac{1}{p^4k^2}\Big[w_sb_rv_ov_t+w_sb_rb_ov_t\Big]C_{si}(k)\Big[\mathfrak{l}(q)-\mathfrak{l}(k)\mathfrak{l}(q)-\mathfrak{l}(q)\mathfrak{l}(p)+\mathfrak{l}(k)\mathfrak{l}(p)\Big]_{ij}C_{tj}(p)(-\delta_{ro}+4\mathfrak{l}_{ro}(p) ) ,
\end{eqnarray}
\begin{eqnarray}\label{3full}
&&-\frac{2}{k^2p^2}\Bigg\{w_sw_rb_ov_t\Big[D_{sri}(k) +C_{si}(k)\frac{k_r}{k^2}\Big] C_{tj}(p)\frac{p_o}{p^2}+w_sa_rv_ov_tC_{si}(k)\frac{k_r}{k^2}\Big[D_{toj}(p)+C_{tj}(p)\frac{p_o}{p^2}\Big]\\
&&+\Big[w_sa_rv_ov_t +w_sa_rb_ov_t +w_sw_rb_ov_t+w_sa_rb_ov_t \Big] C_{si}(k)\frac{k_r}{k^2}C_{tj}(p)\frac{p_o}{p^2}\Big]\Bigg\}\Big[\mathfrak{l}(q)-\mathfrak{l}(k)\mathfrak{l}(q)-\mathfrak{l}(q)\mathfrak{l}(p)+\mathfrak{l}(k)\mathfrak{l}(p)\Big]_{ij}\nonumber.
\end{eqnarray}
We can simplify these expressions by using the following identities 
\begin{eqnarray}
&&(\mathfrak{t}-\mathfrak{l})(k)\Big[\mathfrak{l}(q)-\mathfrak{l}(k)\mathfrak{l}(q)-\mathfrak{l}(q)\mathfrak{l}(p)+\mathfrak{l}(k)\mathfrak{l}(p)\Big](\mathfrak{t}-\mathfrak{l})(p)=\mathfrak{l}(q)-\mathfrak{l}(k)\mathfrak{l}(q)-\mathfrak{l}(q)\mathfrak{l}(p)+\mathfrak{l}(k)\mathfrak{l}(p),\\
&&k\Big[\mathfrak{l}(q)-\mathfrak{l}(k)\mathfrak{l}(q)-\mathfrak{l}(q)\mathfrak{l}(p)+\mathfrak{l}(k)\mathfrak{l}(p)\Big](\mathfrak{t}-\mathfrak{l})(p)=k\Big[\mathfrak{l}(q)-\mathfrak{l}(k)\Big]\mathfrak{l}(p)\nonumber .
\end{eqnarray}
Thus  we have
\begin{eqnarray}\label{11full}
&&-\frac{1}{k^4p^2}w_sw_ra_ov_t\Big[[-2\delta_{ro}+12 \mathfrak{l}_{ro}(k)] \Big[\mathfrak{l}(q)-\mathfrak{l}(k)\mathfrak{l}(q)-\mathfrak{l}(q)\mathfrak{l}(p)+\mathfrak{l}(k)\mathfrak{l}(p)\Big]_{st} +2\delta_{sr}\Big[\mathfrak{l}(k)\mathfrak{l}(q)\mathfrak{l}(p)-\mathfrak{l}(k)\mathfrak{l}(p)\Big]_{ot}\Big]\nonumber\\
&&-\frac{1}{k^4p^2}\Big[w_sa_oa_rv_t\Big][-\delta_{ro}+4\mathfrak{l}_{ro}(k)] \Big[\mathfrak{l}(q)-\mathfrak{l}(k)\mathfrak{l}(q)-\mathfrak{l}(q)\mathfrak{l}(p)+\mathfrak{l}(k)\mathfrak{l}(p)\Big]_{st},
\end{eqnarray}
\begin{eqnarray}\label{12full}
&&-\frac{1}{p^4k^2}w_sv_rb_ov_t\Bigg[\Big[\mathfrak{l}(q)-\mathfrak{l}(k)\mathfrak{l}(q)-\mathfrak{l}(q)\mathfrak{l}(p)+\mathfrak{l}(k)\mathfrak{l}(p)\Big]_{st}[-2\delta_{ro}+12 \mathfrak{l}_{ro}(p) ] +	2[\mathfrak{l}(k)\mathfrak{l}(q)\mathfrak{l}(p)-\mathfrak{l}(k)\mathfrak{l}(p)]_{so}\delta_{rt}\Bigg]\nonumber\\
&&-\frac{1}{p^4k^2}\Big[w_sb_rb_ov_t\Big]\Big[\mathfrak{l}(q)-\mathfrak{l}(k)\mathfrak{l}(q)-\mathfrak{l}(q)\mathfrak{l}(p)+\mathfrak{l}(k)\mathfrak{l}(p)\Big]_{st}[-\delta_{ro}+4\mathfrak{l}_{ro}(p) ],
\end{eqnarray}
\begin{eqnarray}\label{13full}
&&-\frac{2}{k^4p^4}\Bigg\{w_sw_rb_ov_t\delta_{sr}\Big[k[\mathfrak{l}(q)-\mathfrak{l}(k)]\mathfrak{l}(p)\Big]_{t} p_o+w_sa_rv_ov_t\delta_{to}k_r \Big[\mathfrak{l}(k)[\mathfrak{l}(q)-\mathfrak{l}(p)]p\Big]_s\\
&&+\Big[4w_sa_rv_ov_t +w_sa_rb_ov_t +4w_sw_rb_ov_t+w_sa_rb_ov_t \Big] k_rp_o\Big[\mathfrak{l}(q)-\mathfrak{l}(k)\mathfrak{l}(q)-\mathfrak{l}(q)\mathfrak{l}(p)+\mathfrak{l}(k)\mathfrak{l}(p)\Big]_{st}\Bigg\}\nonumber .
\end{eqnarray}

Now collecting terms we find
\begin{eqnarray}\label{21full}
&&-{\rm Tr}[\partial_oSf^{d}\partial_s\partial_rS^{\dagger}](q){\rm Tr}[Sf^{d}\partial_tS^{\dagger }](-q)\frac{\mu^4}{k^6p^4}\times\\
&&\Big[[-2\delta_{ro}+12 \mathfrak{l}_{ro}(k)] \Big[\mathfrak{l}(q)-\mathfrak{l}(k)\mathfrak{l}(q)-\mathfrak{l}(q)\mathfrak{l}(p)+\mathfrak{l}(k)\mathfrak{l}(p)\Big]_{st} +2\delta_{sr}\Big[\mathfrak{l}(k)\mathfrak{l}(q)\mathfrak{l}(p)-\mathfrak{l}(k)\mathfrak{l}(p)\Big]_{ot}\Big]\nonumber\\
&&-{\rm Tr}[\partial_o\partial_rSf^{d}\partial_sS^{\dagger}](q){\rm Tr}[Sf^{d}\partial_tS^{\dagger }](-q)\frac{\mu^4}{k^6p^4}[-\delta_{ro}+4\mathfrak{l}_{ro}(k)] \Big[\mathfrak{l}(q)-\mathfrak{l}(k)\mathfrak{l}(q)-\mathfrak{l}(q)\mathfrak{l}(p)+\mathfrak{l}(k)\mathfrak{l}(p)\Big]_{st}\nonumber,
\end{eqnarray}
\begin{eqnarray}\label{22full}
&&-{\rm Tr}[Sf^{d}\partial_sS^{\dagger}](q){\rm Tr}[\partial_oSf^{d}\partial_r\partial_tS^{\dagger }](-q)\frac{\mu^4}{k^4p^6}\times\\
&&\Bigg[\Big[\mathfrak{l}(q)-\mathfrak{l}(k)\mathfrak{l}(q)-\mathfrak{l}(q)\mathfrak{l}(p)+\mathfrak{l}(k)\mathfrak{l}(p)\Big]_{st}[-2\delta_{ro}+12 \mathfrak{l}_{ro}(p) ] +	2[\mathfrak{l}(k)\mathfrak{l}(q)\mathfrak{l}(p)-\mathfrak{l}(k)\mathfrak{l}(p)]_{so}\delta_{rt}\Bigg]\nonumber\\
&&-{\rm Tr}[Sf^{d}\partial_sS^{\dagger}](q){\rm Tr}[\partial_o\partial_rSf^{d}\partial_tS^{\dagger }](-q)\frac{\mu^4}{k^4p^6}\Big[\mathfrak{l}(q)-\mathfrak{l}(k)\mathfrak{l}(q)-\mathfrak{l}(q)\mathfrak{l}(p)+\mathfrak{l}(k)\mathfrak{l}(p)\Big]_{st}[-\delta_{ro}+4\mathfrak{l}_{ro}(p) ]\nonumber,
\end{eqnarray}
\begin{eqnarray}\label{23full}
&&-\frac{2\mu^4}{k^6p^6}\Bigg\{  {\rm Tr}[Sf^{d}\partial_s\partial_rS^{\dagger}](q){\rm Tr}[\partial_oSf^{d}\partial_tS^{\dagger }](-q) 
\nonumber \\
&& \times 
\Bigg[\delta_{sr}\Big[k[\mathfrak{l}(q)-\mathfrak{l}(k)]\mathfrak{l}(p)\Big]_{t} p_o+4k_rp_o\Big[\mathfrak{l}(q)-\mathfrak{l}(k)\mathfrak{l}(q)-\mathfrak{l}(q)\mathfrak{l}(p)+\mathfrak{l}(k)\mathfrak{l}(p)\Big]_{st}\Bigg]\nonumber\\
&& +{\rm Tr}[\partial_rSf^{d}\partial_sS^{\dagger}](q){\rm Tr}[Sf^{d}\partial_o\partial_tS^{\dagger }](-q)
\nonumber \\
&& \times 
\Bigg[\delta_{to}k_r \Big[\mathfrak{l}(k)[\mathfrak{l}(q)-\mathfrak{l}(p)]p\Big]_s +4k_rp_o\Big[\mathfrak{l}(q)-\mathfrak{l}(k)\mathfrak{l}(q)-\mathfrak{l}(q)\mathfrak{l}(p)+\mathfrak{l}(k)\mathfrak{l}(p)\Big]_{st}\Bigg]\nonumber \\
&&+2\Big[{\rm Tr}[\partial_rSf^{d}\partial_sS^{\dagger}](q){\rm Tr}[\partial_oSf^{d}\partial_tS^{\dagger }](-q) k_rp_o\Big[\mathfrak{l}(q)-\mathfrak{l}(k)\mathfrak{l}(q)-\mathfrak{l}(q)\mathfrak{l}(p)+\mathfrak{l}(k)\mathfrak{l}(p)\Big]_{st}\Bigg\},
\end{eqnarray}
	\begin{eqnarray}
	&&{\rm Tr}[\partial_oSf^{d}\partial_s\partial_rS^{\dagger}](q){\rm Tr}[Sf^{d}\partial_tS^{\dagger }](-q)\frac{\mu^4}{k^4p^4}\times\\	
	&&\Big[\frac{1}{k^2}\Big[D_{sri}(k) +2C_{si}(k)\frac{k_r}{k^2}\Big]\Big[ \mathfrak{t}_{io}(k)k_n+ \mathfrak{t}_{no}(k)k_i\Big]\big[\mathfrak{l}(p)-\mathfrak{l}(q)\big]_{nj}C_{tj}(p)+(p\leftrightarrow k)\Big]\nonumber,
	\end{eqnarray}
	\begin{eqnarray}\label{136}
	&&{\rm Tr}[\partial_rSf^{d}\partial_sS^{\dagger}](q){\rm Tr}[Sf^{d}\partial_o\partial_tS^{\dagger }](-q)\frac{\mu^4}{k^4p^4}\times\\	
	&&\Big[\frac{1}{k^2}C_{si}(k)\Big[ \mathfrak{t}_{ir}(k)k_n+ \mathfrak{t}_{nr}(k)k_i\Big]\big[\mathfrak{l}(p)-\mathfrak{l}(q)\big]_{nj}\Big[D_{toj}(p) +2C_{tj}(p)\frac{p_o}{p^2}\Big]+(p\leftrightarrow k)\Big]\nonumber,
\end{eqnarray}
\begin{eqnarray}
	&&{\rm Tr}[\partial_o\partial_rSf^{d}\partial_sS^{\dagger}](q){\rm Tr}[Sf^{d}\partial_tS^{\dagger }](-q)\frac{\mu^4}{k^4p^4}\times\\		&&\Big[\frac{1}{2k^2}C_{si}(k)\Big[\delta_{on}\big[\delta_{ri}-4\mathfrak{l}_{ri}(k)\big]+\delta_{oi}\big[\delta_{rn}-4\mathfrak{l}_{rn}(k)\big]-2 \mathfrak{l}_{in}(k)\big[\delta_{ro}-4\mathfrak{l}_{ro}(k)\big]\Big]\big[\mathfrak{l}(p)-\mathfrak{l}(q)\big]_{nj}C_{tj}(p)+(p\leftrightarrow k)\Big]\nonumber,
\end{eqnarray}
	\begin{eqnarray}\label{138}
	&&{\rm Tr}[\partial_rSf^{d}\partial_sS^{\dagger}](q){\rm Tr}[\partial_oSf^{d}\partial_tS^{\dagger }](-q)\frac{\mu^4}{k^6p^6}\times\\	
	&&C_{si}(k)\Big[ \mathfrak{t}_{ir}(k)k_n+ \mathfrak{t}_{nr}(k)k_i\Big]\Big[ \mathfrak{t}_{jo}(p)p_n+ \mathfrak{t}_{no}(p)p_j\Big]C_{tj}(p)\nonumber .
\end{eqnarray}
	
	We can simplify these expressions by using the following
	\begin{eqnarray}\label{DS}
	&&D_{sri}(k)+2C_{si}\frac{k_r}{k^2}=4(\mathfrak{t}-\mathfrak{l})_{si}(k)\frac{k_r}{k^2}+\delta_{sr}\frac{k_i}{k^2}, \ \ \ \ \  D_{sri}(k)+C_{si}\frac{k_r}{k^2}=3(\mathfrak{t}-\mathfrak{l})_{si}(k)\frac{k_r}{k^2}+\delta_{sr}\frac{k_i}{k^2},\\
	&&C_{si}(k)[-\delta_{ro}+4\mathfrak{l}_{ro}(k)]+2D_{sri}(k)k_o=-\delta_{si}\delta_{ro}+8\delta_{si}(\mathfrak{t}-\mathfrak{l})_{ro}(k)+[{\rm antisymmetric \ under } (s\leftrightarrow o)],\\
	&&C_{si}(k)\Big[ \mathfrak{t}_{ir}(k)k_n+ \mathfrak{t}_{nr}(k)k_i\Big]= \mathfrak{t}_{sr}(k)k_n \mathfrak{t}_{nr}(k)k_s=\delta_{sr}k_n-\delta_{nr}k_s ,\\
	&&C_{si}(k)\Big[\delta_{on}\big[\delta_{ri}-4\mathfrak{l}_{ri}(k)\big]+\delta_{oi}\big[\delta_{rn}-4\mathfrak{l}_{rn}(k)\big]-2 \mathfrak{l}_{in}(k)\big[\delta_{ro}-4\mathfrak{l}_{ro}(k)\big]\Big]=
2\delta_{rs}(\mathfrak{t}-\mathfrak{l})_{on}(k)+2\delta_{ro} \mathfrak{l}_{sn}(k)\nonumber\\
&&\ \ \ \ \ \ \ \ \ \ \ \ \ \ \ \ \ \ \ \ \ \ \ \ \ \ \ \ +{\rm antisymmetric \ under (o\leftrightarrow r)}		.	\end{eqnarray}	
	
	There is an additional significant simplification. Recall that at the end of the day we need to subtract from these expressions the same expressions with $p\rightarrow -p$. Thus any term which is even under such transformation will cancel in the final answer.
	
	We will further use the model for the calculating the target correlations,
	as explained  in Appendix~\ref{App:Target}. 
	With this averaging procedure denotes by $\langle\dots \rangle_\alpha$, the
	integral involving $q^iq^j$ is always proportional to $\delta^{ij}$. This
	means that it can be calculating considering the trace over the rotational
	indices, and thus we can substitute in all expressions $
	\mathfrak{l}_{ij}(q)\rightarrow 1/2\delta_{ij}$. With this substitution all
	the integrals over $q$ simply yield local condensates of the products of
	the target $S$-matrices.
	Given this, the expressions become quite simple:
	\begin{eqnarray}
	&&
	\langle 
	\langle 
	13\rangle\langle 24
	\rangle
	\rangle_\alpha
	=\int_x{\rm Tr}[Sf^{d}\partial_s\partial_rS^{\dagger}]{\rm Tr}[Sf^{d}\partial_t\partial_oS^{\dagger }](x)\frac{2\mu^4}{k^6p^6} \times X+\\
	\nonumber\\
	&&\int_x{\rm Tr}[\partial_rSf^{d}\partial_sS^{\dagger}]{\rm Tr}[Sf^{d}\partial_o\partial_tS^{\dagger }](x)\frac{\mu^4}{k^4p^4}\times Y+\\
	&&\int_x{\rm Tr}[\partial_rSf^{d}\partial_sS^{\dagger}]{\rm Tr}[\partial_oSf^{d}\partial_tS^{\dagger }](x)\frac{\mu^4}{k^6p^6}\times\\
	&&\Big[\delta_{sr}\delta_{to}k\cdot p-\delta_{to}p_rk_s-\delta_{sr}p_tk_o+\delta_{ro}k_sp_t-2k_rp_o\Big[1-\mathfrak{l}(k)-\mathfrak{l}(p)+2\mathfrak{l}(k)\mathfrak{l}(p)\Big]_{st}\Big]\\
	&&+({\rm symmetric \ under \ p\rightarrow -p})\nonumber\\
	&&=-\frac{1}{4}N_c^3(N_c^2-1)\frac{\mu^4\lambda^4}{p^6k^6}(k\cdot p)\label{169}
		\end{eqnarray}
We did not bother to calculate the first two terms involving $X$ and $Y$, since the condensates that multiply them vanish in our model (see  Appendix~\ref{App:Target}).

\subsection{$\langle 14\rangle\langle 23\rangle$.}
For this term of the calculation we introduce
\begin{equation}
k-l=a;\  \ p-m=b; \  \ -p+l-q=v;\ \  -k+q+m=w
\end{equation}
or 
\begin{equation}
q-l=-(p+v); \ \ \ q+m=k+w; \ \ \ q=k-p+(w+b) .
\end{equation}
With these definitions the expansion now is exactly the same as for the calculation of $\langle 13\rangle\langle 24\rangle$. The only thing that changes is the common factor, which now is
\begin{equation}
{\rm Tr}[S(a)f^{d}S^{\dagger}(v)S(b)f^{d}S^{\dagger }(w)]\frac{\mu^4}{k^2p^2} .
\end{equation}
The effect of this is that all the expansion is the same, but in the final expression the derivatives acting on the two factors $S^\dagger$ have to be interchanged. 
Another major difference is that the momentum of the Fourier transform of the composite operators is not $q$, but $q+p-k$. It also means that we need to consider one extra term, of the type $[1,1,1]$, which will give a contribution proportional to the derivative of the delta function. This last contribution we will calculate last. Meanwhile we can recycle our results.

\begin{eqnarray}
&&{\rm Tr}\Big[[Sf^{d}\partial_s\partial_r\partial_oS^{\dagger}](q+p-k)[Sf^{d}\partial_tS^{\dagger }](-q-p+k)\Big]\frac{\mu^4}{k^4p^4}\times\\
&&	\Big[\frac{1}{k^2}\Big[C_{si}(k)[-\delta_{ro}+4\mathfrak{l}_{ro}(k) ] +	2D_{sri}(k)k_0\Big]\Big[\mathfrak{l}(q)-\mathfrak{l}(k)\mathfrak{l}(q)-\mathfrak{l}(q)\mathfrak{l}(p)+\mathfrak{l}(k)\mathfrak{l}(p)\Big]_{ij}C_{jt}(p)+(p\leftrightarrow k)\Big]\nonumber, 
	\end{eqnarray}
	\begin{eqnarray}
	&&{\rm Tr}\Big[[Sf^{d}\partial_s\partial_rS^{\dagger}](q+p-k)[Sf^{d}\partial_o\partial_tS^{\dagger }](-q-p+k)\Big]\frac{2\mu^4}{k^4p^4}\times\\
	&&\left\{\Big[D_{sri}(k) +C_{si}(k)\frac{k_r}{k^2}\Big] C_{tj}(p)\frac{p_o}{p^2}+C_{si}(k)\frac{k_r}{k^2}\Big[D_{toj}(p)+C_{tj}(p)\frac{p_o}{p^2}\Big]\right\}	
\Big[\mathfrak{l}(q)-\mathfrak{l}(k)\mathfrak{l}(q)-\mathfrak{l}(q)\mathfrak{l}(p)+\mathfrak{l}(k)\mathfrak{l}(p)\Big]_{ij}\nonumber, 
	\end{eqnarray}
	\begin{eqnarray}
	&&{\rm Tr}\Big[[\partial_oSf^{d}\partial_tS^{\dagger}](q+p-k)[Sf^{d}\partial_s\partial_rS^{\dagger }](-q-p+k)\Big]\frac{\mu^4}{k^4p^4}\times\\	
	&&\Big[\frac{1}{k^2}\Big[D_{sri}(k) +2C_{si}(k)\frac{k_r}{k^2}\Big]\Big[ \mathfrak{t}_{io}(k)k_n+ \mathfrak{t}_{no}(k)k_i\Big]\big[\mathfrak{l}(p)-\mathfrak{l}(q)\big]_{nj}C_{tj}(p)+(p\leftrightarrow k)\Big]\nonumber , 
	\end{eqnarray}
	\begin{eqnarray}
	&&{\rm Tr}\Big[[\partial_rSf^{d}\partial_o\partial_tS^{\dagger}](q+p-k)[Sf^{d}\partial_sS^{\dagger }](-q-p+k)\Big]\frac{\mu^4}{k^4p^4}\times\\	
	&&\Big[\frac{1}{k^2}C_{si}(k)\Big[ \mathfrak{t}_{ir}(k)k_n+ \mathfrak{t}_{nr}(k)k_i\Big]\big[\mathfrak{l}(p)-\mathfrak{l}(q)\big]_{nj}\Big[D_{toj}(p) +2C_{tj}(p)\frac{p_o}{p^2}\Big]+(p\leftrightarrow k)\Big]\nonumber , 
\end{eqnarray}
\begin{eqnarray}
	&&{\rm Tr}\Big[[\partial_o\partial_rSf^{d}\partial_tS^{\dagger}](q+p-k)[Sf^{d}\partial_sS^{\dagger }](-q-p+k)\Big]\frac{\mu^4}{k^4p^4}\times\\		&&\Big[\frac{1}{2k^2}C_{si}(k)\Big[\delta_{on}\big[\delta_{ri}-4\mathfrak{l}_{ri}(k)\big]+\delta_{oi}\big[\delta_{rn}-4\mathfrak{l}_{rn}(k)\big]-2 \mathfrak{l}_{in}(k)\big[\delta_{ro}-4\mathfrak{l}_{ro}(k)\big]\Big]\big[\mathfrak{l}(p)-\mathfrak{l}(q)\big]_{nj}C_{tj}(p)+(p\leftrightarrow k)\Big]\nonumber , 
\end{eqnarray}
	\begin{eqnarray}
	&&{\rm Tr}\Big[[\partial_rSf^{d}\partial_tS^{\dagger}](q+p-k)[\partial_oSf^{d}\partial_sS^{\dagger }](-q-p+k)\Big]\frac{\mu^4}{k^6p^6}\times\\	
	&&C_{si}(k)\Big[ \mathfrak{t}_{ir}(k)k_n+ \mathfrak{t}_{nr}(k)k_i\Big]\Big[ \mathfrak{t}_{jo}(p)p_n+ \mathfrak{t}_{no}(p)p_j\Big]C_{tj}(p)\nonumber .
\end{eqnarray}

Now for the $[1,1,1]$ contribution. Since this is going to end up contributing to a derivative with respect to $q$, we only need to keep here the terms that depend on $q$. Thus we have
\begin{equation}
w_sa_rv_t\frac{1}{k^4p^2}C_{si}(k)C_{tj}(p)\big[ \mathfrak{t}_{ri}(k)k_n+ \mathfrak{t}_{rn}(k)k_i\big] \mathfrak{l}_{nj}(q)+w_sb_rv_t\frac{1}{k^2p^4}C_{si}(k)C_{tj}(p) \mathfrak{l}_{in}(q)\big[ \mathfrak{t}_{rj}(p)p_n+ \mathfrak{t}_{rn}(p)p_j\big]
.
\end{equation}

Therefore
\begin{eqnarray}
&&i{\rm Tr}\Big[[\partial_rSf^{d}\partial_tS^{\dagger}](q+p-k)[Sf^{d}\partial_sS^{\dagger }](-q-p+k)\Big]\frac{\mu^4}{k^6p^4}
C_{si}(k)C_{tj}(p)\big[ \mathfrak{t}_{ri}(k)k_n+ \mathfrak{t}_{rn}(k)k_i\big] \mathfrak{l}_{nj}(q)+(p\leftrightarrow k)\nonumber\\
&&=-\partial_o\delta(q+p-k)\int_x{\rm Tr}\Big[\partial_o\left[\partial_rSf^{d}\partial_tS^{\dagger}\right]Sf^{d}\partial_sS^{\dagger }]\Big]\frac{\mu^4}{k^6p^4}
C_{si}(k)C_{tj}(p)\big[ \mathfrak{t}_{ri}(k)k_n+ \mathfrak{t}_{rn}(k)k_i\big] \mathfrak{l}_{nj}(q)+(p\leftrightarrow k)\nonumber\\
&&=\int_x{\rm Tr}\Big[\partial_o\left[\partial_rSf^{d}\partial_tS^{\dagger}\right]Sf^{d}\partial_sS^{\dagger }]\Big]\frac{\mu^4}{k^6p^4(k-p)^2}
\nonumber 
\\&&
\times 
C_{si}(k)C_{tj}(p)\big[ \mathfrak{t}_{ri}(k)k_n+ \mathfrak{t}_{rn}(k)k_i\big]
\big[ \mathfrak{t}_{no}(k-p)(k-p)_j+ \mathfrak{t}_{oj}(k-p)(k-p)_n\big]+(p\leftrightarrow k)
.
\end{eqnarray}

Finally we implement the $\delta$-function approximation and also use Eqs.~(\ref{DS}):
\begin{eqnarray}
&&\langle 14\rangle\langle 23\rangle=
-\int_x{\rm Tr}[\partial_oSf^{d}\partial_tS^{\dagger}Sf^{d}\partial_s\partial_rS^{\dagger }]\frac{\mu^4}{k^6p^4}\times \nonumber \\
&&\Big[[-2\delta_{ro}+12 \mathfrak{l}_{ro}(k)] \Big[\mathfrak{l}(k-p)-\mathfrak{l}(k)\mathfrak{l}(k-p)-\mathfrak{l}(k-p)\mathfrak{l}(p)+\mathfrak{l}(k)\mathfrak{l}(p)\Big]_{st} +2\delta_{sr}\Big[\mathfrak{l}(k)\mathfrak{l}(k-p)\mathfrak{l}(p)-\mathfrak{l}(k)\mathfrak{l}(p)\Big]_{ot}\Big]\nonumber\\
&&-\int_x{\rm Tr}[\partial_o\partial_rSf^{d}\partial_tS^{\dagger}Sf^{d}\partial_sS^{\dagger }]\frac{\mu^4}{k^6p^4}[-\delta_{ro}+4\mathfrak{l}_{ro}(k)] \Big[\mathfrak{l}(k-p)-\mathfrak{l}(k)\mathfrak{l}(k-p)-\mathfrak{l}(k-p)\mathfrak{l}(p)+\mathfrak{l}(k)\mathfrak{l}(p)\Big]_{st}\nonumber
\label{2211full}
\nonumber 
\\
&&-\int_x{\rm Tr}[Sf^{d}\partial_r\partial_tS^{\dagger}\partial_oSf^{d}\partial_sS^{\dagger }]\frac{\mu^4}{k^4p^6}\times \nonumber \\
&&\Bigg[\Big[\mathfrak{l}(k-p)-\mathfrak{l}(k)\mathfrak{l}(k-p)-\mathfrak{l}(k-p)\mathfrak{l}(p)+\mathfrak{l}(k)\mathfrak{l}(p)\Big]_{st}[-2\delta_{ro}+12 \mathfrak{l}_{ro}(p) ] +	2[\mathfrak{l}(k)\mathfrak{l}(k-p)\mathfrak{l}(p)-\mathfrak{l}(k)\mathfrak{l}(p)]_{so}\delta_{rt}\Bigg]\nonumber\\
&&-\int_x{\rm Tr}[Sf^{d}\partial_tS^{\dagger}\partial_o\partial_rSf^{d}\partial_sS^{\dagger }]\frac{\mu^4}{k^4p^6}\Big[\mathfrak{l}(k-p)-\mathfrak{l}(k)\mathfrak{l}(k-p)-\mathfrak{l}(k-p)\mathfrak{l}(p)+\mathfrak{l}(k)\mathfrak{l}(p)\Big]_{st}[-\delta_{ro}+4\mathfrak{l}_{ro}(p) ]\nonumber
\label{2311full}
\nonumber
\\
&&-\int_x \frac{2\mu^4}{k^6p^6}\Bigg\{  {\rm Tr}[Sf^{d}\partial_tS^{\dagger}\partial_oSf^{d}\partial_s\partial_rS^{\dagger }] \Bigg[\delta_{sr}\Big[k[\mathfrak{l}(k-p)-\mathfrak{l}(k)]\mathfrak{l}(p)\Big]_{t} p_o
\nonumber \\&&
+4k_rp_o\Big[\mathfrak{l}(k-p)-\mathfrak{l}(k)\mathfrak{l}(k-p)-\mathfrak{l}(k-p)\mathfrak{l}(p)+\mathfrak{l}(k)\mathfrak{l}(p)\Big]_{st}\Bigg]\nonumber\\
&& +{\rm Tr}[\partial_rSf^{d}\partial_o\partial_tS^{\dagger}Sf^{d}\partial_sS^{\dagger }]\Bigg[\delta_{to}k_r \Big[\mathfrak{l}(k)[\mathfrak{l}(k-p)-\mathfrak{l}(p)]p\Big]_s 
\nonumber \\&&
+4k_rp_o\Big[\mathfrak{l}(k-p)-\mathfrak{l}(k)\mathfrak{l}(k-p)-\mathfrak{l}(k-p)\mathfrak{l}(p)+\mathfrak{l}(k)\mathfrak{l}(p)\Big]_{st}\Bigg]\nonumber \\
&&+2\Big[{\rm Tr}[\partial_rSf^{d}\partial_tS^{\dagger}\partial_oSf^{d}\partial_sS^{\dagger }] k_rp_o\Big[\mathfrak{l}(k-p)-\mathfrak{l}(k)\mathfrak{l}(k-p)-\mathfrak{l}(k-p)\mathfrak{l}(p)+\mathfrak{l}(k)\mathfrak{l}(p)\Big]_{st}\Bigg\} 
\nonumber
\\
&&+\int_x{\rm Tr}\Big[Sf^{d}\partial_s\partial_rS^{\dagger}Sf^{d}\partial_t\partial_oS^{\dagger }\Big]\frac{\mu^4}{k^4p^4}
	\Bigg[6[\mathfrak{l}(k-p)(\mathfrak{t}-\mathfrak{l})(p)+(\mathfrak{t}-\mathfrak{l})(k)\mathfrak{l}(k-p)]_{st}\frac{k_rp_o}{k^2p^2}+\label{2+2}
	\nonumber
	\\
	&&2\delta_{sr}[\mathfrak{l}(k-p)\mathfrak{l}(p)]_{it}\frac{k_ip_o}{k^2p^2}+2\delta_{to}[\mathfrak{l}(k)\mathfrak{l}(k-p)]_{si}\frac{p_ik_r}{k^2p^2}+12\frac{(k\cdot p)k_sk_rp_tp_o}{k^4p^4}-2\delta_{sr}\frac{(k\cdot p)p_tp_o}{k^2p^4}-2\delta_{to}\frac{(k\cdot p)k_sk_r}{k^4p^2}\Bigg]
	\nonumber\\
	&&+{\rm Tr}\Big[\partial_oSf^{d}\partial_tS^{\dagger}Sf^{d}\partial_s\partial_rS^{\dagger }\Big]\frac{\mu^4}{k^4p^4}	\Bigg[\frac{1}{k^2}\Big[4\delta_{so}\frac{k_rk_n}{k^2}-4\delta_{no}\frac{k_sk_r}{k^2}+\delta_{sr} \mathfrak{t}_{no}(k)\Big] \mathfrak{l}_{nj}(k-p)(\mathfrak{t}-\mathfrak{l})_{jt}(p)+(k\rightarrow p)\Bigg]\nonumber\\
	&&+{\rm Tr}\Big[\partial_rSf^{d}\partial_o\partial_tS^{\dagger}Sf^{d}\partial_sS^{\dagger }\Big]\frac{\mu^4}{k^4p^4}\Big[\frac{1}{k^2}\Big[\delta_{sr}k_n-\delta_{nr}k_s\Big]\big[\mathfrak{l}(p)-\mathfrak{l}(k-p)\big]_{nj}\big[4(\mathfrak{t}-\mathfrak{l})_{tj}(p)\frac{p_o}{p^2}+\delta_{to}\frac{p_j}{p^2}\big]+(p\leftrightarrow k)\Big]\nonumber\\
	&&-{\rm Tr}\Big[\partial_o\partial_rSf^{d}\partial_tS^{\dagger}Sf^{d}\partial_sS^{\dagger }\Big]\frac{\mu^4}{k^4p^4}\times\nonumber\\		&&\Big[\frac{1}{2k^2}\Big[\delta_{os}\delta_{rn}+\delta_{on}\delta_{rs}+2\delta_{on} \mathfrak{l}_{rs}(k)+2\delta_{ro} \mathfrak{l}_{sn}(k)
-2\delta_{rn} \mathfrak{l}_{so}(k)-4\delta_{so} \mathfrak{l}_{rn}(k)\Big]\mathfrak{l}(k-p)_{nj}(\mathfrak{t}-\mathfrak{l})_{tj}(p)+(p\leftrightarrow k)\Big]\nonumber\\
	&&+{\rm Tr}\Big[\partial_rSf^{d}\partial_tS^{\dagger}\partial_oSf^{d}\partial_sS^{\dagger }\Big]\frac{\mu^4}{k^6p^6 }\Big[\delta_{sr}\delta_{to}(k\cdot p)-\delta_{sr}k_op_t-\delta_{to}k_sp_r+\delta_{or}k_sp_t\Big]\nonumber\\
	&&+\int_x{\rm Tr}\Big[\partial_o\left[\partial_rSf^{d}\partial_tS^{\dagger}\right]Sf^{d}\partial_sS^{\dagger }]\Big]\frac{\mu^4}{k^4p^4(k-p)^2}\times \nonumber\\
	&&\Bigg[\frac{1}{k^2}\big[\delta_{rs}k_n-\delta_{rn}k_s\big]
\big[ \mathfrak{t}_{no}(k-p)(k-p)_j+ \mathfrak{t}_{oj}(k-p)(k-p)_n\big](\mathfrak{t}-\mathfrak{l})_{jt}(p)+(p\leftrightarrow k)\Bigg]. 
\end{eqnarray}

Now, using the averages from Appendix~\ref{App:Target} we have:
\begin{eqnarray}
&&
\langle 
\langle 14\rangle\langle 23
\rangle
\rangle_\alpha
=
N_c^3(N_c^2-1)S_\perp\frac{\mu^4\lambda^4}{k^4p^4}\Bigg\{\\
&&-7\frac{k\cdot p}{k^2p^2}-3\frac{(k\cdot p)^3}{k^4p^4}+\frac{15}{2}\frac{k\cdot (k-p)p\cdot(k-p)}{k^2p^2(k-p)^2}+\frac{k\cdot p(p\cdot(k-p))^2}{k^2p^4(k-p)^2}+\frac{k\cdot p(k\cdot(k-p))^2}{k^4p^2(k-p)^2}\nonumber\\
&&+\frac{1}{4}\frac{(k\cdot(k-p))^2}{k^2(k-p)^2}\left(\frac{5}{k^2}-\frac{7}{p^2}\right)+\frac{1}{4}\frac{(p\cdot(k-p))^2}{p^2(k-p)^2}\left(\frac{5}{p^2}-\frac{7}{k^2}\right)+\frac{7}{2}\left(\frac{1}{k^2}+\frac{1}{p^2}\right)\frac{k\cdot pk\cdot(k-p)p\cdot(k-p)}{k^2p^2(k-p)^2}\nonumber\\
&&+\frac{3}{8}\left[\frac{k\cdot(k-p)}{k^2(k-p)^2}-\frac{p\cdot(k-p)}{p^2(k-p)^2}\right]\nonumber\Bigg\}. 
\end{eqnarray}

Finally the sum is given by   
\begin{eqnarray}
	&&\label{Eq1234plus1423}
\langle 
\langle 13\rangle\langle 24\rangle+
\langle 14\rangle\langle 23\rangle
\rangle_\alpha
=
N_c^3(N_c^2-1)S_\perp\frac{\mu^4\lambda^4}{k^4p^4}\Bigg\{\\
	&&-\frac{29}{4}\frac{k\cdot p}{k^2p^2}-3\frac{(k\cdot p)^3}{k^4p^4}+\frac{15}{2}\frac{k\cdot (k-p)p\cdot(k-p)}{k^2p^2(k-p)^2}+\frac{k\cdot p(p\cdot(k-p))^2}{k^2p^4(k-p)^2}+\frac{k\cdot p(k\cdot(k-p))^2}{k^4p^2(k-p)^2}\nonumber\\
&&+\frac{1}{4}\frac{(k\cdot(k-p))^2}{k^2(k-p)^2}\left(\frac{5}{k^2}-\frac{7}{p^2}\right)+\frac{1}{4}\frac{(p\cdot(k-p))^2}{p^2(k-p)^2}\left(\frac{5}{p^2}-\frac{7}{k^2}\right)+\frac{7}{2}\left(\frac{1}{k^2}+\frac{1}{p^2}\right)\frac{k\cdot pk\cdot(k-p)p\cdot(k-p)}{k^2p^2(k-p)^2}\nonumber\\
&&+\frac{3}{8}\left[\frac{k\cdot(k-p)}{k^2(k-p)^2}-\frac{p\cdot(k-p)}{p^2(k-p)^2}\right]\nonumber\Bigg\}. 
\end{eqnarray}

\section{Target averaging.}
\label{App:Target}

In this appendix, we compute averages over the target fields
using the averaging procedure explained in the text defined by the two point function in Eq.~\eqref{Eq:Target_Ave}. 
First lets consider 
\begin{equation}
	\langle{\rm Tr}[Sf^{d}\partial_s\partial_rS^{\dagger}]{\rm Tr}[Sf^{d}\partial_t \partial_o S^{\dagger }]\rangle_\alpha
	= \frac{1}{4} \langle E^a_s E^b_r E^c_t E^e_o \rangle_\alpha 
	{\rm Tr}[ f^d \{f^a,f^b\}] {\rm Tr}[ f^d \{f^c,f^e\}]  =0 
\end{equation}
due to ${\rm Tr}[ f^d \{f^a,f^b\}] = 0$. Analogously we have 
\begin{equation}
		\langle{\rm Tr}[\partial_r Sf^{d}\partial_sS^{\dagger}]{\rm Tr}[Sf^{d}\partial_t \partial_o S^{\dagger }]\rangle_\alpha = 0.  
\end{equation}

Proceeding further we get 
\begin{equation}
\langle{\rm Tr}[\partial_rSf^{d}\partial_sS^{\dagger}]{\rm Tr}[\partial_oSf^{d}\partial_tS^{\dagger }]\rangle_\alpha=\frac{1}{4}N_c^3(N_c^2-1)\lambda^4[\delta_{ro}\delta_{st}-\delta_{rt}\delta_{so}]
,
\end{equation}
\begin{equation}
\langle{\rm Tr}\Big[Sf^{d}\partial_s\partial_o\partial_rS^{\dagger}Sf^{d}\partial_tS^{\dagger }\Big]\rangle_\alpha=-\frac{5}{12}N_c^3(N_c^2-1)\lambda^4[\delta_{st}\delta_{ro}+\delta_{sr}\delta_{to}+\delta_{so}\delta_{tr}]
,\end{equation}
\begin{equation}
\langle{\rm Tr}\Big[Sf^{d}\partial_s\partial_rS^{\dagger}Sf^{d}\partial_t\partial_oS^{\dagger }\Big]\rangle_\alpha=-N_c^3(N_c^2-1)\lambda^4\Big[\delta_{sr}\delta_{to}+\frac{1}{8}[\delta_{st}\delta_{ro}+\delta_{so}\delta_{rt}]\Big], 
\end{equation}
\begin{equation}
\langle{\rm Tr}\Big[\partial_oSf^{d}\partial_tS^{\dagger}Sf^{d}\partial_s\partial_rS^{\dagger }\Big]\rangle_\alpha=\frac{1}{2}N_c^3(N_c^2-1)\lambda^4\Big[\delta_{ot}\delta_{sr}+\frac{3}{4}[\delta_{os}\delta_{tr}+\delta_{or}\delta_{ts}]\Big]
,
\end{equation}
\begin{equation}
\langle {\rm Tr}\Big[\partial_rSf^{d}\partial_t\partial_oS^{\dagger}Sf^{d}\partial_sS^{\dagger }\Big]\rangle_\alpha=N_c^3(N_c^2-1)\lambda^4\Big[\delta_{rs}\delta_{ot}+\frac{1}{8}[\delta_{rt}\delta_{os}+\delta_{ro}\delta_{ts}]\Big]
,
\end{equation}

\begin{equation}
	\langle {\rm Tr}\Big[\partial_o\partial_rSf^{d}\partial_tS^{\dagger}Sf^{d}\partial_sS^{\dagger }\Big]\rangle_\alpha=-\frac{1}{2}N_c^3(N_c^2-1)\lambda^4\left[\delta_{or}\delta_{ts}+\frac{3}{4}(\delta_{os}\delta_{rt}+\delta_{ot}\delta_{sr})\right]
,
\end{equation}
\begin{equation}
\langle {\rm Tr}\Big[\partial_rSf^{d}\partial_tS^{\dagger}\partial_oSf^{d}\partial_sS^{\dagger }\Big]\rangle_\alpha=-N_c^3(N_c^2-1)\lambda^4\left[\delta_{rs}\delta_{to}+\frac{1}{4}\delta_{rt}\delta_{os}\right]
,
\end{equation}
\begin{eqnarray}
\langle{\rm Tr}\Big[\partial_o\left[\partial_rSf^{d}\partial_tS^{\dagger}\right]Sf^{d}\partial_sS^{\dagger }\Big]\rangle_\alpha&=&\langle{\rm Tr}\Big[\partial_o\partial_rSf^{d}\partial_tS^{\dagger}Sf^{d}\partial_sS^{\dagger }\Big]\rangle+\langle{\rm Tr}\Big[\partial_rSf^{d}\partial_o\partial_tS^{\dagger}Sf^{d}\partial_sS^{\dagger }\Big]\rangle\nonumber\\
&=&-\frac{1}{2}N_c^3(N_c^2-1)\lambda^4\left[\delta_{or}\delta_{ts}+\frac{3}{4}(\delta_{os}\delta_{rt}+\delta_{ot}\delta_{sr})\right]
\nonumber \\ &&
+\frac{1}{8}N_c^3(N_c^2-1)\lambda^4
\left[\delta_{ro}\delta_{ts}+\delta_{rt}\delta_{os}+8\delta_{rs}\delta_{ot}\right]\nonumber\\
&=&-\frac{1}{8}N_c^3(N_c^2-1)\lambda^4\left[3\delta_{or}\delta_{ts}+2\delta_{os}\delta_{rt}-5\delta_{ot}\delta_{sr}\right]
,
\end{eqnarray}
\begin{equation}
\langle {\rm Tr}\left[[\partial_n\partial_rS]S^\dagger\partial_tS\partial_sS^\dagger\right]\rangle_\alpha=-N_c^2(N_c^2-1)\lambda^4\left[\delta_{rn}\delta_{ts}+\frac{3}{4}\left(\delta_{nt}\delta_{rs}+\delta_{ns}\delta_{rt}\right)\right],
\end{equation}
\begin{equation}
\langle{\rm Tr}\left[[\partial_rS]S^\dagger\partial_tS[\partial_n\partial_sS^\dagger]\right]\rangle_\alpha=N_c^2(N_c^2-1)\lambda^4\left[\delta_{sn}\delta_{tr}+\frac{3}{4}\left(\delta_{nr}\delta_{ts}+\delta_{nt}\delta_{rs}\right)\right],
\end{equation}
\begin{equation}
\langle {\rm Tr}\left[[\partial_rS]S^\dagger[\partial_n\partial_tS]\partial_sS^\dagger\right]	\rangle_\alpha=-N_c^2(N_c^2-1)\lambda^4\left[\delta_{tn}\delta_{rs}+\frac{3}{4}\left(\delta_{ns}\delta_{rt}+\delta_{nr}\delta_{st}\right)\right],
\end{equation}
\begin{eqnarray}
\langle {\rm Tr}\left[\partial_n[\partial_rSS^\dagger]\partial_tS\partial_sS^\dagger\right]\rangle_\alpha&=&\langle
{\rm Tr}\left[[\partial_n\partial_rS]S^\dagger\partial_tS\partial_sS^\dagger\right]\rangle+
\langle {\rm Tr}\left[\partial_rS\partial_nS^\dagger\partial_tS\partial_sS^\dagger\right]\rangle_\alpha\\
&=&-N_c^2(N_c^2-1)\lambda^4\left[\delta_{rn}\delta_{ts}+\frac{3}{4}\left(\delta_{nt}\delta_{rs}+\delta_{ns}\delta_{rt}\right)\right]
\nonumber \\ &&
+N_c^2(N_c^2-1)\lambda^4\left[\delta_{rn}\delta_{ts}+\delta_{nt}\delta_{rs}+\frac{1}{2}\delta_{ns}\delta_{rt}\right]\nonumber\\
&=&\frac{1}{4}N_c^2(N_c^2-1)\lambda^4\left[\delta_{nt}\delta_{rs}-\delta_{ns}\delta_{rt}\right]\nonumber .
\end{eqnarray}

\bibliography{oddcgc}

\begin{thebibliography}{56}
\expandafter\ifx\csname natexlab\endcsname\relax\def\natexlab#1{#1}\fi
\expandafter\ifx\csname bibnamefont\endcsname\relax
  \def\bibnamefont#1{#1}\fi
\expandafter\ifx\csname bibfnamefont\endcsname\relax
  \def\bibfnamefont#1{#1}\fi
\expandafter\ifx\csname citenamefont\endcsname\relax
  \def\citenamefont#1{#1}\fi
\expandafter\ifx\csname url\endcsname\relax
  \def\url#1{\texttt{#1}}\fi
\expandafter\ifx\csname urlprefix\endcsname\relax\def\urlprefix{URL }\fi
\providecommand{\bibinfo}[2]{#2}
\providecommand{\eprint}[2][]{\url{#2}}

\bibitem[{\citenamefont{Khachatryan et~al.}(2010)}]{Khachatryan:2010gv}
\bibinfo{author}{\bibfnamefont{V.}~\bibnamefont{Khachatryan}}
  \bibnamefont{et~al.} (\bibinfo{collaboration}{CMS}), \bibinfo{journal}{JHEP}
  \textbf{\bibinfo{volume}{09}}, \bibinfo{pages}{091} (\bibinfo{year}{2010}),
  \eprint{1009.4122}.

\bibitem[{\citenamefont{Abelev et~al.}(2013)}]{Abelev:2012ola}
\bibinfo{author}{\bibfnamefont{B.}~\bibnamefont{Abelev}} \bibnamefont{et~al.}
  (\bibinfo{collaboration}{ALICE}), \bibinfo{journal}{Phys. Lett.}
  \textbf{\bibinfo{volume}{B719}}, \bibinfo{pages}{29} (\bibinfo{year}{2013}),
  \eprint{1212.2001}.

\bibitem[{\citenamefont{Abelev et~al.}(2014)}]{Abelev:2014mda}
\bibinfo{author}{\bibfnamefont{B.~B.} \bibnamefont{Abelev}}
  \bibnamefont{et~al.} (\bibinfo{collaboration}{ALICE}),
  \bibinfo{journal}{Phys. Rev.} \textbf{\bibinfo{volume}{C90}},
  \bibinfo{pages}{054901} (\bibinfo{year}{2014}), \eprint{1406.2474}.

\bibitem[{\citenamefont{Aad et~al.}(2013{\natexlab{a}})}]{Aad:2012gla}
\bibinfo{author}{\bibfnamefont{G.}~\bibnamefont{Aad}} \bibnamefont{et~al.}
  (\bibinfo{collaboration}{ATLAS}), \bibinfo{journal}{Phys. Rev. Lett.}
  \textbf{\bibinfo{volume}{110}}, \bibinfo{pages}{182302}
  (\bibinfo{year}{2013}{\natexlab{a}}), \eprint{1212.5198}.

\bibitem[{\citenamefont{Aad et~al.}(2013{\natexlab{b}})}]{Aad:2013fja}
\bibinfo{author}{\bibfnamefont{G.}~\bibnamefont{Aad}} \bibnamefont{et~al.}
  (\bibinfo{collaboration}{ATLAS}), \bibinfo{journal}{Phys. Lett.}
  \textbf{\bibinfo{volume}{B725}}, \bibinfo{pages}{60}
  (\bibinfo{year}{2013}{\natexlab{b}}), \eprint{1303.2084}.

\bibitem[{\citenamefont{{The ATLAS collaboration}}(2014)}]{ATLAS:2014dha}
\bibinfo{author}{\bibnamefont{{The ATLAS collaboration}}}
  (\bibinfo{collaboration}{ATLAS}) (\bibinfo{year}{2014}),
  \eprint{ATLAS-CONF-2014-021}.

\bibitem[{\citenamefont{Chatrchyan et~al.}(2013{\natexlab{a}})}]{CMS:2012qk}
\bibinfo{author}{\bibfnamefont{S.}~\bibnamefont{Chatrchyan}}
  \bibnamefont{et~al.} (\bibinfo{collaboration}{CMS}), \bibinfo{journal}{Phys.
  Lett.} \textbf{\bibinfo{volume}{B718}}, \bibinfo{pages}{795}
  (\bibinfo{year}{2013}{\natexlab{a}}), \eprint{1210.5482}.

\bibitem[{\citenamefont{Chatrchyan
  et~al.}(2013{\natexlab{b}})}]{Chatrchyan:2013nka}
\bibinfo{author}{\bibfnamefont{S.}~\bibnamefont{Chatrchyan}}
  \bibnamefont{et~al.} (\bibinfo{collaboration}{CMS}), \bibinfo{journal}{Phys.
  Lett.} \textbf{\bibinfo{volume}{B724}}, \bibinfo{pages}{213}
  (\bibinfo{year}{2013}{\natexlab{b}}), \eprint{1305.0609}.

\bibitem[{\citenamefont{Aad et~al.}(2016)}]{Aad:2015gqa}
\bibinfo{author}{\bibfnamefont{G.}~\bibnamefont{Aad}} \bibnamefont{et~al.}
  (\bibinfo{collaboration}{ATLAS}), \bibinfo{journal}{Phys. Rev. Lett.}
  \textbf{\bibinfo{volume}{116}}, \bibinfo{pages}{172301}
  (\bibinfo{year}{2016}), \eprint{1509.04776}.

\bibitem[{\citenamefont{Aaboud et~al.}(2016)}]{Aaboud:2016yar}
\bibinfo{author}{\bibfnamefont{M.}~\bibnamefont{Aaboud}} \bibnamefont{et~al.}
  (\bibinfo{collaboration}{ATLAS}) (\bibinfo{year}{2016}), \eprint{1609.06213}.

\bibitem[{\citenamefont{Werner et~al.}(2014)\citenamefont{Werner, Bleicher,
  Guiot, Karpenko, and Pierog}}]{Werner:2013ipa}
\bibinfo{author}{\bibfnamefont{K.}~\bibnamefont{Werner}},
  \bibinfo{author}{\bibfnamefont{M.}~\bibnamefont{Bleicher}},
  \bibinfo{author}{\bibfnamefont{B.}~\bibnamefont{Guiot}},
  \bibinfo{author}{\bibfnamefont{I.}~\bibnamefont{Karpenko}}, \bibnamefont{and}
  \bibinfo{author}{\bibfnamefont{T.}~\bibnamefont{Pierog}},
  \bibinfo{journal}{Phys. Rev. Lett.} \textbf{\bibinfo{volume}{112}},
  \bibinfo{pages}{232301} (\bibinfo{year}{2014}), \eprint{1307.4379}.

\bibitem[{\citenamefont{Bozek et~al.}(2013)\citenamefont{Bozek, Broniowski, and
  Torrieri}}]{Bozek:2013ska}
\bibinfo{author}{\bibfnamefont{P.}~\bibnamefont{Bozek}},
  \bibinfo{author}{\bibfnamefont{W.}~\bibnamefont{Broniowski}},
  \bibnamefont{and} \bibinfo{author}{\bibfnamefont{G.}~\bibnamefont{Torrieri}},
  \bibinfo{journal}{Phys. Rev. Lett.} \textbf{\bibinfo{volume}{111}},
  \bibinfo{pages}{172303} (\bibinfo{year}{2013}), \eprint{1307.5060}.

\bibitem[{\citenamefont{Bozek and Broniowski}(2014)}]{Bozek:2014wpa}
\bibinfo{author}{\bibfnamefont{P.}~\bibnamefont{Bozek}} \bibnamefont{and}
  \bibinfo{author}{\bibfnamefont{W.}~\bibnamefont{Broniowski}},
  \bibinfo{journal}{Nucl. Phys.} \textbf{\bibinfo{volume}{A931}},
  \bibinfo{pages}{883} (\bibinfo{year}{2014}), \eprint{1407.6478}.

\bibitem[{\citenamefont{Kozlov et~al.}(2014)\citenamefont{Kozlov, Luzum,
  Denicol, Jeon, and Gale}}]{Kozlov:2014fqa}
\bibinfo{author}{\bibfnamefont{I.}~\bibnamefont{Kozlov}},
  \bibinfo{author}{\bibfnamefont{M.}~\bibnamefont{Luzum}},
  \bibinfo{author}{\bibfnamefont{G.}~\bibnamefont{Denicol}},
  \bibinfo{author}{\bibfnamefont{S.}~\bibnamefont{Jeon}}, \bibnamefont{and}
  \bibinfo{author}{\bibfnamefont{C.}~\bibnamefont{Gale}}
  (\bibinfo{year}{2014}), \eprint{1405.3976}.

\bibitem[{\citenamefont{Bzdak and Ma}(2014)}]{Bzdak:2014dia}
\bibinfo{author}{\bibfnamefont{A.}~\bibnamefont{Bzdak}} \bibnamefont{and}
  \bibinfo{author}{\bibfnamefont{G.-L.} \bibnamefont{Ma}},
  \bibinfo{journal}{Phys. Rev. Lett.} \textbf{\bibinfo{volume}{113}},
  \bibinfo{pages}{252301} (\bibinfo{year}{2014}), \eprint{1406.2804}.

\bibitem[{\citenamefont{Dumitru et~al.}(2011)\citenamefont{Dumitru, Dusling,
  Gelis, Jalilian-Marian, Lappi, and Venugopalan}}]{Dumitru:2010iy}
\bibinfo{author}{\bibfnamefont{A.}~\bibnamefont{Dumitru}},
  \bibinfo{author}{\bibfnamefont{K.}~\bibnamefont{Dusling}},
  \bibinfo{author}{\bibfnamefont{F.}~\bibnamefont{Gelis}},
  \bibinfo{author}{\bibfnamefont{J.}~\bibnamefont{Jalilian-Marian}},
  \bibinfo{author}{\bibfnamefont{T.}~\bibnamefont{Lappi}}, \bibnamefont{and}
  \bibinfo{author}{\bibfnamefont{R.}~\bibnamefont{Venugopalan}},
  \bibinfo{journal}{Phys. Lett.} \textbf{\bibinfo{volume}{B697}},
  \bibinfo{pages}{21} (\bibinfo{year}{2011}), \eprint{1009.5295}.

\bibitem[{\citenamefont{Levin and Rezaeian}(2010)}]{Levin:2010dw}
\bibinfo{author}{\bibfnamefont{E.}~\bibnamefont{Levin}} \bibnamefont{and}
  \bibinfo{author}{\bibfnamefont{A.~H.} \bibnamefont{Rezaeian}},
  \bibinfo{journal}{Phys. Rev.} \textbf{\bibinfo{volume}{D82}},
  \bibinfo{pages}{014022} (\bibinfo{year}{2010}), \eprint{1005.0631}.

\bibitem[{\citenamefont{Kovner and Lublinsky}(2011)}]{Kovner:2010xk}
\bibinfo{author}{\bibfnamefont{A.}~\bibnamefont{Kovner}} \bibnamefont{and}
  \bibinfo{author}{\bibfnamefont{M.}~\bibnamefont{Lublinsky}},
  \bibinfo{journal}{Phys. Rev.} \textbf{\bibinfo{volume}{D83}},
  \bibinfo{pages}{034017} (\bibinfo{year}{2011}), \eprint{1012.3398}.

\bibitem[{\citenamefont{Kovner and Lublinsky}(2013)}]{Kovner:2012jm}
\bibinfo{author}{\bibfnamefont{A.}~\bibnamefont{Kovner}} \bibnamefont{and}
  \bibinfo{author}{\bibfnamefont{M.}~\bibnamefont{Lublinsky}},
  \bibinfo{journal}{Int. J. Mod. Phys.} \textbf{\bibinfo{volume}{E22}},
  \bibinfo{pages}{1330001} (\bibinfo{year}{2013}), \eprint{1211.1928}.

\bibitem[{\citenamefont{Kovchegov and Wertepny}(2013)}]{Kovchegov:2012nd}
\bibinfo{author}{\bibfnamefont{Y.~V.} \bibnamefont{Kovchegov}}
  \bibnamefont{and} \bibinfo{author}{\bibfnamefont{D.~E.}
  \bibnamefont{Wertepny}}, \bibinfo{journal}{Nucl. Phys.}
  \textbf{\bibinfo{volume}{A906}}, \bibinfo{pages}{50} (\bibinfo{year}{2013}),
  \eprint{1212.1195}.

\bibitem[{\citenamefont{Dusling and Venugopalan}(2012)}]{Dusling:2012iga}
\bibinfo{author}{\bibfnamefont{K.}~\bibnamefont{Dusling}} \bibnamefont{and}
  \bibinfo{author}{\bibfnamefont{R.}~\bibnamefont{Venugopalan}},
  \bibinfo{journal}{Phys. Rev. Lett.} \textbf{\bibinfo{volume}{108}},
  \bibinfo{pages}{262001} (\bibinfo{year}{2012}), \eprint{1201.2658}.

\bibitem[{\citenamefont{Dusling and Venugopalan}(2013)}]{Dusling:2013qoz}
\bibinfo{author}{\bibfnamefont{K.}~\bibnamefont{Dusling}} \bibnamefont{and}
  \bibinfo{author}{\bibfnamefont{R.}~\bibnamefont{Venugopalan}},
  \bibinfo{journal}{Phys. Rev.} \textbf{\bibinfo{volume}{D87}},
  \bibinfo{pages}{094034} (\bibinfo{year}{2013}), \eprint{1302.7018}.

\bibitem[{\citenamefont{Dumitru and Giannini}(2015)}]{Dumitru:2014dra}
\bibinfo{author}{\bibfnamefont{A.}~\bibnamefont{Dumitru}} \bibnamefont{and}
  \bibinfo{author}{\bibfnamefont{A.~V.} \bibnamefont{Giannini}},
  \bibinfo{journal}{Nucl. Phys.} \textbf{\bibinfo{volume}{A933}},
  \bibinfo{pages}{212} (\bibinfo{year}{2015}), \eprint{1406.5781}.

\bibitem[{\citenamefont{Borghini et~al.}(2001)\citenamefont{Borghini, Dinh, and
  Ollitrault}}]{Borghini:2001vi}
\bibinfo{author}{\bibfnamefont{N.}~\bibnamefont{Borghini}},
  \bibinfo{author}{\bibfnamefont{P.~M.} \bibnamefont{Dinh}}, \bibnamefont{and}
  \bibinfo{author}{\bibfnamefont{J.-Y.} \bibnamefont{Ollitrault}},
  \bibinfo{journal}{Phys. Rev.} \textbf{\bibinfo{volume}{C64}},
  \bibinfo{pages}{054901} (\bibinfo{year}{2001}), \eprint{nucl-th/0105040}.

\bibitem[{\citenamefont{Dumitru et~al.}(2015)\citenamefont{Dumitru, McLerran,
  and Skokov}}]{Dumitru:2014yza}
\bibinfo{author}{\bibfnamefont{A.}~\bibnamefont{Dumitru}},
  \bibinfo{author}{\bibfnamefont{L.}~\bibnamefont{McLerran}}, \bibnamefont{and}
  \bibinfo{author}{\bibfnamefont{V.}~\bibnamefont{Skokov}},
  \bibinfo{journal}{Phys. Lett.} \textbf{\bibinfo{volume}{B743}},
  \bibinfo{pages}{134} (\bibinfo{year}{2015}), \eprint{1410.4844}.

\bibitem[{\citenamefont{Skokov}(2015)}]{Skokov:2014tka}
\bibinfo{author}{\bibfnamefont{V.}~\bibnamefont{Skokov}},
  \bibinfo{journal}{Phys. Rev.} \textbf{\bibinfo{volume}{D91}},
  \bibinfo{pages}{054014} (\bibinfo{year}{2015}), \eprint{1412.5191}.

\bibitem[{\citenamefont{McLerran and Skokov}(2016)}]{McLerran:2016snu}
\bibinfo{author}{\bibfnamefont{L.}~\bibnamefont{McLerran}} \bibnamefont{and}
  \bibinfo{author}{\bibfnamefont{V.}~\bibnamefont{Skokov}}
  (\bibinfo{year}{2016}), \eprint{1611.09870}.

\bibitem[{\citenamefont{Schenke et~al.}(2015)\citenamefont{Schenke,
  Schlichting, and Venugopalan}}]{Schenke:2015aqa}
\bibinfo{author}{\bibfnamefont{B.}~\bibnamefont{Schenke}},
  \bibinfo{author}{\bibfnamefont{S.}~\bibnamefont{Schlichting}},
  \bibnamefont{and}
  \bibinfo{author}{\bibfnamefont{R.}~\bibnamefont{Venugopalan}},
  \bibinfo{journal}{Phys. Lett.} \textbf{\bibinfo{volume}{B747}},
  \bibinfo{pages}{76} (\bibinfo{year}{2015}), \eprint{1502.01331}.

\bibitem[{\citenamefont{Gotsman et~al.}(2016)\citenamefont{Gotsman, Levin, and
  Maor}}]{Gotsman:2016fee}
\bibinfo{author}{\bibfnamefont{E.}~\bibnamefont{Gotsman}},
  \bibinfo{author}{\bibfnamefont{E.}~\bibnamefont{Levin}}, \bibnamefont{and}
  \bibinfo{author}{\bibfnamefont{U.}~\bibnamefont{Maor}},
  \bibinfo{journal}{Eur. Phys. J.} \textbf{\bibinfo{volume}{C76}},
  \bibinfo{pages}{607} (\bibinfo{year}{2016}), \eprint{1607.00594}.

\bibitem[{\citenamefont{Gotsman and Levin}(2016)}]{Gotsman:2016wtq}
\bibinfo{author}{\bibfnamefont{E.}~\bibnamefont{Gotsman}} \bibnamefont{and}
  \bibinfo{author}{\bibfnamefont{E.}~\bibnamefont{Levin}}
  (\bibinfo{year}{2016}), \eprint{1611.01653}.

\bibitem[{\citenamefont{Gotsman and Levin}(2017)}]{Gotsman:2017zoq}
\bibinfo{author}{\bibfnamefont{E.}~\bibnamefont{Gotsman}} \bibnamefont{and}
  \bibinfo{author}{\bibfnamefont{E.}~\bibnamefont{Levin}}
  (\bibinfo{year}{2017}), \eprint{1705.07406}.

\bibitem[{\citenamefont{Dumitru et~al.}(2008)\citenamefont{Dumitru, Gelis,
  McLerran, and Venugopalan}}]{Dumitru:2008wn}
\bibinfo{author}{\bibfnamefont{A.}~\bibnamefont{Dumitru}},
  \bibinfo{author}{\bibfnamefont{F.}~\bibnamefont{Gelis}},
  \bibinfo{author}{\bibfnamefont{L.}~\bibnamefont{McLerran}}, \bibnamefont{and}
  \bibinfo{author}{\bibfnamefont{R.}~\bibnamefont{Venugopalan}},
  \bibinfo{journal}{Nucl. Phys.} \textbf{\bibinfo{volume}{A810}},
  \bibinfo{pages}{91} (\bibinfo{year}{2008}), \eprint{0804.3858}.

\bibitem[{\citenamefont{Kovner et~al.}(2007)\citenamefont{Kovner, Lublinsky,
  and Wiedemann}}]{Kovner:2007zu}
\bibinfo{author}{\bibfnamefont{A.}~\bibnamefont{Kovner}},
  \bibinfo{author}{\bibfnamefont{M.}~\bibnamefont{Lublinsky}},
  \bibnamefont{and}
  \bibinfo{author}{\bibfnamefont{U.}~\bibnamefont{Wiedemann}},
  \bibinfo{journal}{JHEP} \textbf{\bibinfo{volume}{06}}, \bibinfo{pages}{075}
  (\bibinfo{year}{2007}), \eprint{0705.1713}.

\bibitem[{\citenamefont{Altinoluk
  et~al.}(2009{\natexlab{a}})\citenamefont{Altinoluk, Kovner, Lublinsky, and
  Peressutti}}]{Altinoluk:2009je}
\bibinfo{author}{\bibfnamefont{T.}~\bibnamefont{Altinoluk}},
  \bibinfo{author}{\bibfnamefont{A.}~\bibnamefont{Kovner}},
  \bibinfo{author}{\bibfnamefont{M.}~\bibnamefont{Lublinsky}},
  \bibnamefont{and}
  \bibinfo{author}{\bibfnamefont{J.}~\bibnamefont{Peressutti}},
  \bibinfo{journal}{JHEP} \textbf{\bibinfo{volume}{03}}, \bibinfo{pages}{109}
  (\bibinfo{year}{2009}{\natexlab{a}}), \eprint{0901.2559}.

\bibitem[{\citenamefont{Jalilian-Marian
  et~al.}(1997)\citenamefont{Jalilian-Marian, Kovner, Leonidov, and
  Weigert}}]{JalilianMarian:1997jx}
\bibinfo{author}{\bibfnamefont{J.}~\bibnamefont{Jalilian-Marian}},
  \bibinfo{author}{\bibfnamefont{A.}~\bibnamefont{Kovner}},
  \bibinfo{author}{\bibfnamefont{A.}~\bibnamefont{Leonidov}}, \bibnamefont{and}
  \bibinfo{author}{\bibfnamefont{H.}~\bibnamefont{Weigert}},
  \bibinfo{journal}{Nucl. Phys.} \textbf{\bibinfo{volume}{B504}},
  \bibinfo{pages}{415} (\bibinfo{year}{1997}), \eprint{hep-ph/9701284}.

\bibitem[{\citenamefont{Jalilian-Marian
  et~al.}(1998)\citenamefont{Jalilian-Marian, Kovner, and
  Weigert}}]{JalilianMarian:1997dw}
\bibinfo{author}{\bibfnamefont{J.}~\bibnamefont{Jalilian-Marian}},
  \bibinfo{author}{\bibfnamefont{A.}~\bibnamefont{Kovner}}, \bibnamefont{and}
  \bibinfo{author}{\bibfnamefont{H.}~\bibnamefont{Weigert}},
  \bibinfo{journal}{Phys. Rev.} \textbf{\bibinfo{volume}{D59}},
  \bibinfo{pages}{014015} (\bibinfo{year}{1998}), \eprint{hep-ph/9709432}.

\bibitem[{\citenamefont{Kovner et~al.}(2000)\citenamefont{Kovner, Milhano, and
  Weigert}}]{Kovner:2000pt}
\bibinfo{author}{\bibfnamefont{A.}~\bibnamefont{Kovner}},
  \bibinfo{author}{\bibfnamefont{J.~G.} \bibnamefont{Milhano}},
  \bibnamefont{and} \bibinfo{author}{\bibfnamefont{H.}~\bibnamefont{Weigert}},
  \bibinfo{journal}{Phys. Rev.} \textbf{\bibinfo{volume}{D62}},
  \bibinfo{pages}{114005} (\bibinfo{year}{2000}), \eprint{hep-ph/0004014}.

\bibitem[{\citenamefont{Kovner and Milhano}(2000)}]{Kovner:1999bj}
\bibinfo{author}{\bibfnamefont{A.}~\bibnamefont{Kovner}} \bibnamefont{and}
  \bibinfo{author}{\bibfnamefont{J.~G.} \bibnamefont{Milhano}},
  \bibinfo{journal}{Phys. Rev.} \textbf{\bibinfo{volume}{D61}},
  \bibinfo{pages}{014012} (\bibinfo{year}{2000}), \eprint{hep-ph/9904420}.

\bibitem[{\citenamefont{Weigert}(2002)}]{Weigert:2000gi}
\bibinfo{author}{\bibfnamefont{H.}~\bibnamefont{Weigert}},
  \bibinfo{journal}{Nucl. Phys.} \textbf{\bibinfo{volume}{A703}},
  \bibinfo{pages}{823} (\bibinfo{year}{2002}), \eprint{hep-ph/0004044}.

\bibitem[{\citenamefont{Iancu et~al.}(2001{\natexlab{a}})\citenamefont{Iancu,
  Leonidov, and McLerran}}]{Iancu:2000hn}
\bibinfo{author}{\bibfnamefont{E.}~\bibnamefont{Iancu}},
  \bibinfo{author}{\bibfnamefont{A.}~\bibnamefont{Leonidov}}, \bibnamefont{and}
  \bibinfo{author}{\bibfnamefont{L.~D.} \bibnamefont{McLerran}},
  \bibinfo{journal}{Nucl. Phys.} \textbf{\bibinfo{volume}{A692}},
  \bibinfo{pages}{583} (\bibinfo{year}{2001}{\natexlab{a}}),
  \eprint{hep-ph/0011241}.

\bibitem[{\citenamefont{Iancu et~al.}(2001{\natexlab{b}})\citenamefont{Iancu,
  Leonidov, and McLerran}}]{Iancu:2001ad}
\bibinfo{author}{\bibfnamefont{E.}~\bibnamefont{Iancu}},
  \bibinfo{author}{\bibfnamefont{A.}~\bibnamefont{Leonidov}}, \bibnamefont{and}
  \bibinfo{author}{\bibfnamefont{L.~D.} \bibnamefont{McLerran}},
  \bibinfo{journal}{Phys. Lett.} \textbf{\bibinfo{volume}{B510}},
  \bibinfo{pages}{133} (\bibinfo{year}{2001}{\natexlab{b}}),
  \eprint{hep-ph/0102009}.

\bibitem[{\citenamefont{Ferreiro et~al.}(2002)\citenamefont{Ferreiro, Iancu,
  Leonidov, and McLerran}}]{Ferreiro:2001qy}
\bibinfo{author}{\bibfnamefont{E.}~\bibnamefont{Ferreiro}},
  \bibinfo{author}{\bibfnamefont{E.}~\bibnamefont{Iancu}},
  \bibinfo{author}{\bibfnamefont{A.}~\bibnamefont{Leonidov}}, \bibnamefont{and}
  \bibinfo{author}{\bibfnamefont{L.}~\bibnamefont{McLerran}},
  \bibinfo{journal}{Nucl. Phys.} \textbf{\bibinfo{volume}{A703}},
  \bibinfo{pages}{489} (\bibinfo{year}{2002}), \eprint{hep-ph/0109115}.

\bibitem[{\citenamefont{Kharzeev et~al.}(2005)\citenamefont{Kharzeev, Levin,
  and McLerran}}]{Kharzeev:2004bw}
\bibinfo{author}{\bibfnamefont{D.}~\bibnamefont{Kharzeev}},
  \bibinfo{author}{\bibfnamefont{E.}~\bibnamefont{Levin}}, \bibnamefont{and}
  \bibinfo{author}{\bibfnamefont{L.}~\bibnamefont{McLerran}},
  \bibinfo{journal}{Nucl. Phys.} \textbf{\bibinfo{volume}{A748}},
  \bibinfo{pages}{627} (\bibinfo{year}{2005}), \eprint{hep-ph/0403271}.

\bibitem[{\citenamefont{Kovner and Lublinsky}(2005)}]{Kovner:2005nq}
\bibinfo{author}{\bibfnamefont{A.}~\bibnamefont{Kovner}} \bibnamefont{and}
  \bibinfo{author}{\bibfnamefont{M.}~\bibnamefont{Lublinsky}},
  \bibinfo{journal}{Phys. Rev.} \textbf{\bibinfo{volume}{D71}},
  \bibinfo{pages}{085004} (\bibinfo{year}{2005}), \eprint{hep-ph/0501198}.

\bibitem[{\citenamefont{McLerran and
  Venugopalan}(1994{\natexlab{a}})}]{McLerran:1993ni}
\bibinfo{author}{\bibfnamefont{L.~D.} \bibnamefont{McLerran}} \bibnamefont{and}
  \bibinfo{author}{\bibfnamefont{R.}~\bibnamefont{Venugopalan}},
  \bibinfo{journal}{Phys. Rev.} \textbf{\bibinfo{volume}{D49}},
  \bibinfo{pages}{2233} (\bibinfo{year}{1994}{\natexlab{a}}),
  \eprint{hep-ph/9309289}.

\bibitem[{\citenamefont{McLerran and
  Venugopalan}(1994{\natexlab{b}})}]{McLerran:1993ka}
\bibinfo{author}{\bibfnamefont{L.~D.} \bibnamefont{McLerran}} \bibnamefont{and}
  \bibinfo{author}{\bibfnamefont{R.}~\bibnamefont{Venugopalan}},
  \bibinfo{journal}{Phys. Rev.} \textbf{\bibinfo{volume}{D49}},
  \bibinfo{pages}{3352} (\bibinfo{year}{1994}{\natexlab{b}}),
  \eprint{hep-ph/9311205}.

\bibitem[{\citenamefont{Kovner and Lublinsky}(2006)}]{Kovner:2006wr}
\bibinfo{author}{\bibfnamefont{A.}~\bibnamefont{Kovner}} \bibnamefont{and}
  \bibinfo{author}{\bibfnamefont{M.}~\bibnamefont{Lublinsky}},
  \bibinfo{journal}{JHEP} \textbf{\bibinfo{volume}{11}}, \bibinfo{pages}{083}
  (\bibinfo{year}{2006}), \eprint{hep-ph/0609227}.

\bibitem[{\citenamefont{Kovner et~al.}(2006)\citenamefont{Kovner, Lublinsky,
  and Weigert}}]{Kovner:2006ge}
\bibinfo{author}{\bibfnamefont{A.}~\bibnamefont{Kovner}},
  \bibinfo{author}{\bibfnamefont{M.}~\bibnamefont{Lublinsky}},
  \bibnamefont{and} \bibinfo{author}{\bibfnamefont{H.}~\bibnamefont{Weigert}},
  \bibinfo{journal}{Phys. Rev.} \textbf{\bibinfo{volume}{D74}},
  \bibinfo{pages}{114023} (\bibinfo{year}{2006}), \eprint{hep-ph/0608258}.

\bibitem[{\citenamefont{Altinoluk
  et~al.}(2009{\natexlab{b}})\citenamefont{Altinoluk, Kovner, and
  Lublinsky}}]{Altinoluk:2009jf}
\bibinfo{author}{\bibfnamefont{T.}~\bibnamefont{Altinoluk}},
  \bibinfo{author}{\bibfnamefont{A.}~\bibnamefont{Kovner}}, \bibnamefont{and}
  \bibinfo{author}{\bibfnamefont{M.}~\bibnamefont{Lublinsky}},
  \bibinfo{journal}{JHEP} \textbf{\bibinfo{volume}{03}}, \bibinfo{pages}{110}
  (\bibinfo{year}{2009}{\natexlab{b}}), \eprint{0901.2560}.

\bibitem[{\citenamefont{Bartels et~al.}(2000)\citenamefont{Bartels, Lipatov,
  and Vacca}}]{Bartels:1999yt}
\bibinfo{author}{\bibfnamefont{J.}~\bibnamefont{Bartels}},
  \bibinfo{author}{\bibfnamefont{L.~N.} \bibnamefont{Lipatov}},
  \bibnamefont{and} \bibinfo{author}{\bibfnamefont{G.~P.} \bibnamefont{Vacca}},
  \bibinfo{journal}{Phys. Lett.} \textbf{\bibinfo{volume}{B477}},
  \bibinfo{pages}{178} (\bibinfo{year}{2000}), \eprint{hep-ph/9912423}.

\bibitem[{\citenamefont{Kovchegov et~al.}(2004)\citenamefont{Kovchegov,
  Szymanowski, and Wallon}}]{Kovchegov:2003dm}
\bibinfo{author}{\bibfnamefont{Y.~V.} \bibnamefont{Kovchegov}},
  \bibinfo{author}{\bibfnamefont{L.}~\bibnamefont{Szymanowski}},
  \bibnamefont{and} \bibinfo{author}{\bibfnamefont{S.}~\bibnamefont{Wallon}},
  \bibinfo{journal}{Phys. Lett.} \textbf{\bibinfo{volume}{B586}},
  \bibinfo{pages}{267} (\bibinfo{year}{2004}), \eprint{hep-ph/0309281}.

\bibitem[{\citenamefont{Golec-Biernat and
  Wusthoff}(1998)}]{GolecBiernat:1998js}
\bibinfo{author}{\bibfnamefont{K.~J.} \bibnamefont{Golec-Biernat}}
  \bibnamefont{and} \bibinfo{author}{\bibfnamefont{M.}~\bibnamefont{Wusthoff}},
  \bibinfo{journal}{Phys. Rev.} \textbf{\bibinfo{volume}{D59}},
  \bibinfo{pages}{014017} (\bibinfo{year}{1998}), \eprint{hep-ph/9807513}.

\bibitem[{\citenamefont{Golec-Biernat and
  Wusthoff}(1999)}]{GolecBiernat:1999qd}
\bibinfo{author}{\bibfnamefont{K.~J.} \bibnamefont{Golec-Biernat}}
  \bibnamefont{and} \bibinfo{author}{\bibfnamefont{M.}~\bibnamefont{Wusthoff}},
  \bibinfo{journal}{Phys. Rev.} \textbf{\bibinfo{volume}{D60}},
  \bibinfo{pages}{114023} (\bibinfo{year}{1999}), \eprint{hep-ph/9903358}.

\bibitem[{\citenamefont{Altinoluk et~al.}(2015)\citenamefont{Altinoluk,
  Armesto, Beuf, Kovner, and Lublinsky}}]{Altinoluk:2015uaa}
\bibinfo{author}{\bibfnamefont{T.}~\bibnamefont{Altinoluk}},
  \bibinfo{author}{\bibfnamefont{N.}~\bibnamefont{Armesto}},
  \bibinfo{author}{\bibfnamefont{G.}~\bibnamefont{Beuf}},
  \bibinfo{author}{\bibfnamefont{A.}~\bibnamefont{Kovner}}, \bibnamefont{and}
  \bibinfo{author}{\bibfnamefont{M.}~\bibnamefont{Lublinsky}},
  \bibinfo{journal}{Phys. Lett.} \textbf{\bibinfo{volume}{B751}},
  \bibinfo{pages}{448} (\bibinfo{year}{2015}), \eprint{1503.07126}.

\bibitem[{\citenamefont{Dumitru and Skokov}(2015)}]{Dumitru:2014vka}
\bibinfo{author}{\bibfnamefont{A.}~\bibnamefont{Dumitru}} \bibnamefont{and}
  \bibinfo{author}{\bibfnamefont{V.}~\bibnamefont{Skokov}},
  \bibinfo{journal}{Phys. Rev.} \textbf{\bibinfo{volume}{D91}},
  \bibinfo{pages}{074006} (\bibinfo{year}{2015}), \eprint{1411.6630}.

\bibitem[{\citenamefont{Aad et~al.}(2014)}]{Aad:2014lta}
\bibinfo{author}{\bibfnamefont{G.}~\bibnamefont{Aad}} \bibnamefont{et~al.}
  (\bibinfo{collaboration}{ATLAS}), \bibinfo{journal}{Phys. Rev.}
  \textbf{\bibinfo{volume}{C90}}, \bibinfo{pages}{044906}
  (\bibinfo{year}{2014}), \eprint{1409.1792}.

\end{thebibliography}

		\end{document}